\documentclass[onecolumn]{emulateapj}
\usepackage{natbib}
\usepackage{graphicx}
\usepackage{multirow}
\begin{document}
\title{Relativistic Magnetohydrodynamics: Renormalized eigenvectors \\
and full wave decomposition Riemann solver} 

\author{Luis Ant\'on\altaffilmark{1}, 
Juan A. Miralles\altaffilmark{2}, 
Jos\'e M$^{\underline{\mbox{a}}}$ Mart\'{\i}\altaffilmark{1}, 
\\ 
Jos\'e M$^{\underline{\mbox{a}}}$ Ib\'a\~nez\altaffilmark{1}, 
Miguel A. Aloy\altaffilmark{1}, Petar Mimica\altaffilmark{1}}

\altaffiltext{1}{Departamento de Astronom\'{\i}a y Astrof\'{\i}sica,
Universidad de Valencia, 46100 Burjassot (Valencia), Spain}
\altaffiltext{2}{Departament de F\'{\i}sica Aplicada,
Universitat d'Alacant, Ap. Correus 99,
03080 Alacant, Spain}


\markboth{Ant\'on et al.: RMHD: Renormalized eigenvectors and full
  wave decomposition Riemann solver}{} 

\begin{abstract}
  We obtain renormalized sets of right and left eigenvectors of the flux vector Jacobians of the relativistic MHD equations, which are regular and span a complete basis in any physical state including degenerate ones. The renormalization procedure relies on the characterization of the degeneracy types in terms of the normal and tangential components of the magnetic field to the wavefront in the fluid rest frame. Proper expressions of the renormalized eigenvectors in conserved variables are obtained through the corresponding matrix transformations. Our work completes previous analysis that present different sets of right eigenvectors for non-degenerate and degenerate states, and can be seen as a relativistic generalization of earlier work performed in classical MHD. Based on the full wave decomposition (FWD) provided by the the renormalized set of eigenvectors in conserved variables, we have also developed a linearized (Roe-type) Riemann solver. Extensive testing against one- and two-dimensional standard numerical problems allows us to conclude that our solver is very robust. When compared with a family of simpler solvers that avoid the knowledge of the full characteristic structure of the equations in the computation of the numerical fluxes, our solver turns out to be less diffusive than HLL and HLLC, and comparable in accuracy to the HLLD solver. The amount of operations needed by the FWD solver makes it less efficient computationally than those of the HLL family in one-dimensional problems. However its relative efficiency increases in multidimensional simulations.
\end{abstract}


\section{Introduction}
\label{intro}
  Astrophysical plasma is probably the most exciting, complex, physically entangled state of matter in Nature.  Even under the restrictive assumption that such plasma can be modeled as a fluid, rather than as a statistical ensemble of interacting particles, the diversity of physical phenomena it can host deserves a careful study. This is particularly true if the magnetic fields providing the coupling for the plasma constituents become dynamically
important. Additional challenges for the physical modeling of astrophysical plasma pop up if its constituents are relativistic, if the plasma moves either at bulk speeds very close to the speed of light, or if it is embedded in ultra-intense gravitational fields. This is so because relativistic flows in association with intense gravitational and magnetic fields are commonly linked up to extremely energetic phenomena in the Universe, viz. pulsar winds
\citep{RG74}, anomalous X-ray pulsars \citep{Paradijsetal95}, soft gamma-ray repeaters (SGR; \citealp{kouveliotou}), gamma-ray bursts (GRBs; \citealp{Piran04}), relativistic jets in active galactic nuclei
(AGNs; \citealp{BBR84}), etc.

  Neutron stars are among the astrophysical objects, which may harbor ultra-intense magnetic fields. Dipole magnetic fields in excess of $10^{12}\,$G have been inferred from the spin-down rates of radio pulsars \citep{TS86,Kulkarni92}. As the total magnetic energy of the star is probably much greater than the exterior dipole component, it is expected that the neutron star interior will be pierced by even larger field values ($\sim 10^{14}\,$G). Also huge magnetic fields ($\sim 10^{14}\,$G) are attributed to anomalous X-ray pulsars \citep{kouveliotou}. Extreme magnetars as SGRs host dipole magnetic fields as large as $10^{14}-10^{15}\,$G, and interior fields in the range $(5 - 10)\times 10^{15}\,$G, \citep{TD96,WT06}. The mechanism to build up such superlative fields is still disputed, but they shall be produced in the course of the collapse of massive stellar cores, not only as a result of the compression of the stellar progenitor field, but also, most likely, because of the efficient field amplification by means of the magneto-rotational instability (e.g., \citealp{Akiyamaetal03,Obergaulingeretal09}) or the dynamo action of in the proto neutron star \citep{TD93}.

  Low--mass X--ray binaries containing a black-hole candidate display high-frequency quasi-periodic oscillations (QPOs) in their spectra. The QPO activity has been theoretically attributed to the normal modes of oscillation of thick accretion disks (tori) girding stellar-mas black holes \citep{Rezzollaetal03}. Recently, \cite{Monteroetal07} have shown that such oscillations also happen if the tori (assumed to obey an ideal gas equation of state) are pierced by toroidal magnetic fields of non-negligible strength; the fundamental mode and its overtones being very similar to those of unmagnetized tori (but see \citealp{BB09}, for the damping properties that a realistic equation of estate and more elaborated microphysics may have on the normal modes of oscillation). QPOs have also been observed in the X--ray tail of giant flares of SGRs \citep{Israeletal05,SW05}. These flares are attributed to crust-quakes in magnetars \citep{DT96}. Recent general relativistic magnetohydrodynamic (GRMHD) studies show that the QPOs can be created by the oscillation modes of the crust \citep{SA07} coupled to Alfv\'en oscillations of the interior \citep{CSF09}.

  Some models of GRB progenitors link them with the formation of rapidly rotating, highly magnetized neutron stars \citep{Usov94}. Rotational and/or magnetic energy (a few $10^{51}\,$erg), decimated in scales of seconds, in a pulsar-like mechanism is released in the form of an initially opaque fireball. The $\gamma-$radiation appears via plasma instabilities when the MHD approximation of the pulsar wind breaks down at scales of $\sim 10^{13}\,$cm
\citep{Thompson94}. Alternatively, the rotational energy of a newly-born stellar-mass black hole can be tap into the plasma above the poles of the black hole by gigantic magnetic fields threading its ergosphere \citep{blandford:77}. This mechanism has been proven to work on a number of GRMHD simulations (e.g.,
\citealp{McKinney06,BK08,BK09,TMN08}) provided that both the specific angular momentum of the black hole, and the magnetic field strength and topology are adequate (see also \citealp{AO07}).  For the subclass of short-duration GRBs a merger of compact objects in a close binary is the progenitor of choice \citep[][, Aloy, Janka \& M\"uller 2005]{Paczynski86,Goodman86,Mochkovitch93}. Very recently, the role played by the magnetic field in the merger of neutron stars has started to be investigated by means of global GRMHD simulations
\citep{PR06,Rosswog07,Liuetal08,Andersonetal08,GRB09}. These simulations suggest that intense magnetic fields, in some cases exceeding by far $10^{15}\,$G, may develop because of the Kelvin-Helmholtz instabilities rising form the contact surface between the two neutron stars. However, the resolution of the aforementioned global approaches is too coarse to extract definitive conclusions, as demonstrated by local MHD simulations \citep{OAM09}.

  Magnetic fields might also be non-negligible, even at large distances from the formation site, in relativistic outflows associated to pulsar winds, AGNs, microquasars and GRBs. Even if the MHD approximation may break down when modeling pulsar magnetospheres, it is a good starting point for their study \citep{Komissarov06}. This has motivated a number of works where relativistic magneto-hydrodynamic (RMHD) simulations have been employed as tools to increase our understanding of the flow structure, observed emission, polarization and spectral properties (e.g., \citealp{Bucciantini06,Komissarov06}). The plasma flowing along the channels drilled by relativistic jets in AGNs, microquasars and GRBs should be magnetized to some extent, since we observe non-thermal spectra that can be partly produced by the synchrotron radiation emitted by leptons gyrating in a magnetic field (see, e.g., \citealp{Ferrari98} for a review). The high degree of polarization of many AGNs at parsec scale (e.g., \citealp{Gabuzda00}) advocates in favor of magnetic fields not being (only) randomly oriented. At larger scales linear polarization maps reveal a certain ordered topology, which varies from source to source (e.g.,
\citealp{WBU87,Kigureetal04}, etc.). For the galactic microquasar GRS1915+105, \cite{RM99} find magnetic fields $\sim 0.05\,$G, to account for the observed spectrum. Likewise, the ejecta associated to GRBs might be moderately or highly magnetized. The paucity of optical flashes in the afterglow of most GRBs can be readily explained if the magnetization of the flow (measured by the Poynting to inertial flux ratio) is close to unity (e.g., \citealp{GMA08,MGA09}).


  The necessity to model the aforementioned astrophysical scenarios in the framework of (general) RMHD, together with the fast increase in computing power, is pushing towards the development of more and more accurate and efficient numerical algorithms. In the last years, considerable progress has been achieved in numerical special RMHD, mainly trough the extension of high-resolution shock-capturing (HRSC) methods, after the success of this kind of techniques in special relativistic hydrodynamics (see, e.g., \citealp{martilr:03}). HRSC methods are written in conservation form and the time evolution of zone averaged state variables is governed by some functions, the
numerical fluxes, evaluated at zone interfaces. In the so called Godunov-type methods, an important subsample of HRSC methods, numerical fluxes are evaluated through the exact or approximate solution of the Riemann problem, the decay of an initial discontinuity between constant states. Despite the fact that such an analytical solution in RMHD is known (\citealp{romero} for some particular orientation of the magnetic field and the fluid velocity field; \citealp{GR06}), approximate algorithms are usually preferred (because of their larger numerical efficiency). \citet{komissarov99}, \citet{Balsara01}, and \citet{koldoba} developed independent algorithms ({\it Roe-type}
algorithms; see \citealp{martilr:03}) based on linearized Riemann solvers relying on the characteristic structure of the RMHD equations.

  More recently, \citet{delzanna03} have developed a third order HRSC scheme for RMHD which does not require the knowledge of the characteristic structure of the equations to obtain the numerical fluxes. The algorithm follows the \citeauthor{harten:83} (1983; HLL) approach that approximates the full structure of the Riemann solution (with seven waves separating the two initial states plus six intermediate states) by just two limiting waves with a single intermediate state. The algorithms developed by \cite{Leismann05} and \cite{HKM08} are also based in the simple HLL approach. In an effort to improve the accuracy of the HLL strategy, \cite{MB06} and \cite{HJ07} have extended to RMHD the so called HLLC scheme (initially devised for the Euler equations by \cite{TSS94}, and extended by \cite{Gurski04} and \cite{Li05} to classical MHD). In the HLLC scheme, the middle (entropy) wave is also captured. Finally, \citeauthor{MUB09} (2009; MUB09) have developed a five-wave HLL solver for RMHD to capture Alfv\'en discontinuities.


  Following the trail of special relativistic MHD, numerical GRMHD is blossoming. \cite{komissarov05} has extended his code based in the linearized Riemann solver to GRMHD using the approach described in \cite{pons:98}.  The codes developed by \cite{gammie:03}, \cite{Mizunoetal06}, \cite{TMN07}, and \cite{delzanna07} are based in
the HLL approach. The interested reader is addressed to the review of \cite{fontlr:08} for an up-to-date list and description of codes in both general relativistic hydrodynamics and magnetohydrodynamics, including approaches not mentioned in this introduction, as symmetric TVD schemes \citep{koide99} or artificial viscosity methods
\cite{devilliers:03}.


  The purpose of the present paper is twofold. On one hand, to present a {\it renormalized} set of right and left eigenvectors of the flux vector Jacobians of the RMHD equations which are regular and span a complete basis in any physical state, including degenerate states. On the other hand, to evaluate numerically the performance of a RMHD Riemann solver based on the aforementioned spectral decomposition. Both the theoretical analysis and the numerical applications presented in this paper are based on the work developed by \cite{Anton08}. Moreover, the numerical procedure described in this paper was already used in GRMHD calculations by \cite{Antonetal06} using the method described in \cite{pons:98} to extend the RMHD solver to GRMHD. Our numerical approach deviates in several aspects from previous works based on linearized Riemann solver approaches \citep[i.e.,][]{komissarov99, Balsara01, koldoba}. First, numerical fluxes are computed from the spectral decomposition in conserved variables. Second, we present explicit expressions also for the left eigenvectors. Third, and more important, we have extended \cite{BW88} classical MHD strategy to relativistic flows, giving sets of right and left eigenvectors which are well defined through degenerate states.


  The organization of the paper is as follows. In \S~2 the equations of RMHD are presented and the suitable definitions of variables given. In \S~3 the mathematical structure of the equations is reviewed. Special attention in this section is paid to the characterization of degeneracies and the procedure of renormalization of the right and left eigenvectors in covariant variables. Sect.~4 includes a brief note concerning the non-convex character of relativistic MHD. Sects.~5 and 6 are devoted to obtain, respectively, the renormalized right and left eigenvectors in conserved variables. Sect.~7 summarizes the properties of the numerical code based in the full wave decomposition Riemann solver built from the renormalized eigenvectors presented in the preceding sections. A thorough study of the performance of the code, based on a battery of one and two-dimensional tests and the comparison with other numerical strategies, is presented in \S~8. Sect.~9 presents the conclusions of our work.

\section{The equations of ideal relativistic magnetohydrodynamics}

  Let $J^{\mu}$, $T^{\mu \nu}$ and $^*F^{\mu \nu}$ ($\mu, \nu=0,1,2,3$) be the components of the rest-mass current density, the energy--momentum tensor and the Maxwell dual tensor of an ideal magneto-fluid, respectively

\begin{equation}
J^\mu = \rho u^\mu
\end{equation}

\begin{equation}
T^{\mu \nu} = \rho h^* u^\mu u^\nu + g^{\mu \nu} p^* - b^\mu b^\nu
\end{equation}

\begin{equation}
^*F^{\mu \nu} = u^\mu b^\nu - u^\nu b^\mu,
\end{equation}

\noindent
where $\rho$ is the proper rest-mass density, $h^* =1 + \epsilon + p/\rho + b^2/\rho$ is the specific enthalpy including the contribution from the magnetic field ($b^2$ stands for $b^\mu b_\mu$), $\epsilon$ is the specific internal energy, $p$ the thermal pressure, $p^* = p + b^2/2$ the total pressure, and $g^{\mu \nu}$ the metric of the space-time where the fluid evolves. Throughout the paper we use units in which the speed of light is $c=1$ and the $(4 \pi)^{1/2}$ factor is absorbed in the definition of the magnetic field.

  The four-vectors representing the fluid velocity, $u^\mu$, and the magnetic field measured in the fluid rest frame, $b^\mu$, satisfy the conditions $u^\mu u_\mu = -1$ and $u^\mu b_\mu = 0$, and there is an equation of state relating the thermodynamics variables, $p$, $\rho$ and $\epsilon$,  $p = p(\rho, \epsilon)$. All the discussion will be valid for a general equation of state but results will be shown for an ideal gas, for which $p = (\gamma - 1) \rho \epsilon$, where $\gamma$ is the adiabatic exponent.

  The equations of ideal RMHD correspond to the conservation of rest-mass and energy-momentum, and the Maxwell equations. In a flat space-time and Cartesian coordinates, these equations read:

\begin{equation}
\label{cont}
J^\mu_{\,\,\,\,,\mu} = 0
\end{equation}

\begin{equation}
\label{e-mom}
T^{\mu \nu}_{\,\,\,\,\,\,,\mu} = 0
\end{equation}

\begin{equation}
\label{Maxwell}
^*F^{\mu \nu}_{\,\,\,\,\,\,\,\,,\mu} = 0,
\end{equation}

\noindent
where subscript $_{,\mu}$ denotes partial derivative with respect to the corresponding coordinate, $(t,x,y,z)$, and the standard Einstein sum convention is assumed. Greek indices will run from 0 to 3 (or from $t$ to $z$) while Roman run from 1 to 3 (or from $x$ to $z$). 

  The above system can be written as a system of conservation laws as follows

\begin{equation}
\frac{\partial {\bf U}}{\partial t} +
\frac{\partial {\bf F}^{i}}{\partial x^{i}} = 0,
\label{eq:fundsystem}
\end{equation}

\noindent
where the state vector, ${\bf U}$, and the fluxes, ${\bf F}^i$ ($i=1,2,3$ or $i=x,y,z$), are the following column vectors,

\begin{eqnarray}
{\bf U} & = & \left(\begin{array}{c}
  D    \\
  S^j  \\
  \tau \\
  B^k
\end{array}\right),
\label{state_vector}
\end{eqnarray}

\begin{eqnarray}
{\mathbf F}^i & =& \left(\begin{array}{c}
  D v^i \\
  S^j v^i + p^{*} \delta^{ij} - b^j B^i/W \\
  \tau v^i + p^{*} v^i - b^0 B^i/W \\
  v^i B^k - v^k B^i
\end{array}\right).
\label{flux2}
\end{eqnarray}

  In the preceding equations, $D$, $S^j$ and $\tau$ stand, respectively for the rest-mass density, the momentum density of the magnetized
fluid in the $j$-direction and its total energy density as measured in the laboratory (i.e., Eulerian) frame,

\begin{equation}
\label{eq:D}
  D = \rho W,
\end{equation}

\begin{equation}
\label{eq:Sj}
  S^j = \rho h^* W^2 v^j - b^0 b^j,
\end{equation}

\begin{equation}
\label{eq:tau}
  \tau = \rho h^* W^2 - p^* - (b^0)^2 - D.
\end{equation}

\noindent
Quantities $v^i$ stand for the components of the fluid velocity trivector as measured in the laboratory frame. They are related with the components of the fluid four-velocity according to the following expression $u^\mu = W(1, v^x, v^y, v^z)$, where $W$ is the flow Lorentz factor, $W^2=1/(1-v^i v_i)$.

  The following fundamental relations hold between the components of the magnetic field four-vector in the comoving frame and the three vector components $B^i$ measured in the laboratory frame,

\begin{eqnarray}
\label{b0}
  b^0 & = & W\, {\bf B} \cdot {\bf v} \ , \\
  \label{bi}
  b^i & = & \frac{B^i}{W} + b^0 v^i \ ,
\end{eqnarray}
${\bf v}$ and ${\bf B}$ being, respectively, the tri-vectors $(v^x,v^y,v^z)$ and $(B^x,B^y,B^z)$.

Finally, the square of the modulus of the magnetic field can be written as

\begin{equation}
  b^2 = \frac{{\bf B}^2}{W^2} + ({\bf B} \cdot {\bf v})^2 \ .
\end{equation}

  The preceding system must be complemented with the time component of equation~(\ref{Maxwell}), that becomes the usual divergence constraint

\begin{equation}
  \label{eq:divb}
  \frac{\partial B^i}{\partial x^i} = 0\;, 
\end{equation}

\noindent
which should be fulfilled at all times.

  Fluxes ${\mathbf F}^i$ ($i = x,y,z$) are functions of the conserved variables, ${\mathbf U}$, although for the RMHD this dependence, in general, can not be expressed explicitly. It is therefore necessary to introduce another set of variables, the so-called {\it primitive variables}, derived from the conserved ones, in terms of which the fluxes can be computed explicitly. The set of primitive variables used along this work is written as the column vector

\begin{equation}
\label{eq:primitive}
  {\bf V} = (\rho, p, v^x, v^y, v^z, B^x, B^y, B^z)^T,
\end{equation}

\noindent
where the superscript $^{\rm T}$ stands for the transposition. The transformation between conserved and primitive variables is done from the following expressions (see Komissarov 1999, Leismann et al. 2005), obtained,  after some algebra, from equations~(\ref{eq:Sj}), (\ref{eq:tau}), (\ref{b0}) and (\ref{bi}), 

\begin{equation}
  {\bf S}^2 = (Z + {\bf B}^2)^2 \frac{W^2-1}{W^2} - (2Z + {\bf B}^2) \frac {({\bf B} \cdot {\bf S})^2}{Z^2},
\label{eq:s2}
\end{equation}

\begin{equation}
  \tau = Z + {\bf B}^2 - p - \frac{{\bf B}^2}{2W^2} - \frac{({\bf B} \cdot {\bf S})^2}{2Z^2} - D,
\label{eq:tauZ}
\end{equation}

\noindent
where ${\bf S}$ is the tri-vector $(S^x,S^y,S^z)$ and $Z = \rho h W^2$.

  Equations~(\ref{eq:s2}) and (\ref{eq:tauZ}), together with the definitions of $D$ (equation \ref{eq:D}) and $Z$, form a system for the unknowns $\rho$, $p$ and $W$, assuming the function $h = h(\rho, p)$ is provided. In our calculations, since we restrict ourselves to an ideal gas equation of state, $h = 1 + \gamma p/\rho(\gamma-1)$.

\section{Characteristic structure of the RMHD equations} \label{s:csrmhde}

 The hyperbolicity of the equations of RMHD including the derivation of wavespeeds and the corresponding eigenvectors, and the analysis of various degeneracies has been studied by \cite{AP87} and reviewed by \cite{anile}. These authors did their analysis in a covariant framework, using a set of variables, the so-called covariant variables, in which, the vector of unknows, 

\begin{equation}
  {\bf \tilde{U}} = (u^\mu, b^\mu, p, s)^T,
\end{equation}

\noindent
is augmented to 10 variables, where $s$ is the specific entropy. The discussion has been more recently remasterized and cast in a form more suitable for numerical applications by \cite{komissarov99} (see also \citealp{Balsara01}). We shall review it here.

  In terms of variables ${\bf \tilde{U}}$ \citep[see][]{anile}, the system of RMHD equations can be written as a quasi-linear system of the form

\begin{equation}
  {\cal A}^\mu {\bf \tilde{U}}_{; \mu}= 0, 
\label{amuab}
\end{equation}

\noindent
where $_{; \mu}$ stands for the covariant derivative, and the $10\times 10$ matrices ${\cal A}^{\mu}$ are given by

\begin{eqnarray}
  {\cal A}^{\mu} = \left( \begin{array}{cccc} {\cal E}
  u^\mu \delta^{\alpha}_{ \beta}\; & -b^{\mu}\delta^{\alpha}_{\beta} +
  P^{\alpha\mu}b_\beta & l^{\alpha\mu} & 0^{\alpha\mu} \\ 
  b^\mu \delta^{\alpha}_{ \beta} & -u^\mu\delta^{\alpha}_{\beta} &
  f^{\mu\alpha} & 0^{\alpha\mu} \\ 
  \rho h \delta^{\mu}_\beta & 0^{\mu}_\beta & u^\mu/c_s^2 & 0^\mu \\
  0^{\mu}_\beta & 0^{\mu}_\beta & 0^\mu & u^\mu \\ 
\end{array} \right),
\label{amu}
\end{eqnarray} 

\noindent
where $c_s$ stands for the speed of sound 

\begin{eqnarray}
c_s^2 =\left(\frac{\partial p}{\partial e}\right)_s,
\end{eqnarray}

\noindent
$e$ being the mass-energy density of the fluid $e=\rho(1+\epsilon)$. In equation~(\ref{amu}) the following definitions are introduced:

\begin{eqnarray}
{\cal E}&=&\rho h + b^2,
\\
P^{\alpha\mu}&=&g^{\alpha\mu}+2 u^\alpha u^\mu, 
\\
l^{\mu\alpha}&=&(\rho h g^{\mu\alpha}+(\rho h -b^2/c_s^2) u^\mu u^\alpha)/ (\rho h), 
\\ 
f^{\mu\alpha}&=&(u^\alpha b^\mu/c_s^2- u^\mu b^\alpha)/(\rho h),
\\ \nonumber
\end{eqnarray}

\noindent
as well as the notation

\noindent
\begin{equation}
0^\mu = 0, \,\,\,\, 0^{\alpha \mu} = (0,0,0,0)^{\rm T}, \,\,\,\, 0^\mu_\beta = (0,0,0,0).
\end{equation}

\noindent
It is important to remark that the 10 covariant variables we have used to write the system of equations are not independent, since they are related by the constraints 

\begin{equation}
   u^{\alpha} u_{\alpha} = -1, 
\end{equation}

\begin{equation}
b^{\alpha} u_{\alpha} = 0,  
\end{equation}

\noindent 
and
\begin{equation}
\partial_\alpha (u^\alpha b^0 - u^0 b^\alpha) = 0,
\end{equation}

\noindent
which reduces to the usual divergence constraint.

\subsection{Wavespeeds}
\label{ss:ws}

  If $\phi(x^\mu)=0$ defines a characteristic hypersurface of the above system (\ref{amuab}), the characteristic matrix, given by ${\cal A^\alpha}\phi_\alpha$ can be written as

\begin{eqnarray}
\label{ch_matrix}
{\cal A}^{\alpha} \phi_{\alpha} = \left( \begin{array}{cccc}
{\cal E} a \delta^{\mu}_{ \nu}  &  m^{\mu}_{\nu} & l^{\mu} & 0^{\mu} \\
\mathcal{B} \delta^{\mu}_{ \nu}  & a \delta^{\mu}_{\nu} & f^{\mu} & 0^{\mu} \\
\rho h \phi_{\nu} &  0_{\nu} &  a/c_s^2 & 0 \\
 0_{\nu} &  0_{\nu} & 0 &  a \\ 
\end{array}\right) \\ \nonumber
\end{eqnarray}

\noindent 
where $ \phi_\mu = \partial_\mu \phi$, $a = u^{\alpha} \phi_{\alpha}$, $\mathcal{B} = b^{\alpha} \phi_{\alpha}$, $G = \phi^{\alpha} \phi_{\alpha}$, $l^{\mu} = l^{\mu\alpha} \phi_\alpha=\phi^\mu+(\rho h - b^2/c_s^2) a u^\mu /(\rho h) + \mathcal{B} b^\mu/(\rho h)$, $f^{\mu} = f^{\mu\alpha}\phi_\alpha = (a b^\mu/c_s^2-\mathcal{B} u^\mu)/(\rho h)$, and $m^\mu_\nu = (\phi^\mu+2au^\mu)b_\nu-\mathcal{B}\delta^\mu_\nu$. Since $\phi(x^\mu) = 0$ is a characteristic surface, the determinant of the matrix (\ref{ch_matrix}) must vanish, i.e.

\begin{equation}
{\rm det}({\cal A}^{\alpha} \phi_{\alpha})={\cal E}\,a^2 \mathcal{A}^2 {\cal N}_4 = 0 \ ,
\label{eq:det}
\end{equation}

\noindent
where
 
\begin{eqnarray}
 {\cal A} &=& {\cal E} a^2 -\mathcal{B}^2, \\
\label{n4}
 {\cal N}_4 &=& \rho h \left( \frac{1}{c_s^2} -1 \right) a^4 - \left(\rho h +\frac{b^2}{c_s^2} \right) a^2 G +\mathcal{B}^2 G \,. \\ \nonumber
\end{eqnarray}

  The above equations, valid for a general space-time, will be applied to obtain the wavespeeds in a flat space-time in Cartesian coordinates. To this end, let us consider a planar wave propagating in a given direction, that we choose as the $x$-axis, with speed $\lambda$. The normal to the characteristic hypersurface describing this wave is given by the four-vector 

\begin{equation}
\label{ppp}
\phi_\mu=(-\lambda,1,0,0), 
\end{equation}

\noindent
and by substituting equation~(\ref{ppp}) in equation~(\ref{eq:det}) we obtain the so called {\it characteristic polynomial}, whose zeroes give the characteristic speed of the waves propagating in the $x$-direction. Three different kinds of waves can be obtained according to which factor in equation~(\ref{eq:det}) becomes zero. For entropic waves $a=0$, for Alfv\'en waves $\mathcal{A}=0$, and for magnetosonic waves ${\cal N}_4=0$.  

  The characteristic speed $\lambda$ of the entropic waves propagating in the $x$-direction, given by the solution of the equation $a=0$, is the following

\begin{equation}
  \lambda = v^x (= \lambda_e). 
\end{equation}

\noindent
For Alfv\'en waves, given by $\mathcal{A}=0$, there are two solutions corresponding, in general, to different speeds of the waves,

\begin{equation}
\label{alfven}
  \lambda=\frac{b^x \pm \sqrt{{\cal E}}u^x }{b^0 \pm \sqrt{{\cal E}}W} (= \lambda_{a,\pm}).
\end{equation}

\noindent
In the case of magnetosonic waves there are four solutions, two of them corresponding to the slow magnetosonic waves and the other two to the fast magnetosonic waves. It is however not possible, in general, to obtain simple expressions for their speeds since they are given by the solutions of the quartic equation ${\cal N}_4=0$ with $a$, ${\cal B}$ and $G$ in equation~(\ref{n4}) explicitly written in terms of $\lambda$ as

\begin{eqnarray}
 a & = & W(-\lambda + v^x), \\
 {\cal B} & = & b^x - b^0 \lambda, \label{defcalB}\\
 G & = & 1 - \lambda^2. \\ \nonumber
\end{eqnarray}
  
  It is worth noticing that just as in the classical case, the seven eigenvalues corresponding to the entropic (1), Alfv\'en (2), slow magnetosonic (2) and fast magnetosonic waves (2) can be ordered  as follows

\begin{equation}
\label{order}
\lambda^-_f \le \lambda^-_a \le \lambda^-_s \le \lambda_e \le \lambda^+_s \le \lambda^+_a \le \lambda^+_f,
\end{equation}

\noindent
where the subscripts $e$, $a$, $s$ and $f$ stand for {\it entropic}, {\it Alfv\'en}, {\it slow magnetosonic} and {\it fast magnetosonic} respectively, and the superscript $-$ or $+$ refer to the lower or higher value of each pair. This fact allows us to group the Alfv\'en and magnetosonic eigenvalues in two classes separated by the entropic eigenvalue. It should be also emphasized that the superscript does not correspond to the sign on the left hand side of equation~(\ref{alfven}). We will use subscripts to refer to this sign whenever needed.

  Let us remark that in the previous discussion about the roots of the characteristic polynomial we have omitted the fact that the entropy waves as well as the Alfv\'en waves appear as double roots. These superfluous eigenvalues are associated with unphysical waves and are the result of working with the unconstrained system of equations. We note that \citet{vanputten91} derived a different augmented system of RMHD equations in constrained-free form with different unphysical waves. Any attempt to develop a numerical procedure to solve the RMHD equations based on the wave structure of the  RMHD equations must remove these unphysical waves (i.e. the corresponding eigenvectors) from the wave decomposition. \citet{komissarov99} and \citet{koldoba} eliminate the unphysical eigenvectors by demanding the waves to preserve the values of the invariants $u^\mu u_\mu = -1$ and $u^\mu b_\mu = 0$ as suggested by \citet{anile}. Correspondingly, \citet{Balsara01} selects the physical eigenvectors by comparing with the equivalent expressions in the nonrelativistic limit.

  In Appendix~\ref{app_A} we discuss our results obtained in the representation of the characteristic speeds for fluids under different thermodynamical conditions and states of motion. These diagrams show the normal speed of planar wavefronts propagating in different directions (phase speed diagrams). These diagrams are the generalization to relativistic, arbitrarily moving flows of the original diagrams introduced by Friedrichs (see \citep{JT64}). In a recent paper, \cite{KM08} have revisited the theory for linear RMHD wave propagation showing the equivalence with the characteristic speed approach, and the connection between the phase and group speed diagrams by means of a Huygens construction, in arbitrary reference frames.

\subsection{Degeneracies}
\label{ss:degs}

  As in the case of classical magnetohydrodynamics, the equations of RMHD present degeneracies in the sense that two or more eigenvalues of the system of equations coincide and, therefore, the strict hyperbolicity of the system breaks down. The conditions for that to happen have been analyzed in the context of RMHD by \citet{komissarov99} and we summarize them here. Komissarov's analysis is done for the characteristic wave speeds in the fluid frame and leads to the conclusion that the conditions of degeneracy in RMHD are the same as in classical MHD. Degeneracies are encountered for waves propagating perpendicular to the magnetic field direction (Type I) and for waves propagating along the magnetic field direction (Type II).  Finally, a particular subcase of Type II degeneracy appears when the sound speed is equal to $c_a \equiv \sqrt{b^2/\cal E}$. We will refer to this special subcase as Type II$^\prime$. 

  For Type I degeneracy, the two Alfv\'en waves, the entropic wave and the two slow magnetosonic waves propagate at the same speed ($\lambda_a^-=\lambda_s^-= \lambda_e=\lambda_s^+=\lambda_a^+$). For Type II degeneracy, an Alfv\'en wave and a magnetosonic wave (slow or fast) of the same class propagate at the same speed ($\lambda_f^-=\lambda_a^-$ or $\lambda_a^-=\lambda_s^-$ or $\lambda_s^+=\lambda_a^+$ or $\lambda_a^+=\lambda_f^+$). In the special Type II$^\prime$ subcase, an Alfv\'en wave and both the slow and fast magnetosonic waves of the same class propagate at the same speed ($\lambda_f^-=\lambda_a^-=\lambda_s^-$ or $\lambda_s^+=\lambda_a^+=\lambda_f^+$). We have to emphasize that in classical MHD, if Type II degeneracy appears, both Alfv\'en waves exhibit such a degeneracy. In RMHD however, due to aberration, the condition for degeneracy can be fulfilled for one Alfv\'en wave but not for the other. Only if the tangential component of the fluid velocity vanishes, we recover the classical behaviour. In Appendix~\ref{app_A} we present a series diagrams displaying the characteristic wave speeds as a function of direction for fluid with different thermodynamical conditions and states of motion, paying attention to the manifestation of the various types of degeneracy in the laboratory frame.

  \cite{komissarov99} also gives a covariant characterization of the different types of degeneracy. Type I degeneracy is characterized by ${\cal B}$, equation~(\ref{defcalB}), being zero for the case of the Alfv\'en wave, whereas Type II degeneracy arises when the quantity ${\cal B}^2/(G + a^2)$ for the Alfv\'en wave equals to $b^2$. We have made a step forward concerning the characterization of both types of degeneracy by introducing the components of the magnetic field parallel and perpendicular to the Alfv\'en wave in the comoving frame. To this end, let $\nu^\alpha$ be the unitary vector, normal to the front wave as described by the comoving observer with the fluid \citep[see][]{anile}. To obtain this vector, first we project the vector $\phi^\alpha$ onto the orthogonal space to the fluid four velocity and, second, we normalize this projection. It is easy to see that this vector is given by

\begin{equation}
\label{eq:nu}
\nu^\alpha = \frac{\phi^\alpha + a u^\alpha}{\sqrt{G + a^2}},
\end{equation}

\noindent
which is well defined even for the degenerate cases.

  Now, we can characterize both types of degeneracies in terms of the magnetic field components normal and tangential to the Alfv\'en wave front in the comoving frame, $b^\alpha_n$ and $b^\alpha_t$ respectively. The magnetic field $b^\alpha$ can be written as

\begin{equation}
\label{eq:bdecomp}
  b^\alpha=b^\alpha_n+b^\alpha_t=(b_\beta \nu^\beta_a)\nu^\alpha_{a}+b^\alpha_t,
\end{equation}

\noindent
where $\nu_a$ denotes the normal to the Alfv\'en front wave. Using this decomposition and the definition of $\nu_{a}$, equation~(\ref{eq:nu}) particularized to the Alfv\'en wave, it is easy to check that

\begin{equation}
b_n^2 = b_{n\alpha}b^\alpha_n=(b_{\alpha}\nu_a^\alpha)^2=\frac{{\cal B}^2}{G+a^2}.
\end{equation}

\noindent
  Now, Type I degeneracy reduces to the case when $b^\alpha_n = (0,0,0,0)$, whereas Type II degeneracy is characterized by $b_n^2 = b^2$, i.e., Type I degeneracy occurs whenever the component of $b^\alpha$ along the direction of propagation of the Alfv\'en wave (normal to the Alfv\'en wave front), $b^\alpha_n$ is equal to zero, whereas Type II degeneracy occurs when the component of the magnetic field tangential to the Alfv\'en wave front, $b^\alpha_t$, is equal to zero. Finally, let us note that the condition $b^\alpha_n=(0,0,0,0)$ leads, for a wave
propagation along the $x$ direction in the laboratory frame, to $B^x=0$.

\subsection{Renormalized right eigenvectors}
\label{ss:renom_eigen}

  Let us start with the right eigenvectors for the eigenvalue problem $({\cal A}^{\mu} \phi_{\mu}){\bf r}=0$, as derived by Anile (1989),

{\it Entropy eigenvector}:

\begin{equation}
{\bf r}_e = (0^\alpha, 0^\alpha, 0, 1)^{\rm T}.
\label{Eeig}
\end{equation}

{\it Alfv\'en eigenvectors}:

\begin{equation}
\label{Eq:Aeig}
{\bf r}_{a,\pm} = (a \varepsilon^\alpha_{\,\,\beta \gamma \delta} u^\beta \phi^\gamma b^\delta, {\cal B} \varepsilon^\alpha_{\,\,\beta \gamma \delta}
u^\beta \phi^\gamma b^\delta, 0, 0)^{\rm T},
\end{equation}

\noindent
where the dependence in the corresponding wavespeeds (eigenvalues), $\lambda_{a,\pm}$, is hidden in the quantities $a$, ${\cal B}$ and $\phi^\alpha$, and $\varepsilon^{\alpha\beta\gamma\delta}$ is the Levi-Civita alternating tensor.

{\it Magnetosonic eigenvectors}:

\begin{equation}
\label{Eq:meig}
{\bf r}_{m,\pm} = (a d^\alpha, {\cal B} d^\alpha + a {\cal A} f^\alpha, a^2 {\cal A}, 0)^{\rm T}, 
\end{equation}

\noindent 
($m = s, f$) where

\begin{equation}
d^\alpha = \left(\frac{c_s^2 - 1}{c_s^2}\right) \left(\frac{a^4}{G} (\phi^\alpha + a u^\alpha) - \frac{a^2 {\cal B}}{\rho h} b^\alpha \right),
\end{equation}

\begin{equation}
 f^\alpha = \frac{1}{\rho h} \left(\frac{a}{c_s^2} b^\alpha - {\cal B} u^\alpha \right),
\end{equation}

\noindent
where, again, the dependence in the corresponding characteristic wavespeeds, $\lambda_{s,\pm}$ and $\lambda_{f,\pm}$, is hidden in the quantities $a$, ${\cal B}$, $G$, ${\cal A}$ and $\phi^\alpha$. Eigenvalues $\lambda_{m,\pm}$ are defined as belonging to the same class as $\lambda_{a,\pm}$, i.e.,

$$
\lambda_{m,\pm} = \left\{\begin{array}{ll}
                                \lambda_{m}^\pm, & \mbox{if $\lambda_{a,\pm} = \lambda_{a}^\pm$} \\
                                \lambda_{m}^\mp, & \mbox{if $\lambda_{a,\pm} = \lambda_{a}^\mp$}.
                                          \end{array}
                                \right.
$$

  As it is well known, Alfv\'en and magnetosonic eigenvectors just presented have a pathological behaviour at degeneracies, since they become zero or linearly dependent and they do not form a basis. In the next we describe a procedure to renormalize them.

\subsubsection{Renormalized Alfv\'en Eigenvectors}

  We start by renormalizing the Alfv\'en eigenvectors. From equation~(\ref{Eq:Aeig}) we see that the eigenvectors corresponding to the Alfv\'en waves, ${\bf r}_{a,\pm}$, become zero for Type I and Type II degeneracies. To avoid this problem, let us first consider Type I degeneracy. In this case, $\lambda_{a,\pm} = \lambda_e$, leading to $a = 0$ and ${\cal B} = 0$, which is the reason for the Alfv\'en eigenvectors to become zero. However, the quantity ${\cal B}/a$ is well defined for $\lambda_{a,\pm}\ne \lambda_e$, 

\begin{equation}
\left(\frac{\cal B}{a}\right)_{a,\pm} = \mp \sqrt{{\cal E}},
\label{bovera}
\end{equation}

\noindent
and we can use this value to define the function ${\cal B}/a$ at $\lambda_{a,\pm}=\lambda_e$. Hence, we can obtain new Alfv\'en eigenvectors just dividing the previous ones by $a$. These new eigenvectors are well defined even if we have a state in which the Type I degeneracy condition is fulfilled, and can be written as follows:

\begin{equation}
{\bf r}_{a,\pm} = (\varepsilon^\alpha_{\,\,\beta \gamma \delta} u^\beta \phi^\gamma b^\delta, \mp \sqrt{{\cal E}}
\varepsilon^\alpha_{\,\,\beta \gamma \delta} u^\beta \phi^\gamma b^\delta, 0, 0)^{\rm T}.
\end{equation}

  Let us now consider degeneracy of Type II. The new Alfv\'en eigenvectors just derived are proportional to

\begin{equation}
\label{eq:vareps}
  \varepsilon^\alpha = \varepsilon^\alpha_{\,\,\beta \gamma \delta} u^\beta \phi^\gamma b^\delta.
\end{equation}

\noindent
This four-vector is orthogonal to $u^\alpha$, $\phi^\alpha$ and $b^\alpha$. As discussed in the previous subsection, in the case of Type II degeneracy, $b^\alpha$ can be written as 

\begin{equation}
  b^\alpha = (b_\beta\nu^\beta_a)\nu^\alpha_a = (b_\beta\nu^\beta_a)\frac{\phi^\alpha+a u^\alpha}{\sqrt{G+a^2}}.
\end{equation}

\noindent
From this expression, it is apparent that $b^\alpha$ becomes a linear combination of $\phi^\alpha$ and $u^\alpha$ and, therefore, the four-vector $\varepsilon^\alpha$ is equal to zero for Type II degeneracy and so are the Alfv\'en eigenvectors defined above.

  To avoid this problem, we proceed as follows. Using equation~(\ref{eq:bdecomp}), we see that only the component of $b^\alpha$ tangential to the wave front, $b_t^\alpha$, has a nonzero contribution to the four-vector $\varepsilon^\alpha$, i.e., 

\begin{equation}
\label{eq:vareps_bt}
  \varepsilon^\alpha = \varepsilon^\alpha_{\,\,\beta \gamma \delta} u^\beta \phi^\gamma b_t^\delta.
\end{equation}

\noindent
If we are considering an Alfv\'en wave propagating in the $x$-direction, the vectors $u^\alpha$, $\nu^\alpha$, $\tau^\alpha_y=(0,0,1,0)$ and $\tau^\alpha_z=(0,0,0,1)$ form a basis that we can use to decompose the vector $b_t^\alpha$:

\begin{equation}
b_t^\alpha=g_1\tau_y^\alpha+g_2\tau_z^\alpha+g_3 u^\alpha+g_4\nu^\alpha.
\end{equation}

\noindent 
Substituting this decomposition of $b_t^\alpha$ in equation~(\ref{eq:vareps_bt}), we obtain

\begin{equation}
\label{eq:vareps_alpha}
\varepsilon^\mu = g_1 \alpha_1^\mu + g_2 \alpha_2^\mu,
\end{equation}

\noindent
where

\begin{equation}
  \alpha_1^\mu = \varepsilon^\mu_{\,\,\beta \gamma \delta} u^\beta \phi^\gamma \tau_y^\delta =   W (v^z, \lambda v^z, 0, 1 - \lambda v^x),
\end{equation}

\begin{equation}
  \alpha_2^\mu = \varepsilon^\mu_{\,\,\beta \gamma \delta} u^\beta \phi^\gamma \tau_z^\delta = -W (v^y, \lambda v^y, 1 - \lambda v^x, 0).
\end{equation}

\noindent
Coefficients $g_1$ and $g_2$ follow from (\ref{eq:vareps_alpha}) once $\varepsilon^\mu$ is computed from its definition (equation \ref{eq:vareps}),

\begin{equation}
g_1 =\frac{1}{W}\left( B^y+\frac{\lambda v^y}{1-\lambda v^x} B^x\right),
\end{equation}

\begin{equation}
g_2 = \frac{1}{W}\left(B^z+\frac{\lambda v^z}{1-\lambda v^x} B^x\right). 
\end{equation}

  Vectors $\alpha_1^\mu$, $\alpha_2^\mu$ are clearly well defined through Type II degeneracy. On the other hand, since they form a basis of the orthogonal space to $u^\alpha$ and $\nu^\alpha$ they can be used to express, as we have done with the vector $\varepsilon^\mu$, the tangential magnetic field $b_t$ as a linear combination of them,

\begin{equation}
\label{eq:btalpha}
  b_t^\mu = C_1\alpha^\mu_1 + C_2\alpha_2^\mu.
\end{equation}

\noindent
Coefficients $C_1$ and $C_2$ are related to $g_1$ and $g_2$ as follows,

\begin{eqnarray}
C_1 = \frac{g_1\alpha_{12}+g_2\alpha_{22}}{\alpha_{11}\alpha_{22}-\alpha_{12}^2}W(1-\lambda v^x), 
\end{eqnarray}

\begin{eqnarray}
C_2 = -\frac{g_1\alpha_{11}+g_2\alpha_{12}}{\alpha_{11}\alpha_{22}-\alpha_{12}^2}W(1-\lambda v^x), 
\\ \nonumber
\end{eqnarray}

\noindent
where $\alpha_{11} = \alpha_1^\mu\alpha_{1\mu}$, $\alpha_{12} = \alpha_1^\mu\alpha_{2\mu}$ and $\alpha_{22} = \alpha_2^\mu\alpha_{2\mu}$. From here it is easy to check that the necessary and sufficient condition for the tangential field component $b^\alpha_t$ to vanish is $g_1 = g_2 = 0$. We use this fact to renormalize the Alfv\'en right eigenvectors. If we divide by $\sqrt{g_1^2+g_2^2}$, the right eigenvectors for Alfv\'en waves read

\begin{equation}
  {\bf r}_{a,\pm} = \Big(f_1 \alpha_1^\mu + f_2 \alpha_2^\mu, \mp \sqrt{{\cal E}} (f_1 \alpha_1^\mu + f_2 \alpha_2^\mu), 0, 0\Big)^{\rm T},
\label{RAeig}
\end{equation}

\noindent
where

\begin{equation}
\label{eq:f12}
  f_{1,2} = \frac{g_{1,2}}{\sqrt{g_1^2 + g_2^2}},
\end{equation}

\noindent 
and we take, at points where $g_1=g_2=0$, which corresponds to the Type II degeneracies, the following prescription

\begin{equation}
 f_{1,2}=\frac{1}{\sqrt{2}}.
\end{equation}

  These renormalized Alfv\'en right eigenvectors are a linear combination of the ones proposed by \cite{komissarov99} for the Type II degeneracy case. However, contrary to the Komissarov's choice, our expressions are free of pathologies not only in the Type II degeneracy but also in the Type I degeneracy case. Note that the renormalization of the Alfv\'en right eigenvectors just described is the relativistic generalization of the \cite{BW88} method for classical MHD and the limiting valuse of $f_{1,2}$ are chosen as in that work.

\subsubsection{Renormalized Magnetosonic Eigenvectors}

  Let us consider the eigenvectors corresponding to magnetosonic waves. Again, if $B_x = 0$ (Type I degeneracy), provided that $a(\lambda)$ and ${\cal B}(\lambda)$ vanish for $\lambda = \lambda_e$, we have that $\lambda_{s,\pm} = \lambda_e$ are solutions for the characteristic wavespeeds of slow magnetosonic waves. Since $a = 0$ and ${\cal B} = 0$, the corresponding right eigenvectors ${\bf r}_{s,\pm}$ defined in (\ref{Eq:meig}) are zero. This pathological behaviour is removed if we divide the eigenvectors by $a^4$ and take the appropriate value of the quantity ${\cal B}/a$ when $a=0$. A simple calculation establishes that for the magnetosonic eigenvalues $\lambda_{m, \pm}$ the following relation is fulfilled

\begin{equation}
  \left(\frac{\cal B}{a}\right)_{m,\pm} = \mp \sqrt{\left(\rho h + \frac{b^2}{c_s^2}\right) - \rho h \left(\frac{1}{c_s^2} - 1 \right) \frac{a^2}{G}},
\label{bas}
\end{equation}

\noindent 
where $m = s, f$. 

  After dividing by $a^4$ and taking into account equation~(\ref{bas}), the new eigenvectors for the magnetosonic waves read

\begin{eqnarray}
\label{msrenorm}
{\bf r}_{m,\pm} = \left( \frac{a}{G}(\phi^\mu + au^\mu) - \frac{b^\mu}{\rho h} \left(\frac{\cal B}{a}\right)_{m,\pm}, \right. \nonumber \\ 
\left.  \left(\frac{\cal B}{a}\right)_{m,\pm} \left(\phi^\mu \frac{a}{G} + \left(\frac{2a^2}{G} - \frac{b^2}{\rho h}\right)
  u^\mu\right) - b^\mu \left(1 + \frac{a^2}{G} \right),
  b^2 - \rho h \frac{a^2}{G}, 0 \right)^{\rm T},
\end{eqnarray}

\noindent
where we have further divided by $(c_s^2 - 1)/c_s^2$.

  These magnetosonic eigenvectors are well behaved at Type I degeneracy as we can see by direct substitution of $a=0$ in the above equation, indicating that the pathological behaviour of the slow magnetosonic eigenvectors has been removed. Actually, for Type I degeneracy, the slow magnetosonic eigenvectors are given by

\begin{eqnarray}
  {\bf r}_{s,\pm}= \Bigg(\mp \frac{b^{\mu}}{\rho h} \sqrt{ \rho h + \frac{b^2}{c^2_s} }, \mp \sqrt{\rho h + \frac{b^2}{c^2_s}} \left(\frac{b^2}{\rho h} \right)   u^{\mu} - b^{ \mu}, b^2 ,0 \Bigg)^{\rm T},
\end{eqnarray}

\noindent
which are a linear combination of the ones proposed by \citet{komissarov99}.

  We now turn to Type II degeneracy. First, we note that the above magnetosonic eigenvectors, equation~(\ref{msrenorm}), have all their components proportional to $|b_t| \equiv \sqrt{b_t^\nu b_{t\, \nu}}$ (see Appendix~\ref{app_B}). Hence the components of the eigenvectors associated to characteristic waves for which $b_t^\nu = (0,0,0,0)$ (Type II degeneracy) will be equal to zero. To avoid this problem we can renormalize them by dividing by $|b_t|$. If we denote by 

\begin{eqnarray}
\label{eq:rm}
{\bf  r}_{m,\pm} = (e^\nu, L^\nu, {\cal C}, 0)^T
\\ \nonumber
\end{eqnarray}

\noindent
the renormalized eigenvectors, from the expressions in Appendix~\ref{app_B}), we have

\begin{eqnarray}
\label{eq:rme}
  e^\nu = - \frac{a {\cal C}}{\rho h c_s^2 (G + a^2)} (\phi^\nu + a u^\nu) - \left(\frac{\cal B}{a}\right)_{m,\pm} \frac{1}{\rho h} \frac{b_t^\nu}{|b_t|}, 
\end{eqnarray}

\begin{eqnarray}
\label{eq:rmL}
  L^\nu =  - \left(\frac{\cal B}{a}\right)_{m,\pm} \frac{{\cal C}}{\rho h} u^\nu - \left(1 + \frac{a^2}{G}\right) \frac{b_t^\nu}{|b_t|}, 
\end{eqnarray}

\begin{eqnarray}
\label{eq:rmC}
  {\cal C} = - \frac{(G + a^2) c_s^2}{a^2 -(G+ a^2)c_s^2} |b_t|. 
\\ \nonumber
\end{eqnarray}

\noindent
Taking into account equation~(\ref{eq:btalpha}), the quantity $b_t^\nu/|b_t|$ appearing in the definition of $e^\nu$ and $L^\nu$ can be written as

\begin{eqnarray}
\label{ubt}
  \frac{ b_t^{\mu} }{ \mid b_t \mid  } = \frac{ (f_1 \alpha_{12} + f_2 \alpha_{22} ) \, \alpha_1^{\mu} - (f_1 \alpha_{11} + f_2 \alpha_{12} ) \, \alpha_2^{\mu}}{\big[ (\alpha_{11} \alpha_{22} -\alpha_{12}^2) (f_1^2 \alpha_{11} +2 f_1 f_2 \alpha_{12} + f_2^2 \alpha_{22} ) \big]^{1/2}}. 
\\ \nonumber
\end{eqnarray}

\noindent
where $f_{1,2}$ are given by equation~(\ref{eq:f12}) with $f_{1,2} = 1/\sqrt{2}$ if $g_1 = g_2 = 0$.

  The renormalization just described can be applied to the four magnetosonic eigenvectors: corresponding to two slow and two fast magnetosonic waves. However, in order to maintain the independence of the eigenvectors when we approach the degeneracy, we only apply the above renormalization to two magnetosonic eigenvectors, one of each class, those whose eigenvalues are closer to the Alfv\'en eigenvalue of the same class. For them, we use expression~(\ref{eq:rm}). For the  particular subcase Type II$^\prime$, corresponding to
$\lambda_{a,\pm} =\lambda_{s,\pm} = \lambda_{f,\pm}$, $\lambda_{a,+} \ne \lambda_{a,-}$, for which the denominator of equation~(\ref{eq:rmC}) vanishes\footnote{This result is easily derived from the following conditions: i) for Alfv\'en eigenvalues, $a^2/{\cal B}^2 = {\cal E}$; ii) for the degenerate eigenvalues in the Type II degeneracy, ${\cal B}^2/(a^2 + G) = b^2$; iii) the definition of Type II$^\prime$ degeneracy, $c_s^2 = b^2/{\cal E}$.}, we take ${\cal C} = 0$ and the problem is removed. 

  For the remaining two magnetosonic eigenvectors we obtain renormalized expressions by dividing the eigenvectors  (\ref{msrenorm}) by $\rho h a^2/G- b^2$. After some algebra, if we denote by ${\bf  r}_{m,\pm} = (e^\nu, L^\nu, {\cal C}, 0)^{\rm T}$ the renormalized eigenvectors, we get:

\begin{eqnarray}
  e^\nu = \frac{a}{\rho h c_s^2 (G + a^2)} (\phi^\nu + a u^\nu) - \left(\frac{\cal B}{a}\right)_{m,\pm} \frac{G b_t^\nu}{\rho h(\rho h a^2 - b^2G)}, 
\end{eqnarray}

\begin{eqnarray}
  L^\nu = \left(\frac{\cal B}{a}\right)_{m,\pm} \frac{1}{\rho h} u^\mu - \left(1 + \frac{a^2}{G}\right) \frac{G b_t^\nu}{\rho h a^2 - b^2G}, 
\end{eqnarray}

\begin{eqnarray}
  {\cal C} = -1, 
\\ \nonumber
\end{eqnarray}

\noindent
with the following prescription for the particular degeneracy subcase Type II$^\prime$ (for which ${a^2}/{G} = b^2/\rho h$)\footnote{It follows easily from i) $a^2 - (G+ a^2) c_s^2 = 0$, and ii) the definition of Type II$^\prime$ degeneragy, $c_s^2 = b^2/{\cal E}$.},

\begin{eqnarray}
  \frac{b_t^\nu}{\rho h a^2 - b^2 G} = 0. 
\\ \nonumber
\end{eqnarray}

  This procedure guaranties to have a complete set of right eigenvectors linearly independent for all possible states.

\subsection{Renormalized left eigenvectors}
\label{ss:renorm_leigen}

  We start with the vectors, ${\bf l}$,\footnote{These vectors are related with those presented by Anile (1989) for the problem ${\hat {\bf l}}({\cal A}^\mu\phi_\mu) = 0$, with $\phi_\mu=(-\lambda,1,0,0)$, through ${\bf l}={\hat{\bf l}}{\cal A}^0$.}

{\it Entropic wave}:

\begin{eqnarray}  \label{vec:entrol}
  {\bf l}_e = (0_{\alpha}, 0_{\alpha}, 0, u^0),
\end{eqnarray}

{\it Alfv\'en waves}:

\begin{eqnarray} \label{vec:Ala0}
  {\bf l}_{a,\pm} = \left( \begin{array}{c}
\left( {\cal E} u^0 - \displaystyle{\frac{{\cal E}}{({\mathcal B}/a)_{a,\pm}}} b^0 \right) \, \varepsilon_{\alpha \beta \gamma \delta} \phi^{\beta}  u^{\gamma} b^{\delta} \\
\\ 
\left(-b^0 + \displaystyle{\frac{{\cal E} }{({\mathcal B}/a)_{a,\pm}}} u^0 \right) \varepsilon_{\alpha \beta \gamma \delta} \phi^{\beta} u^{\gamma} b^{\delta} + (\varepsilon^{0}_{ \beta \gamma \delta} \phi^{\beta} u^{\gamma} b^{\delta}) b_{\alpha}  \\
\\
\varepsilon^{0}_{ \beta \gamma \delta} \phi^{\beta} u^{\gamma} b^{\delta}  \\
\\
0 \\
  \end{array} \right)^T, 
\\ \nonumber
\end{eqnarray}

\noindent
where $\displaystyle{({\mathcal B}/a)_{a,\pm}}$ is given by equation~(\ref{bovera}).

{\it Magnetosonic waves}:

\begin{eqnarray} \label{vec:Ama0}
  {\bf l}_{m,\pm} = \left( \begin{array}{c}
\phi_{\nu} \left({\cal E} u^0 -\displaystyle{\left(\frac{\mathcal{B}}{a}\right)_{m,\pm}} b^0 \right) + b_{\nu} \left( \displaystyle{\frac{G}{a}} +2 a \right) \left(b^0  - \displaystyle{\left(\frac{\mathcal{B}}{a}\right)_{m,\pm}} u^0 \right) - \displaystyle{\frac{\mathcal A }{a}} \,\delta^0_{\nu} \\
\\
\phi_{\nu} \left(- b^0 + \displaystyle{\left(\frac{\mathcal B}{a}\right)_{m,\pm}} u^0\right) + \left (\phi^0 - \displaystyle{\frac{G }{a}} u^0\right) b_{\nu} \\
\\
\phi^0 - a u^0 \displaystyle{\left(\frac{1}{c^2_s} - 1 \right)} + \displaystyle{\frac{G}{ \rho h a}} \left(\displaystyle{\frac{b^2}{c^2_s}} u^0 - \displaystyle{\left(\frac{\mathcal B }{a}\right)_{m,\pm}} b^0 \right) \\
\\
0 \\
  \end{array} \right)^T, 
\\ \nonumber
\end{eqnarray}

\noindent
where $\displaystyle{({\mathcal B}/a)_{m,\pm}}$ is given by equation~(\ref{bas}).

  Following the same procedure we have used to renormalize the right eigenvectors, we can obtain left eigenvectors well behaved for degenerate states. Alfv\'en left eigenvectors take the form

\begin{eqnarray}\label{lvecm}
  {\bf l}_{a,\pm} = \left( \begin{array}{c}
({\cal E} u^0 \pm b^0 \sqrt{{\cal E}}) ( f_1 \alpha_{1 \mu} + f_2 \alpha_{2 \mu}) \\
\\ 
(-b^0 \mp \sqrt{{\cal E}} u^0) ( f_1 \alpha_{1 \mu} + f_2 \alpha_{2 \mu}) + ( f_1 \alpha^0_{1} + f_2 \alpha^0_{2} ) b_{\mu} \\
\\
f_1 \alpha^0_{1} + f_2 \alpha^0_{2} \\
\\
0 \\
  \end{array} \right)^T. 
\\ \nonumber
\end{eqnarray}

\noindent
where $f_{1,2}$ have been defined in equation~(\ref{eq:f12}) and we take again $f_{1,2}=1/\sqrt{2}$ when $g_1=g_2=0$.

  The magnetosonic left eigenvectors given by (\ref{vec:Ama0}) are not well defined for Type I degeneracy since $a={\cal B}=0$. To get rid of this problem we multiply the eigenvectors by $a$ so that

\begin{eqnarray} \label{de1}
  {\bf l}_{m,\pm} = \left( \begin{array}{c}
\phi_{\nu} ({\cal E} u^0 a - \mathcal{B} b^0 ) + b_{\nu} (G + 2 a^2) \left(b^0 - \displaystyle{\left( \frac{ {\mathcal B } }{a} \right)_{m,\pm}} u^0 \right) - {\mathcal A } \, \delta^0_{\nu} \\
\\
\phi_{\nu} (-a b^0 + {\mathcal B} u^0) + (a \phi^0 - G u^0) b_{\nu} \\
\\
a \phi^0 - a^2 u^0 \displaystyle{\left(\frac{1}{c^2_s} - 1 \right)} + \displaystyle{\frac{G}{ \rho h}}
  \left( \displaystyle{\frac{ b^2 u^0 }{c^2_s}} - \displaystyle{\left( \frac{ {\mathcal B } }{a} \right)_{m,\pm}} b^0 \right) \\
  \\
  0 \\
  \end{array} \right)^T .
\end{eqnarray}

\section{A note on relativistic MHD convexity}
\label{s:convex}

  In a convex system, all the characteristic fields are genuinely non-linear or linearly degenerate. Brio \& Wu (1988) noted that the equations of classical MHD are non-convex since at degenerate states magnetosonic waves change from genuinely non-linear to linearly degenerate. We have checked the non-convex character of the relativistic MHD and found that the relativistic counterpart is also non-convex with the magnetosonic waves changing character also at degenerate states. We refer the reader to \cite{Anton08}, where an analysis of the characteristic fields of RMHD in covariant variables is presented.

  The non-convex character of the classical (and relativistic) MHD equations is source of several pathological behaviours, as the development of the so-called compound waves. The shock tube problem discussed in Sect.~ \ref{ss:st1} (proposed by Brio \& Wu 1988 in classical MHD and adapted to relativistic MHD by van Putten 1993) displays the propagation of one of such compound waves. 

\section{Right eigenvectors in conserved variables}\label{s:recv}

\subsection{Transformation matrix}

  The renormalized eigenvectors obtained in Sect.~3 are now transformed to conserved variables. To this aim, we must construct the transformation matrix between the set of covariant variables, ${\tilde{\bf U}}$, and the set of the conserved ones, ${\bf U}$, namely $(\partial {\bf U}/\partial {\tilde{\bf U}})$. At this point, let us recall that our aim is to build up a Riemann solver based on the spectral decomposition of the flux vectors Jacobians of the system in conservation form. Since the Riemann solver is used to compute the numerical fluxes for the advance in time in a dimensional-splitting manner, only a one-dimensional version of system (\ref{eq:fundsystem}) need to be considered. Consistently with the choice made in Sect.~3, we particularize our discussion to the $x$-direction. Along this direction, the evolution equation for $B^x$ is $\partial B^x/\partial t = 0$, and it can be simply removed from the system. Hence, the desired spectral decomposition will be directly worked out for the shrunk, $7\times7$ Jacobian of the flux vector along the $x$-direction. Accordingly, in what follows the set of conserved variables will contain only seven variables and the aforementioned matrix will be of dimension $7 \times 10$. Its elements, corresponding to the partial derivatives of the conserved variables with respect to the covariant ones, are the following\footnote{The transformation matrix is not unique since the constraints can be used in different ways to give different functional dependences. However, the resulting transformed eigenvectors are independent of the matrix used.} 

\begin{eqnarray}
\left(\frac{\partial {\bf U}}{\partial {\tilde{\bf U}}} \right) = 
= \left( \begin{array}{cccccc}      
  \rho & 0 & 0 & 0 & 0 & 0  \\
\\
  {\cal E} u^x & {\cal E} u^0 & 0 & 0 & -2 b^0 u^0 u^x - b^x & 2 b^x u^0 u^x - b^0 \\
\\
  {\cal E} u^y & 0 & {\cal E} u^0 & 0 & -2 b^0 u^0 u^y - b^y & 2 b^x u^0 u^y \\ 
\\
  {\cal E} u^z & 0 & 0 & {\cal E} u^0 & -2 b^0 u^0 u^z - b^z & 2 b^x u^0 u^z \\
\\
  2 {\cal E} u^0 & 0 & 0 & 0 & -2 b^0 u^0 u^0 - b^0 & 2 b^x (u^0)^2 - b^x \\
\\
  b^y & 0 & -b^0 & 0 & -u^y & 0 \\
\\
  b^z & 0 & 0 & -b^0 & -u^z & 0 \\ 
  \end{array} \right. \nonumber 
\end{eqnarray}
		
\begin{eqnarray}
\label{proyeM}
\left. \begin{array}{cccc}      
0 & 0 & \partial_p \rho u^0 & \partial_s \rho u^0  \\
\\
2 b^y u^0 u^x & 2 b^z u^0 u^x & \partial_p (\rho h) u^0 u^x & \partial_s (\rho h) u^0 u^x \\
\\
2 b^y u^0 u^z - b^0 & 2 b^z u^0 u^z & \partial_p (\rho h) u^0 u^y & \partial_s (\rho h) u^0 u^y \\ 
\\
2 b^y u^0 u^z & 2 b^z u^0 u^z - b^0 & \partial_p (\rho h) u^0 u^z & \partial_s (\rho h) u^0 u^z \\ 
\\
2 b^y (u^0)^2 -b^y & 2 b^z (u^0)^2 - b^z & \partial_p (\rho h) (u^0)^2-1 & \partial_s (\rho h) (u^0)^2 \\
\\
u^0 & 0 & 0 & 0 \\
\\
0 & u^0 & 0 & 0 \\ 
  \end{array} \right)
\end{eqnarray}
\noindent
where
\begin{displaymath}
  \partial_{p} = \left( \frac{\partial  }{ \partial p}  \right)_s    
  \qquad \partial_{ s} = \left( \frac{\partial  }{ \partial s}  \right)_p.
\end{displaymath}

\subsection{Right eigenvectors in conserved variables}

  The right eigenvectors in conserved variables, ${\bf R}$, are computed as follows

\begin{equation}
\label{eq:R}
  {\bf R} = \left(\frac{\partial {\bf U}}{\partial {\tilde{\bf U}}} \right) \, \, {\bf r},
\end{equation}

\noindent
where ${\bf r}$ are the corresponding (renormalized) eigenvectors in covariant variables defined in equation~(\ref{Eeig}) --entropic vector--, equation~(\ref{RAeig}) --Alfv\'en vectors-- and equation~(\ref{eq:rm}) --magnetosonic vectors, with the corresponding definitions of $e^\nu$, $L^\nu$ and ${\cal C}$, for each pair of vectors--.

\subsubsection{Entropic eigenvector}

  The entropic eigenvector in conserved variables is
		  
\begin{equation} 
\label{eq:entro.cons}
  {\bf R}_e = u^0 \Big(\partial_s \rho, u^x\partial_s (\rho h), u^y\partial_s (\rho h), u^z\partial_s (\rho h), u^0\partial_s (\rho h), 0, 0\Big)^T.
\end{equation}
		
\subsubsection{Alfv\'en eigenvectors}

  Alfv\'en eigenvectors are

\begin{equation}
\label{eq:alfven.cons}
  {\bf R}_{a, \pm} = f_1 {\bf V}_{a, 1, \pm} + f_2 {\bf V}_{a, 2, \pm},
\end{equation}           
		  
\noindent
where

\begin{eqnarray}
  {\bf V}_{a, 1, \pm } = 
  \left( \begin{array}{c}
\rho u^z \\
\\
2 u^z \big( {\cal E} u^x  \pm \sqrt{{\cal E}}  b^x \big) \\
\\
{\cal E} u^y u^z  \pm \sqrt{{\cal E}} \,  b^y u^z  \\
\\
{\cal E} \Big( (u^0)^2+  (u^z)^2 - (u^x)^2 \Big) \pm \sqrt{{\cal E}} \big( b^z u^z + b^0 u^0 -b^x u^x \big)  \\
\\ 
2 u^z  \big( {\cal E} u^0   \pm  \sqrt{{\cal E}} \, b^0 \big)  \\
\\
b^y u^z \pm \sqrt{{\cal E}} u^y u^z  \\
\\
-b^y u^y \mp \sqrt{{\cal E}} \big(  1 +(u^y)^2 \big)   \\
  \end{array}  \right).
\end{eqnarray}

\begin{eqnarray}
  {\bf V}_{a, 2, \pm } =- 
  \left( \begin{array}{c}
\rho u^y \\
\\
2u^y ({\cal E} u^x  \pm  \sqrt{{\cal E}} b^x) \\
\\
{\cal E}  \Big( (u^0)^2 + (u^y)^2 - (u^x)^2  \Big) \pm \sqrt{{\cal E}} \big( b^y u^y + b^0 u^0 - b^x u^x \big) \\
\\
{\cal E} u^y u^z  \pm \sqrt{{\cal E}} \,  b^z u^y \\
\\
2 u^y \big( {\cal E} u^0   \pm  \sqrt{{\cal E}} \,   b^0 \big) \\
\\
-b^z u^z  \mp \sqrt{{\cal E}} \big(  1 + (u^z)^2  \big) \\
\\
b^z u^y \pm \sqrt{{\cal E}} u^y u^z \\
  \end{array}  \right), 
\\ \nonumber
\end{eqnarray}

\noindent
Vectors ${\bf V}_{a, 1, \pm}$ and ${\bf V}_{a, 2 , \pm}$ are well defined in Type I degeneracy. In the case of the Type II degeneracy, coefficients $f_1$ and $f_2$ in (\ref{eq:alfven.cons}) must be taken equal to $1/\sqrt{2}$.

\subsubsection{Magnetosonic eigenvectors}

  Finally, in terms of $e^\nu$, $L^\nu$ ($\nu = 0, x, y, z$) and ${\mathcal C}$, the components of the magnetosonic eigenvectors in covariant variables defined in equations~(\ref{eq:rme})-(\ref{eq:rmC}) respectively, the  magnetosonic eigenvectors in conserved variables are given by 

\begin{eqnarray}
  {\bf R}_{m,\pm} = \left( \begin{array}{c}
\rho e^0 +  { \mathcal C} u^0 \partial_p\rho \\
\\
{\cal E} (u^x e^0 + u^0 e^x ) + 2 u^0 u^x b_{\alpha} L^{\alpha} - b^x L^0 - b^0 L^x + u^0 u^x { \mathcal C} \partial_p (\rho h) \\
\\  
{\cal E} (u^y e^0 + u^0 e^y ) + 2 u^0 u^y b_{\alpha} L^{\alpha} - b^y L^0 - b^0 L^y + u^0 u^y { \mathcal C} \partial_p (\rho h) \\
\\
{\cal E} (u^z e^0 + u^0 e^z ) + 2 u^0 u^z b_{\alpha} L^{\alpha} - b^z L^0 - b^0 L^z + u^0 u^z { \mathcal C} \partial_p (\rho h) \\
\\
2 {\cal E} u^0 e^0  + \Big(2 (u^0)^2 -1\Big) b_{\alpha} L^{\alpha} - 2 b^0 L^0 + \Big( \partial_p (\rho h)   (u^0)^2 -1\Big) { \mathcal C} \\   
\\
b^y e^0 - b^0 e^y - u^y L^0 + u^0 L^y \\
\\
b^z e^0 - b^0 e^z - u^z L^0 + u^0 L^z \\
  \end{array}  \right). \\ \nonumber
\end{eqnarray}

\noindent
According to the renormalization recipes given in Sect.~3, the components of the magnetosonic eigenvectors corresponding to the eigenvalue closer to the Alfv\'en one within each class, are

\begin{eqnarray}
  R_{m,\pm, \, D} = \frac{a |b_t| }{ h (a^2  -c^2_s  (G +a^2) )} ( \phi^{0} + a u^{0}) - \frac{ ({\cal B}/a)_{m,\pm} }{ h } \frac{ b^{0}_t }{ |b_t| }  - |b_t| \, \frac{ (G+a^2)c_s^2 }{ a^2 - (G+a^2) c_s^2}  u^0 \partial_p \rho, 
\end{eqnarray}
		 
\begin{eqnarray}
  R_{m,\pm, \, S^x} = \left( 1 + \frac{a^2}{G} \right) \bigg\{\frac{ |b_t| }{a^2 -c_s^2 (G +a^2) } \Big( a (u^x \phi^0 + u^0 \phi^x) (1-c^2_s)  + 2 u^0 u^x c_s^2 G \Big)  {}
\nonumber \\
  {} - \left(\frac{\cal B}{a}\right)_{m,\pm} \left( \frac{ b_t^x u^0 }{|b_t|} + \frac{ b_t^0 u^x }{|b_t|}   \right) + \left( \frac{ b_t^x b^0 }{|b_t|} + \frac{ b_t^0 b^x }{|b_t|}   \right) \bigg\} + u^0 u^x { \mathcal C}  \partial_p (\rho h), 
\end{eqnarray}
		     
\begin{eqnarray}
  R_{m,\pm, \, S^j} = \left( 1 + \frac{a^2}{G} \right) \bigg\{\frac{ u^j |b_t| }{a^2 -c_s^2 (G +a^2) } \Big(a \phi^0 (1-c^2_s) + 2 u^0 c_s^2 G \Big)  {}
\nonumber \\
  {} - \left(\frac{\cal B}{a}\right)_{m,\pm} \left( \frac{ b_t^j u^0 }{|b_t| } + \frac{ b_t^0 u^j }{|b_t| } \right) + \left( \frac{ b_t^j b^0 }{|b_t|} + \frac{ b_t^0 b^j }{|b_t|} \right) \bigg\} + u^0 u^j { \mathcal C}  \partial_p (\rho h), 
\end{eqnarray}
	     
\begin{eqnarray}
  R_{m,\pm, \, \tau} = \left(1 + \frac{a^2}{G} \right) \bigg\{\frac{ 2 u^0 |b_t|}{a^2 - c_s^2 (G +a^2)} \Big(a \phi^0 (1-c^2_s) + u^0 c_s^2 G \Big) {} 
\nonumber \\
{} + |b_t| + \left( b^0 - \left(\frac{\cal B}{a}\right)_{m,\pm} u^0 \right) \frac{2  b_t^0}{|b_t|} \bigg\} + {\mathcal C} \Big( \partial_p (\rho h)  (u^0)^2 -1 \Big), 
\end{eqnarray}
		   
\begin{eqnarray}
  R_{m,\pm,  \, B^j} = \frac{  u^x \lambda_{m, \pm}  -u^0}{ G} \frac{ b^j_t}{|b_t| } + u^j \frac{ b^0_t}{|b_t|}, 
\\ \nonumber
\end{eqnarray}

\noindent
where in the preceding expressions, $j$ stands for $y$ and $z$, and $a$, $G$ and $\phi_\mu$ are computed for the corresponding magnetosonic eigenvalue, $\lambda_{m,\pm}$.

  These eigenvectors are already well defined in Type I degeneracy. Quantity $ b_t^{\mu} /|b_t|$, depending on $f_{1,2}$, was defined in equation~(\ref{ubt}). In the case of Type II degeneracy, coefficients $f_1$ and $f_2$ must be taken equal to $1/\sqrt{2}$. 
 
  For the components of the two remaining magnetosonic eigenvectors we have

\begin{eqnarray}
  R_{m,\pm,  \, D} = \frac{a }{ h (G +a^2) c_s^2 } ( \phi^{0} + a u^{0}) -\frac{ ({\cal B}/a)_{m,\pm} }{ h} \frac{ b^{0}_t G }{ \rho h a^2 -{b}^2 G } - u^0 \partial_p \rho, 
\end{eqnarray}
		 
\begin{eqnarray}
  R_{m,\pm,  \, S^x} = \frac{1}{G c_s^2} \Big( a (1-c^2_s) (u^x \phi^0 + u^0 \phi^x)  +2 u^0 u^x c_s^2 G \Big) {}
\nonumber \\
{}  + \frac{G+a^2}{\rho h a^2 - {b}^2 G} \left( b_t^x b^0 + b_t^0 b^x - (b_t^x u^0 + b_t^0 u^x) \left(\frac{\cal B}{a}\right)_{m,\pm} \right) - u^0 u^x \partial_p (\rho h),  
\end{eqnarray}
		     
\begin{eqnarray}
  R_{m,\pm,  \, S^j } = \frac{u^j}{G c^2_s} \Big( a \phi^0 (1-c^2_s) + 2 u^0  c_s^2 G \Big) {}
\nonumber \\
{} + \frac{G+a^2}{\rho h a^2 -b^2 G } \left( b_t^j b^0 + b_t^0 b^j - (b_t^j u^0 + b_t^0 u^j) \left(\frac{\cal B}{a}\right)_{m,\pm}\right) - u^0 u^j \partial_p (\rho h), 
\end{eqnarray}
		     
\begin{eqnarray}
  R_{m,\pm,  \, \tau} = \frac{1}{G c^2_s} \Bigg( 2 u^0 \Big( a \phi^0 (1-c^2_s) + u^0 c_s^2 G \Big) + a^2 - c^2_s (G+a^2) \Bigg) {}
\nonumber \\  
{} + \frac{2 b_t^0 (G+a^2)}{\rho h a^2 -b^2 G} \left(b^0 - \left(\frac{\cal B}{a}\right)_{m,\pm} u^0 \right) - \partial_p (\rho h) (u^0)^2 + 1, 
\end{eqnarray}
		   
\begin{eqnarray}
  R_{m,\pm,  \, B^j} = \frac{1}{ \rho h a^2 - b^2 G} \Big( (u^x \lambda_{m, \pm} -u^0) b^j_t + G u^j  b^0_t \Big), 
\\ \nonumber
\end{eqnarray}

\noindent
where, again, $j=y,z$, and $a$, $G$ and $\phi_\mu$ are computed for the corresponding magnetosonic eigenvalue, $\lambda_{m,\pm}$. In the special subcase of Type II$^\prime$ degeneracy, $b_t^\nu / (\rho h a^2 - b^2 G)$ must be taken equal to 0.

\section{Left eigenvectors in conserved variables}\label{s:lecv}

\subsection{Transformation matrices}

  The calculation of the left eigenvectors in conserved variables is far more involved than that for the right eigenvectors. One possibility would be to obtain them by direct inversion of the matrix of right eigenvectors. However, in order to obtain tractable expressions, we have proceeded in a two-step process starting from the left eigenvectors in covariant variables presented in Section 3. First we get the left eigenvectors, ${\bar {\bf l}}$, in the so-called {\it reduced system of covariant variables}, ${\tilde {\bf V}} = (u^x, u^y, u^z, b^y, b^z, p, \rho)$, through the transformation

\begin{equation}
\label{eq:l_bar}
  {\bar {\bf l}} = {\bf l} \, \, \left(\frac{\partial {\tilde {\bf U}}}{\partial {\tilde{\bf V}}} \right),
\end{equation}   

\noindent
where $(\partial {\tilde {\bf U}}/\partial {\tilde{\bf V}})$ is the $10 \times 7$ matrix built with the partial derivatives of the covariant variables as functions of the covariant variables in the reduced system.

  In a second step, the left eigenvectors in the conserved variables, ${\bf L}$, are obtained from those in the reduced system of covariant variables through

\begin{equation}
\label{eq:L}
  {\bf L} = {\bar {\bf l}} \, \, \left(\frac{\partial {\tilde {\bf V}}}{\partial {\bf U}} \right).
\end{equation}

  We give now the expressions of the two matrices. For the first one we have

\begin{eqnarray}
\label{mat:u_tilde-v_tilde}
\left(\frac{\partial {\tilde {\bf U}}}{\partial {\tilde{\bf V}}} \right) = \displaystyle{\left( \begin{array}{ccccccc}      

  \displaystyle{\frac{\partial u^0}{\partial u^x}} & \displaystyle{\frac{\partial u^0}{\partial u^y}} & 
\displaystyle{\frac{\partial u^0}{\partial u^z}} & 0 & 0 & 0 & 0 \\
\\
  1 & 0 & 0 & 0 & 0 & 0 & 0 \\
\\
  0 & 1 & 0 & 0 & 0 & 0 & 0 \\
\\
  0 & 0 & 1 & 0 & 0 & 0 & 0 \\
\\
  \displaystyle{\frac{\partial b^0}{\partial u^x}} & \displaystyle{\frac{\partial b^0}{\partial u^y}} & 
\displaystyle{\frac{\partial b^0}{\partial u^z}} & \displaystyle{\frac{\partial b^0}{\partial b^y}} & 
\displaystyle{\frac{\partial b^0}{\partial b^z}} & 0 & 0 \\
\\
  \displaystyle{\frac{\partial b^x}{\partial u^x}} & \displaystyle{\frac{\partial b^x}{\partial u^y}} & 
\displaystyle{\frac{\partial b^x}{\partial u^z}} & \displaystyle{\frac{\partial b^x}{\partial b^y}} & 
\displaystyle{\frac{\partial b^x}{\partial b^z}} & 0 & 0 \\
\\
  0 & 0 & 0 & 1 & 0 & 0 & 0 \\
\\
  0 & 0 & 0 & 0 & 1 & 0 & 0 \\
\\
  0 & 0 & 0 & 0 & 0 & 1 & 0 \\
\\
  0 & 0 & 0 & 0 & 0 & \displaystyle{\left(\frac{\partial s}{\partial p}\right)_\rho} & 
\displaystyle{\left(\frac{\partial s}{\partial \rho}\right)_p} \\
  \end{array} \right).}
\end{eqnarray}

\noindent
In the previous expression,

\begin{eqnarray}  \label{parci1}
  \frac{\partial u^0}{ \partial u^i} = \frac{u^i}{ u^0}  \qquad  i=x,y,z ,
\end{eqnarray}          
			    
\begin{eqnarray} \label{parci2}
  \frac{\partial b^0}{ \partial b^i } = \frac{ u^i u^0 }{ (u^0)^2 - (u^x)^2 }   \qquad  i=y,z ,
\end{eqnarray}
			    
\begin{eqnarray} \label{parci3}
   \frac{\partial b^0}{ \partial u^x } = \frac{b^x}{u^0} ,
\end{eqnarray}

\begin{eqnarray}
  \frac{\partial b^0}{ \partial u^i } =
  \frac{1}{(u^0)^2 -(u^x)^2 }  \left( B^i - b^x \frac{u^x u^i}{u^0}
  \right)  \qquad  i=y,z,
\end{eqnarray}

\begin{eqnarray}
  \frac{\partial b^x}{\partial b^i} =  
  \frac{u^x u^i }{(u^0)^2 -(u^x)^2 } 
  \qquad  i=y,z,
\end{eqnarray}                      
					       
\begin{eqnarray}
  \frac{\partial b^x}{\partial u^x} = 
  \frac{b^0}{u^0}, 
\end{eqnarray}                      

\begin{eqnarray} \label{parci7}
  \frac{\partial b^x}{\partial u^i} = 
  \frac{1 }{(u^0)^2 -(u^x)^2 } \left(\frac{B^i u^x}{u^0} - b^x u^i \right) 
  \quad i=y,z. 
\\ \nonumber
\end{eqnarray}  

  The second matrix, $(\partial {\tilde {\bf V}} / \partial {\bf U})$, involves the derivatives of covariant variables with respect to the conserved ones. In order to obtain general expressions, we use the variable $Z = \rho h (u^0)^2$, introduced in Sect.~ 2. Once the function $h = h(\rho, p)$ is provided through the equation of state, the full system is closed and the partial derivatives of $Z$ can be calculated. 

  As a first step, we write $u^0$, $b^0$ and $b^x$ as functions of the conserved variables and $Z$:

\begin{equation}
\label{eq:u02}
  u^0 = \frac{ Z +{\bf B}^2}{ \left((Z+{\bf B}^2)^2 -{\bf S}^2 
  - \displaystyle{\frac{ ({\bf S \cdot B})^2 }{Z^2}} (2 Z +{\bf B}^2) \right)^{1/2}},
\end{equation}

\begin{equation}
\label{eq:b0} 
  b^0 =  ({\bf S \cdot B })  \frac{u^0}{Z},
\end{equation}
		     
\begin{eqnarray}
  b^x = \frac{B^x}{ u^0} + \frac{ S^x u^0 + b^0 B^x }{ Z+ {\bf B}^2} 
  \frac{b^0}{u^0}. 
\\ \nonumber
\end{eqnarray}

\noindent
From these expressions, the corresponding partial derivatives with respect to the conserved variables can be written in terms of the derivatives of $Z$. Then, the covariant variables in ${\tilde {\bf V}}$ are written in terms of the conserved variables, $u^0$, $b^0$ and $b^x$, and $Z$

\begin{equation}
  u^i = \frac{ u^0 S^i + b^0 B^i}{Z +{\bf B}^2} \,\,\,\,\,\,\,\,\,\, (i = x,y,z), 
\end{equation}

\begin{equation}
  b^i = \frac{ B^i + b^0 u^i}{u^0} \,\,\,\,\,\,\,\,\,\, (i = y,z),
\end{equation}

\begin{equation}
  p = Z -\tau  + {\bf B}^2 \left( 1 - \frac{1}{2 (u^0)^2} \right) - 
  \frac{ ({\bf S} \cdot {\bf B}) }{2 Z^2},
\end{equation}

\begin{equation}
  \rho = \frac{D}{u^0}.
\end{equation}   

\noindent
The elements of the transformation matrix are obtained by derivation of the previous expressions and substitution, leading to:

\begin{eqnarray} \label{eq:duidd}
  \frac{\partial u^i}{\partial D} = \frac{-u^0}{Z +{\bf B}^2} \left( u^0 u^i + \frac{b^0 b^i }{Z} \right) \frac{\partial Z }{\partial D},
\end{eqnarray}  
 		     
\begin{eqnarray}
  \frac{   \partial u^i}{ \partial S^j} = \frac{u^0}{Z+{\bf B}^2} \Bigg\{ \delta^i_j - \left(u^0 u^i + \frac{b^0 b^i}{Z} \right) \frac{\partial Z}{\partial S^j} + \left(  u_j u^i + \frac{B_j b^i u^0}{Z} \right) \Bigg\},
\end{eqnarray}  
		       
\begin{eqnarray}
  \frac{\partial u^i}{ \partial \tau} = \frac{-u^0}{Z +{\bf B}^2} \left( u^0 u^i + \frac{b^0 b^i }{Z} \right) \frac{\partial Z }{\partial \tau},
\end{eqnarray}  
	   
\begin{eqnarray}
  \frac{\partial u^i}{ \partial B^j} = \frac{1}{Z + {\bf B}^2} \Bigg\{ b^0 \delta^i_j - u^0 \left( u^0 u^i + \frac{b^0 b^i} {Z} \right) \frac{\partial Z }{\partial B^j} - 2 (u^0)^2 B_j u^i +b^0 u^i u_j  + \frac{ (u^0)^2 S_j b^i }{Z } \Bigg\},               
\end{eqnarray} 

\begin{eqnarray}
  \frac{\partial b^i}{\partial D} = \frac{1}{Z + {\bf B}^2} \left( b^i \Big((u^0)^2 -1\Big) - u^0 u^i b^0 \frac{2Z+ b^2}{Z} \right) \frac{\partial Z}{\partial D},
\end{eqnarray}
		     
\begin{eqnarray}
  \frac{ \partial b^i}{\partial S^j} = \frac{ 1}{Z + {\bf B}^2} \Bigg\{\bigg( b^i \Big( (u^0)^2 -1\Big) - u^0 u^i b^0 \frac{2Z+ b^2}{Z} \bigg) \frac{\partial Z}{\partial S^j }   \nonumber \\
  +  u^i \left( B_j \left(1+ \frac{ b^2 (u^0)^2 }{Z} \right) + b^0 u_j \right) - B^i u_j + \frac{b^0}{u^0} \delta^i_j \Bigg\}, 
\end{eqnarray}                  
		    
\begin{eqnarray}
  \frac{\partial b^i}{\partial \tau} = \frac{1}{Z + {\bf B}^2} \left( b^i \Big((u^0)^2 -1\Big) - u^0 u^i b^0 \frac{2Z+ b^2}{Z} \right) \frac{\partial Z}{\partial \tau},
\end{eqnarray}                     
	   
\begin{eqnarray}
  \frac{\partial b^i}{\partial B^j} = \frac{\delta^i_j}{u^0} + \frac{1}{Z+{\bf B}^2} \Bigg\{\bigg( b^i \Big((u^0)^2 -1\Big) - u^0 u^i b^0 \frac{2Z + b^2}{Z} \bigg) \frac{\partial Z}{\partial B^j }   {}
\nonumber \\
  + \frac{(b^0)^2}{u^0} \delta^i_j + S_j \left(1+ \frac{ b^2 (u^0)^2 }{Z} \right) u^i + \frac{u_j b^0 }{u^0} \Big( b^0 u^i - B^i \Big) - 2 B_j \left( u^0 b^i -\frac{B^i}{u^0} \right)  \Bigg\},
\end{eqnarray}

\begin{eqnarray}
  \frac{ \partial p}{\partial D} = \frac{ Z +b^2 }{Z +{\bf B}^2} \, \frac{\partial Z }{\partial D},
\end{eqnarray}                              
			 
\begin{eqnarray}
  \frac{ \partial p}{\partial S^j} = \frac{ Z + b^2 }{Z +{\bf B}^2} \, \frac{\partial Z}{\partial S^j} + \frac{1 }{ (Z +{\bf B}^2) u^0} \bigg( {\bf B}^2 u_j - b^0 B_j \bigg),
\end{eqnarray}
	     
\begin{eqnarray}
  \frac{ \partial p}{\partial \tau} = \frac{ Z +b^2 }{Z +{\bf B}^2} \, \frac{\partial Z }{\partial \tau} - 1,                  
\end{eqnarray}                              
	     	  
\begin{eqnarray}
  \frac{ \partial p}{\partial B^j} = \frac{1 }{Z +{\bf B}^2} \Bigg\{ (Z +b^2) \frac{\partial Z }{\partial B^j} 
- \frac{ b^0 S_j }{u^0 } + \frac{ B^2 b_j }{u^0 } + \left( 2 - \frac{1}{(u^0)^2}  \right) Z B_j \Bigg\},
\end{eqnarray}

\begin{equation}
  \frac{\partial \rho}{\partial D} = \frac{1}{u^0} - \frac{D}{ u^0(Z+{\bf B}^2)}\left(1 -(u^0)^2 - \frac{(b^0)^2}{Z} \right) \frac{\partial Z}{\partial D},
\end{equation}
				 
\begin{eqnarray}
  \frac{\partial \rho}{\partial S^j} = \frac{-D}{u^0 (Z+{\bf B}^2)} \bigg\{ \left(1 -(u^0)^2 - \frac{(b^0)^2}{Z} \right) \frac{\partial Z}{\partial S_j} +
u^0 \left(u_j + \frac{b^0 B_j }{Z}  \right) \bigg\},
\end{eqnarray}
				 
\begin{equation}
  \frac{\partial \rho}{\partial \tau} = \frac{-D}{u^0 (Z+{\bf B}^2)}\left(1 -(u^0)^2 - \frac{(b^0)^2}{Z} \right) \frac{ \partial Z}{\partial \tau},
\end{equation}
		 
\begin{eqnarray} \label{eq:drhodbj}
  \frac{\partial \rho }{ \partial B^j} = \frac{ -D}{ u^0 ( Z+{\bf B}^2)}\Bigg\{  \left( 1 -(u^0)^2 - \frac{(b^0)^2}{Z}  \right)
\frac{ \partial Z}{\partial B^j}  +  2 B_j \Big(1-(u^0)^2\Big) + b^0  \left(   \frac{ u^0 S_j}{Z}  +  u_j  \right)  \Bigg\}.
\end{eqnarray}

\subsection{Left eigenvectors in the reduced system of covariant variables}

\subsubsection{Entropic eigenvector}

  Multiplication of the entropic eigenvector in covariant variables and the transformation matrix $(\partial {\tilde {\bf U}}/\partial {\tilde{\bf V}})$, defined in (\ref{mat:u_tilde-v_tilde}), leads to the entropic vector in the reduced system of covariant variables

\begin{equation}
 \bar{{\bf l}}_e = \left(0, 0, 0, 0, 0, u^0 \left( \frac{ \partial s }{\partial p} \right)_{\rho}, u^0 \left( \frac{ \partial s }{\partial \rho} \right)_p \right).
\end{equation}

\subsubsection{Alfv\'en eigenvectors}

  Performing the same operation with the renormalized Alfv\'en eigenvectors in covariant variables, we get

\begin{equation}
\label{eq:l.alfven.red}
  \bar{{\bf l}}_{a, \pm} = f_1 \bar{{\bf w}}_{a, 1, \pm} + f_2 \bar{{\bf w}}_{a, 2, \pm},
\end{equation}
	      
\noindent  
where

\begin{eqnarray}
  \bar{{\bf w}}_{a, 1, \pm} = \left(\begin{array} {c}
2 ( -{\cal E} a + b^{0} {\mathcal B} /u^0 ) u^z \\
\\
\Big( -{\cal E} u^0 u^y \pm \sqrt{{\cal E}} (b^y u^0 -2b^0 u^y) \Big) u^z/u^0  \\
\\
{\cal E} \Big(1+ (u^y)^2\Big) \pm \sqrt{{\cal E}} \bigg( b^0\Big( u^0- 2(u^z)^2/u^0\Big) +b^z u^z -b^x u^x \bigg)  \\
\\
u^z ( \pm \sqrt{{\cal E}} u^y + b^y )  \\
\\
-b^y u^y \mp \sqrt{{\cal E}} \Big(1 + (u^y)^2\Big) \\
\\
u^z \\
\\
0 \\
  \end{array}  \right)^T
\end{eqnarray}

\noindent
and 

\begin{eqnarray}
  \bar{{\bf w}}_{a, 2, \pm} =- \left(\begin{array} {c}
2 ( -{\cal E} a + b^{0} {\mathcal B} /u^0 ) u^y \\
\\
{\cal E} \Big(1+ (u^z)^2\Big) \pm \sqrt{{\cal E}} \bigg( b^0\Big( u^0- 2(u^y)^2/u^0\Big) + b^y u^y -b^x u^x \bigg) \\
\\
\Big( -{\cal E} u^0 u^z \pm \sqrt{{\cal E}} (b^z u^0 -2b^0 u^z) \Big) u^y/u^0 \\
\\
-b^z u^z \mp \sqrt{{\cal E}} \Big(1 + (u^z)^2\Big)  \\
\\
u^y ( \pm \sqrt{{\cal E}} u^z + b^z )  \\
\\
u^y \\
\\
0 \\
  \end{array}  \right)^T. 
\\ \nonumber
\end{eqnarray}

\noindent
Vectors $\bar{{\bf w}}_{a, 1, \pm}$ and $\bar{{\bf w}}_{a, 2, \pm}$ are well defined in Type I degeneracy. In the case of the Type II degeneracy, coefficients $f_1$ and $f_2$ in (\ref{eq:l.alfven.red}) must be taken equal to $1/\sqrt{2}$.

\subsubsection{Magnetosonic eigenvectors}

  Finally, multiplication of the left eigenvectors in covariant variables and the transformation matrix $(\partial {\tilde {\bf U}}/\partial {\tilde{\bf V}})$, defined in (\ref{mat:u_tilde-v_tilde}), and further algebraic manipulation lead to the following expressions for the components of the magnetosonic eigenvectors in terms of ${\mathcal A}$, $g_1$, $g_2$ and $b_t^\alpha$ (see Ant\'on 2008),
		 
\begin{eqnarray} \label{lm1nue} 
  \bar{l}_{m, \pm,  u^x} = \frac{1}{u^0} \left( \frac{ \mathcal{A} }{a} \Big( (u^0)^2-(u^x)^2 \Big) + 2 B^x b_t^0 ( G+ a^2 ) \right),
\end{eqnarray}

\begin{eqnarray} \label{lm2nue}
  \bar{l}_{m, \pm,  u^y} = \frac{u^y}{u^0} \bigg( \frac {2 B^x \lambda_{m, \pm} b^0_t (G + a^2)}{u^0 - \lambda_{m, \pm} u^x} - \frac{ {\mathcal A} }{a} u^x \bigg) +  \nonumber \\
  g_1 \left\{ (G + 2 a^2) \left( b^0 - \left(\frac{\cal B}{a}\right)_{m,\pm} u^0 \right) - b^x \lambda_{m, \pm} + b^0 \right\},
\end{eqnarray}

\begin{eqnarray}
  \bar{l}_{m, \pm,  u^z} = \frac{u^z}{u^0} \bigg( \frac {2 B^x \lambda_{m, \pm} b^0_t (G + a^2)}{u^0 - \lambda_{m, \pm} u^x} - \frac{ {\mathcal A} }{a} u^x \bigg) + \nonumber \\
  g_2 \left\{ (G + 2 a^2) \left( b^0 - \left(\frac{\cal B}{a}\right)_{m,\pm} u^0 \right) - b^x \lambda_{m, \pm} + b^0 \right\},
\end{eqnarray}

\begin{eqnarray}
  \bar{l}_{m, \pm,  b^y} = -g_1 (u^0 - \lambda_{m, \pm} u^x), 
\end{eqnarray}

\begin{eqnarray}
  \bar{l}_{m, \pm,  b^z} = -g_2 (u^0 - \lambda_{m, \pm} u^x),
\end{eqnarray}
	   
\begin{eqnarray}   
  \bar{l}_{m, \pm,  p} = \frac{ b_t^2 G }{\rho h \Big(a^2 - (G + a^2)c_s^2\Big)} ( \lambda_{m, \pm} u^x - u^0 ) - \frac{G}{ \rho h} \left(\frac{\cal B}{a}\right)_{m,\pm}  b^0_t,
\end{eqnarray}

\begin{eqnarray} \label{lm7nue}
  \bar{l}_{m, \pm,  \rho} = 0. 
\\ \nonumber
\end{eqnarray}

  The inspection of the previous components allows us to conclude that all of them are proportional either to ${\mathcal A}$, $g_1$, $g_2$ or $b_t^\alpha$, which are all zero in the Type II degeneracy. Then the magnetosonic left eigenvectors just obtained are not well suited for Type II degeneracy and must be renormalized.

  In a first step, we eliminate ${\mathcal A}$ taking into acount that, for the magnetosonic eigenvectors\footnote{The way to obtain this expression is a bit tedious: substitute ${\cal B}^2$ in  equation~(\ref{tranv1}), valid for the magnetosonic eigenvectors, by ${\cal B}^2 = (G + a^2) ((\rho h + b_t^2) a^2 -{\cal A})/G$. Previously, obtain this expression starting from the definition of ${\cal B}^2$, equation~(\ref{defcalB}), then substitute $b^2$  in ${\cal E}$ in terms of its normal and tangential parts according to equation~(\ref{apB:b2}), and work out the value of ${\cal B}^2$ from the resulting expression.}

\begin{equation}
  {\mathcal A} = \frac{a^2 (G + a^2) (1 - c_s^2)}{a^2 - (G+a^2) c^2_s} b_t^2.
\end{equation}

  Similarly to the case of the right magnetosonic eigenvectors, those eigenvectors, one of each class, whose eigenvalues are closer of the Alfv\'en eigenvalues, are then divided by $|b_t|$. The new eigenvectors have components:

\begin{eqnarray}
  \bar{l}_{m, \pm,  u^x} = \frac{G+a^2}{u^0} \left( \frac{ (1 - c_s^2 ) |b_t| a }{a^2 - (G+a^2) c^2_s} \Big( (u^0)^2 - (u^x)^2 \Big) + 2 B^x \frac{ b_t^0 }{ |b_t| } \right),
\end{eqnarray}        

\begin{eqnarray}
  \bar{l}_{m, \pm,  u^y} = \frac{ u^y (G+a^2) }{u^0 (u^0 -\lambda_{m, \pm} u^x)} \left( \frac{ (1 - c_s^2 ) |b_t| a u^x}{a^2 - (G+a^2) c^2_s} (u^x \lambda_{m, \pm} - u^0) - 2 B^x \lambda_{m, \pm} \frac{b^0_t}{ |b_t|} \right) {}
\nonumber \\ 
{} + \frac{ g_1 }{ |b_t| } \left\{ (G + 2 a^2) \left( b^0 - \left(\frac{\cal B}{a}\right)_{m,\pm} u^0 \right) - b^x \lambda_{m, \pm} + b^0 \right\},
\end{eqnarray}
 
\begin{eqnarray}
  \bar{l}_{m, \pm,  u^z} = \frac{ u^z (G+a^2) }{u^0 (u^0 -\lambda_{m, \pm} u^x)} \left( \frac{ (1 - c_s^2 ) |b_t|  a u^x}{a^2 - (G+a^2) c^2_s} (u^x \lambda_{m, \pm} -u^0) - 2 B^x \lambda_{m, \pm} \frac{ b^0_t }{ |b_t| } \right) {}
\nonumber \\ 
{} + \frac{g_2}{|b_t|} \left\{  (G + 2 a^2) \left( b^0 - \left(\frac{\cal B}{a}\right)_{m,\pm} u^0 \right) - b^x \lambda_{m, \pm} + b^0 \right\},
\end{eqnarray}

\begin{eqnarray}
  \bar{l}_{m, \pm,  b^y} = -\frac{g_1}{ |b_t| } (u^0 -\lambda_{m, \pm} u^x),
\end{eqnarray}                 
		   
\begin{eqnarray}
  \bar{l}_{m, \pm,  b^z} = -\frac{g_2}{ |b_t| } (u^0 -\lambda_{m, \pm} u^x),
\end{eqnarray}                 
		 
\begin{eqnarray}   
  \bar{l}_{m, \pm,  p} = \frac{ |b_t| \, G } {\rho h (a^2 -(G+a^2)c_s^2)} ( \lambda_{m, \pm} u^x - u^0 ) - \frac{G}{ \rho h} \left(\frac{\cal B}{a}\right)_{m,\pm} \frac{b^0_t}{|b_t|},
\end{eqnarray}

\begin{eqnarray}
  \bar{l}_{m, \pm,  \rho} = 0. 
\\ \nonumber
\end{eqnarray}

  In the previous expressions, $b_t^{\mu} / |b_t|$ follows from equation~(\ref{ubt}), and
			    	      
\begin{eqnarray} \label{g12}
\label{eq:gbt}
  \frac{ g_{1,2} }{ |b_t| } = \frac{ f_{1,2} \, \bigg( (u^0 - \lambda_{m, \pm} u^x)^2 - (1-\lambda_{m, \pm}^2)\Big( (u^y)^2 + (u^z)^2 \Big)^{1/2} }{\left( f_2^2 \alpha_{22} +2 f_1 f_2 \alpha_{12} + f_1^2 \alpha_{11} \right)^{1/2}  }. 
\\ \nonumber
\end{eqnarray}

\noindent
In the Type II degeneracy, coefficients $f_1$ and $f_2$, in (\ref{ubt}) and (\ref{eq:gbt}), must be taken equal to $1/\sqrt{2}$.

  For the remaining two eigenvectors, those of each class whose eigenvalues are farther from the corresponding Alfv\'en eigenvalue, we propose to divide the original components by the expression $b_t^2 / (a^2 - (G + a^2) c_s^2)$, and make the substitution 

\begin{eqnarray}
  \frac{ b_t^2 }{ a^2 - (G + a^2) c_s^2} = \frac{\rho h a^2 - b^2 G }{ G (G+a^2) c_s^2}, 
\\ \nonumber
\end{eqnarray}

\noindent
valid for the magnetosonic eigenvectors. The renormalized eigenvectors are

\begin{eqnarray}
  \bar{l}_{m, \pm,  u^x}  =\frac{1}{u^0} \left(\frac{(1 - c_s^2) a}{c_s^2} \Big( (u^0)^2-(u^x)^2 \Big) + \frac{2 B^x b_t^0  G}
{\rho h a^2 - b^2 G} \right),
\end{eqnarray}

\begin{eqnarray}
  \bar{l}_{m, \pm,  u^y} = \frac{ u^y }{u^0 (u^0 -\lambda_{m, \pm} u^x)} \left( \frac{(1 - c_s^2) a u^x}{c_s^2} (u^x \lambda_{m, \pm} -u^0) - \frac{2 B^x \lambda_{m, \pm} b^0_t G}{\rho h a^2 - b^2 G} \right) {}
\nonumber \\ 
{} + \frac{ g_1 G}{\rho h a^2 - b^2 G } \left\{ (G + 2 a^2) \left( b^0 - \left(\frac{\cal B}{a}\right)_{m,\pm} u^0 \right) - b^x \lambda_{m, \pm} + b^0 \right\},
\end{eqnarray}

\begin{eqnarray}
  \bar{l}_{m, \pm,  u^z} = \frac{ u^z }{u^0 (u^0 -\lambda_{m, \pm} u^x)} \left( \frac{(1- c_s^2) a u^x}{c_s^2} (u^x \lambda_{m, \pm} -u^0) - \frac{2 B^x \lambda_{m, \pm} b^0_t G}{\rho h a^2 - b^2 G} \right) {}
\nonumber \\ 
{} + \frac{ g_2 G}{ \rho h a^2 - b^2 G } \left\{ (G + 2 a^2) \left( b^0 - \left(\frac{\cal B}{a}\right)_{m,\pm} u^0 \right)
  - b^x \lambda_{m, \pm} + b^0 \right\},
\end{eqnarray}
			    
\begin{eqnarray}
  \bar{l}_{m, \pm,  b^y} = - \frac{g_1 G}{\rho h a^2 - b^2 G } (u^0 -\lambda_{m, \pm} u^x),
\end{eqnarray}                 

\begin{eqnarray}
  \bar{l}_{m, \pm,  b^z} = - \frac{g_2 G}{\rho h a^2 - b^2 G } (u^0 -\lambda_{m, \pm} u^x),
\end{eqnarray}                 

\begin{eqnarray}   
  \bar{l}_{m, \pm,  p} = \frac{ (\lambda_{m, \pm} u^x - u^0 ) G }{\rho h (G+a^2)c_s^2}   
  - \frac{G^2}{ \rho h} \frac{b^0_t}{\rho h a^2 -b^2 G } \left(\frac{\cal B}{a}\right)_{m,\pm},
\end{eqnarray}
			   
\begin{eqnarray}
   \bar{l}_{m, \pm,  \rho} = 0. 
\\ \nonumber
\end{eqnarray}

\noindent
In the case of Type II$^\prime$ degeneracy, quantity $b_t^\nu / (\rho h a^2 - b^2 G)$ must be taken equal to 0.

  We note that in all the expressions of this section, quantities depending on the eigenvalues, have been computed in terms of the corresponding $\lambda_{m, \pm}$.

\subsection{Left eigenvectors in conserved variables}

\subsubsection{Entropic eigenvector}

  The last step involves the multiplication of the entropic eigenvector in the reduced system of covariant variables by the matrix $(\partial {\tilde {\bf V}} / \partial {\bf U})$, whose elements have been defined through the expressions (\ref{eq:duidd})-(\ref{eq:drhodbj}). The resulting eigenvector in conserved variables is

\begin{equation}
  L_{e, \, D} = \frac{1}{W} \frac{\partial s }{ \partial \rho} + \frac{1}{Z+ {\bf B}^2} \bigg\{ (Z+b^2) \frac{\partial s}{\partial p} - \rho \left(1 - W^2 - \frac{(b^0)^2}{Z} \right) \frac{\partial s}{\partial \rho} \bigg\} \frac{\partial Z}{\partial D},
\end{equation}

\begin{eqnarray}
  L_{e, \, S^i} = \frac{1}{Z + {\bf B}^2} \Bigg\{\bigg( (Z+b^2) \frac{\partial s}{\partial p} - \rho \left(1 - W^2 - \frac{(b^0)^2}{Z} \right) \frac{\partial s}{\partial \rho} \bigg) \frac{\partial Z}{\partial S^i} + \frac{ B^2 u^i - b^0 B^i}{W} \frac{\partial s }{ \partial p}  \nonumber \\
  -D \left( u^i + \frac{ b^0 B^i}{Z} \right) \frac{\partial s }{ \partial \rho} \Bigg\},
\end{eqnarray}
 
\noindent
with $i = x,y,z$,

\begin{equation}
  L_{e, \, \tau} = - \frac{\partial s}{\partial p} + \frac{1}{Z + {\bf B}^2} \bigg\{ (Z + b^2) \frac{\partial s}{\partial p} - \rho \left(1 - W^2 - \frac{(b^0)^2}{Z} \right) \frac{\partial s}{\partial \rho} \bigg\} \frac{\partial Z}{\partial \tau},
\end{equation}
	     
\begin{eqnarray}
  L_{e, \, B^i} = \frac{1}{Z + {\bf B}^2} \Bigg\{ \bigg( (Z + b^2) \frac{\partial s}{\partial p} - \rho \left(1 - W^2 - \frac{(b^0)^2}{Z} \right) \frac{\partial s}{\partial \rho} \bigg) \frac{\partial Z}{\partial B^i} {}
\nonumber \\
{} - \rho \left( 2 B^i (1 - W^2) + b^0 \left( \frac{W S^i}{Z} + u^i \right) \right) \frac{\partial s}{\partial \rho} + \left( \frac{ B^2 b^i - b^0 S^i}{W} + \left(2 - \frac{1}{W^2} \right) Z B^i \right) \frac{\partial s}{\partial p}  \Bigg\}, 
\end{eqnarray}     
	  
\noindent  
with $i = y,z$.

\subsubsection{Alfv\'en eigenvectors}

  The same operation performed now with the Alfv\'en eigenvectors leads to the corresponding eigenvectors in conserved variables

\begin{eqnarray}
\label{eq:l.alf.cons}
  {\bf L}_{a, \pm} = f_1 {\bf W}_{a, 1, \pm} + f_2 {\bf W}_{a, 2, \pm}, 
\end{eqnarray}

\noindent				 
where the components of vectors ${\bf W}_{a, 1, \pm}$ and ${\bf W}_{a, 2, \pm}$ are  

\begin{eqnarray}
  W_{a, 1, \pm, D} = C_{a, 1, \pm} \frac{\partial Z}{\partial D},
\end{eqnarray}   
			   
\begin{eqnarray}
  W_{a, 1, \pm, S^x} = C_{a, 1, \pm} \frac{\partial Z}{\partial S^x} + ({\bf B}^2 u^x - b^0 B^x) \frac{u^z}{u^0} - u^x b^y (B^y u^z - B^z u^y) {}
\nonumber \\
{} + ({\cal E} u^0 \pm \sqrt{{\cal E}} b^0 ) u^z (u^x - 2 a) \pm \sqrt{{\cal E}} \Big( u^x B^z - u^z B^x - u^y u^x (u^z B^y - u^y B^z) \Big),
\end{eqnarray}   
			   
\begin{eqnarray}
  W_{a, 1, \pm, S^y} = C_{a, 1, \pm} \frac{\partial Z}{\partial S^y} + ({\bf B}^2 u^y - b^0 B^y ) \frac{u^z}{u^0}  + b^0 b^y u^z - u^y b^y (B^y u^z - B^z u^y) {}
\nonumber \\
{} \pm \sqrt{{\cal E}} \bigg( b^y u^z u^0 + \Big(1 + (u^y)^2 \Big) (u^y B^z - u^z B^y) \bigg),
\end{eqnarray}           

\begin{eqnarray}
  W_{a, 1, \pm, S^z} = C_{a, 1, \pm} \frac{\partial Z}{\partial S^z} + ({\bf B}^2 u^z - b^0 B^z) \frac{u^z}{u^0} - b^0 b^y u^y - u^z b^y (B^y u^z - B^z u^y) {}
\nonumber \\
{} + {\cal E} (u^0)^2 (u^0 - \lambda_{a \pm} u^x) \pm \sqrt{{\cal E}} \Big(b^0 u^x a + b^z u^z u^0 - u^z u^y (u^z B^y - u^y B^z) \Big),
\end{eqnarray}  
			   
\begin{eqnarray}
  W_{a, 1, \pm, \tau} = C_{a, 1, \pm} \frac{\partial Z}{\partial \tau} - u^z (Z + {\bf B}^2),
\end{eqnarray}   
	      
\begin{eqnarray}
  W_{a, 1, \pm, B^y} = C_{a, 1, \pm} \frac{\partial Z}{\partial B^y}
+ \left\{ \frac{{\bf B}^2 b^y - b^0 S^y}{u^0} + \left(2 - \frac{1}{(u^0)^2} \right) Z B^y \right\} u^z \nonumber \\
+ b^y (B^z u^y - B^y u^z) \left( \frac{u^y b^0}{u^0} + 2 B^y \frac{1-(u^0)^2}{u^0} \right) {}
\nonumber \\
{} + ({\cal E} u^0 \pm b^0 \sqrt{{\cal E}} ) \left( \frac{B^y u^z}{u^0} + (1 - au^x)\Big(b^0 u^y -2 (u^0)^2 B^y\Big) \frac{u^z}{u^0} -\frac{S^y u^0 }{Z} b^x u^z a \right) {}
\nonumber \\  
{} \mp \sqrt{{\cal E}} \left\{ S^y u^z \left(1 + \frac{b^x B^x u^0}{Z} \right) - \left( {\cal E} u^0 - \frac{(b^0)^2}{u^0} \right) u^y u^z + \Big(b^0 u^y - 2 (u^0)^2 B^y \Big) \frac{u^z u^x B^x}{u^0}  \right. {}
\nonumber \\
{} \left. - \Big(B^z + (B^z u^y - B^y u^z ) u^y\Big) \left( \frac{u^y b^0 }{u^0} + 2 B^y \frac{1 - (u^0)^2}{u^0} \right) \right\},
\end{eqnarray}

\begin{eqnarray}
  W_{a, 1, \pm, B^z} = C_{a, 1, \pm}  \frac{\partial Z}{\partial B^z} 
+ \left\{ \frac{{\bf B}^2 b^z - b^0 S^z}{u^0} + \left(2 - \frac{1}{(u^0)^2} \right) Z B^z \right\} u^z  \nonumber \\
+b^y (B^z u^y - B^y u^z) \left( \frac{u^z b^0}{u^0} + 2 B^z \frac{1 - (u^0)^2}{u^0} \right)  \nonumber \\
+ ({\cal E}u^0 \pm b^0 \sqrt{{\cal E}} ) \bigg( \frac{B^z u^z }{u^0} + (1 - au^x)\Big(b^0 u^z -2 (u^0)^2 B^z \Big) \frac{u^z}{u^0} - \frac{S^z u^0 }{Z} b^x u^z a \bigg)  \nonumber \\
\mp \sqrt{{\cal E}}  \left\{ S^z u^z \left(1 + \frac{b^x B^x u^0}{Z} \right)  + \left( {\cal E} u^0  - \frac{(b^0)^2}{u^0} \right) \Big( (u^0)^2 - \lambda_{a \pm}  u^0 u^x - (u^z)^2 \Big) + \right.  \nonumber \\
 \left. \Big(b^0 u^z - 2 (u^0)^2 B^z \Big) \frac{u^z u^x B^x}{u^0} 
 - \Big(B^z + (B^z u^y - B^y u^z ) u^y\Big) \left( \frac{u^z b^0 }{u^0} + 2 B^z \frac{1 - (u^0)^2}{u^0} \right) \right\}.
\end{eqnarray}

\begin{eqnarray}
  W_{a, 2, \pm, D} = C_{a, 2, \pm} \frac{\partial Z}{\partial D},
\end{eqnarray}   

\begin{eqnarray}
  W_{a, 2, \pm, S^x} = C_{a, 2, \pm}  \frac{\partial Z}{\partial S^x} + ({\bf B}^2 u^x - b^0 B^x) \frac{u^y}{u^0} - u^x b^z (B^z u^y -B^y u^z) {}
\nonumber \\
{} + ({\cal E}u^0 \pm \sqrt{{\cal E}} b^0 ) u^y(u^x - 2 a) \pm \sqrt{{\cal E}} \Big( u^x B^y - u^y B^x - u^z u^x (u^y B^z - u^z B^y) \Big),   
\end{eqnarray}   

\begin{eqnarray}
  W_{a, 2, \pm, S^y} = C_{a, 2, \pm} \frac{\partial Z}{\partial S^y} + ({\bf B}^2 u^y - b^0 B^y) \frac{u^y}{u^0} - b^0 b^z u^z - 
u^y b^z (B^z u^y - B^y u^z) {}
\nonumber \\
{} + {\cal E} (u^0)^2 (u^0 - \lambda_{a \pm} u^x) \pm \sqrt{{\cal E}} \Big(b^0 u^x a + b^y u^y u^0 - u^z u^y (u^y B^z - u^z B^y) \Big),
\end{eqnarray}  
			
\begin{eqnarray}
  W_{a, 2, \pm, S^z} = C_{a, 2, \pm} \frac{\partial Z}{\partial S^z} + ({\bf B}^2 u^z - b^0 B^z ) \frac{u^y}{u^0} + b^0 b^z u^y - u^z b^z (B^z u^y -B^y u^z) {}
\nonumber \\
{} \pm \sqrt{{\cal E}} \bigg( b^z u^y u^0 + \Big(1 + (u^z)^2\Big) (u^z B^y - u^y B^z) \bigg),
\end{eqnarray}           
	   
\begin{eqnarray}
  W_{a, 2, \pm, \tau} = C_{a, 2, \pm} \frac{\partial Z}{\partial \tau} - u^y (Z + {\bf B}^2),
\end{eqnarray}
			     
\begin{eqnarray}
  W_{a, 2, \pm, B^y} =  C_{a, 2, \pm}  \frac{\partial Z}{\partial B^y}
+ \left\{ \frac{{\bf B}^2 b^y -b^0 S^y}{u^0}  + \left(2 - \frac{1}{(u^0)^2} \right) Z B^y \right\} u^y \nonumber \\
+ b^z (B^y u^z -B^z u^y) \left( \frac{u^y b^0}{u^0} + 2 B^y \frac{1 - (u^0)^2}{u^0} \right) {}
\nonumber \\
{} + ({\cal E}u^0 \pm b^0 \sqrt{{\cal E}} ) \bigg( \frac{B^y u^y }{u^0} + (1 - au^x) \Big( b^0 u^y -2 (u^0)^2 B^y \Big) 
\frac{u^y}{u^0} -\frac{S^y u^0 }{Z} b^x u^y a \bigg) {}
\nonumber \\
{}  \mp \sqrt{{\cal E}} \left\{ S^y u^y \left(1 + \frac{b^x B^x u^0}{Z} \right) + \left( {\cal E} u^0 - \frac{(b^0)^2}{u^0}
\right) \Big( (u^0)^2 - \lambda_{a \pm} u^0 u^x - (u^y)^2 \Big) \right. 
\nonumber \\
\left. + \Big(b^0 u^y - 2 (u^0)^2 B^y \Big) \frac{u^y u^x B^x}{u^0} \right. {}
\nonumber \\
{} \left.  - \Big(B^y + (B^y u^z - B^z u^y ) u^z\Big) \left( \frac{u^y b^0 }{u^0} + 2 B^y \frac{1 - (u^0)^2}{u^0} \right) \right\},
\end{eqnarray}

\begin{eqnarray}
  W_{a, 2, \pm, B^z} = C_{a, 2, \pm} \frac{\partial Z}{\partial B^z} 
 + \left\{ \frac{{\bf B}^2 b^z -b^0 S^z}{u^0} + \left(2 - \frac{1}{(u^0)^2} \right) Z B^z \right\} u^y \nonumber \\
 + b^z (B^y u^z - B^z u^y) \left( \frac{u^z b^0}{u^0} + 2 B^z \frac{1 - (u^0)^2}{u^0} \right) {}
\nonumber \\
{} + ({\cal E} u^0 \pm b^0 \sqrt{{\cal E}} ) \bigg( \frac{B^z u^y }{u^0} + (1 - a u^x)\Big( b^0 u^z -2 (u^0)^2 B^z \Big) \frac{u^y}{u^0} -\frac{S^z u^0 }{Z} b^x u^y a \bigg) {}
\nonumber \\
{} \mp \sqrt{{\cal E}}  \left\{S^z u^y \left(1 + \frac{b^x B^x u^0}{Z} \right) - \left( {\cal E} u^0 - \frac{(b^0)^2}{u^0} \right) u^y u^z + \Big(b^0 u^z -2 (u^0)^2 B^z \Big) \frac{u^y u^x B^x}{u^0} \right. {}
\nonumber \\
{} \left. - \Big(B^y + (B^y u^z -B^z u^y ) u^z\Big) \left( \frac{u^z b^0 }{u^0} + 2 B^z \frac{1-(u^0)^2}{u^0} \right)  \right\}.
\end{eqnarray}
			
\noindent
In the previous expressions,

\begin{eqnarray}
  C_{a, 1, \pm} = -{\cal E} u^0 u^z (1 - a u^x) - ({\cal E} u^0 \pm \sqrt{{\cal E}} b^0) \frac{b^0}{Z} \Big(b^z (1 + a u^x) + u^y (b^z u^y - b^y u^z)  -2 b^x a u^z \Big) \nonumber \\
+ (Z + b^2) u^z 
-b^y (b^z u^y - b^y u^z)  \Big( (u^0)^2 - 1\Big) \nonumber \\
\pm \sqrt{{\cal E}} \left\{ u^z \left((u^0)^2 u^x \mathcal{B} + 
\frac{b^0 b^x B^x}{Z} \right) - \Big((u^0)^2 - 1\Big) \Big(b^z + u^y(b^z u^y - b^y u^z)\Big) \right\},
\end{eqnarray}

\begin{eqnarray}
  C_{a, 2, \pm} = -{\cal E} u^0 u^y (1 - a u^x) - ({\cal E} u^0 \pm \sqrt{{\cal E}} b^0) \frac{b^0}{Z} \Big(b^y (1 + a u^x) + u^z (b^y u^z - b^z u^y)  -2 b^x a u^y\Big) \nonumber \\
  + (Z + b^2) u^y {}
{} -b^z (b^y u^z - b^z u^y)  ( (u^0)^2 - 1) \nonumber \\
\pm \sqrt{{\cal E}} \left\{ u^y \left((u^0)^2u^x \mathcal{B} + 
\frac{b^0 b^x B^x}{Z} \right) - \Big((u^0)^2 - 1\Big) \Big(b^y + u^z(b^y u^z - b^z u^y)\Big) \right\}.
\end{eqnarray}  
		 
  Vectors ${\bf W}_{a, 1, \pm}$ and ${\bf W}_{a, 2, \pm}$ are well defined in Type I degeneracy. In the case of the Type II degeneracy, coefficients $f_1$ and $f_2$ in (\ref{eq:l.alf.cons}) must be taken equal to $1/\sqrt{2}$.

\subsubsection{Magnetosonic eigenvectors}

  We present now the magnetosonic eigenvectors in conserved variables from the transformation of the corresponding eigenvectors in the reduced system of covariant variables before the renormalization process for Type II degeneracy, those defined through expressions (\ref{lm1nue})-(\ref{lm7nue}). We start by defining the quantities

\begin{eqnarray}
  H_{m, \pm} = (G+2a^2) \left(b^0 - \left(\frac{\cal B}{a}\right)_{m,\pm}  \, u^0 \right) - (b^x \lambda_{m, \pm} - b^0),
\end{eqnarray} 

\begin{eqnarray}      
  C_{m, \pm} = \frac{ b^2_t}{a^2 -(G+a^2) c^2_s}\times \hspace{2cm}\nonumber \\
  \times \left\{(c^2_s-1) (G+a^2) a u^0 \left(u^x + \frac{b^0 B^x}{ Z} \right) + \left( (u^0)^2 + \frac{b^2}{\rho h} \right) (\lambda_{m, \pm} u^x - u^0) G \right\} {}
\nonumber \\
{} - b^0_t \left\{ \frac{ 2 (G+a^2) B^x u^0 }{u^0 -\lambda_{m, \pm} u^x} \left( u^0 a + \lambda_{m, \pm} + \frac{ b^0 \mathcal{B} }{Z} \right)  + \Big( (u^0)^2 + \frac{b^2}{\rho h} \Big)  G \left(\frac{\cal B}{a}\right)_{m,\pm} \right\} {}
\nonumber \\
{} - g_2 \left\{ u^0 H_{m, \pm} \left( u^0 u^y + \frac{b^0 b^y}{Z} \right) + \bigg(b^y \Big( (u^0)^2 -1 \Big)  - u^0 u^y b^0 \frac{2Z +b^2}{Z} \bigg) (u^0 -\lambda_{m, \pm} u^x) \right\}  {}
\nonumber \\
{} - g_1 \left\{ u^0 H_{m, \pm} \left( u^0 u^z + \frac{b^0 b^z}{Z} \right) + \bigg( b^z \Big( (u^0)^2 -1 \Big) -u^0 u^z b^0 \frac{2Z +b^2}{Z} \bigg) (u^0 -\lambda_{m, \pm} u^x)  \right\}. 
\end{eqnarray}

  Now, the components of the magnetosonic eigenvectors read,

\begin{equation}
  L_{m ,\pm, D} = C_{m,\pm} \frac{\partial Z }{\partial D},
\end{equation}

\begin{eqnarray}
  L_{m ,\pm, S^x} = C_{m,\pm} \frac{\partial Z }{\partial S^x} 
+ \frac{b^2_t}{a^2 -(G+a^2) c^2_s} \times \nonumber \\
\times \bigg\{ (1-c^2_s) a (G+a^2) \left( (u^0)^2 + \frac{ (B^x)^2}{\rho h}  \right) + \frac{ B^2 u^x - b^0 B^x}{ \rho h u^0} (\lambda_{m, \pm} u^x - u^0) G \bigg\} {}
\nonumber \\
{} + b^0_t \bigg\{ \frac{2(G+a^2) B^x }{u^0 -\lambda_{m, \pm} u^x} \left( (1+a u^x) u^0 + \frac{ B^x \mathcal{B} }{\rho h } \right) - \frac{ B^2 u^x -b^0 B^x}{ \rho h u^0} \,\left(\frac{\cal B}{a}\right)_{m,\pm}\, G \bigg\} {}
\nonumber \\
{} + g_1 \bigg\{ u^0 H_{m, \pm} \left(u^x u^y + \frac{B^x b^y}{\rho h u^0} \right) - (u^0 -\lambda_{m, \pm} u^x) \bigg( u^y \left( b^x u^0 + B^x \frac{b^2}{\rho h} \right) - B^y u^x \bigg) \bigg\} {}
\nonumber \\
{} + g_2 \bigg\{ u^0 H_{m, \pm} \left(u^x u^z + \frac{B^x b^z}{\rho h u^0} \right) - (u^0 - \lambda_{m, \pm} u^x) \bigg( u^z \left( b^x u^0 + B^x \frac{b^2} {\rho h} \right) - B^z u^x \bigg) \bigg\},
\end{eqnarray}
			   
\begin{eqnarray}
  L_{m ,\pm, S^y} = C_{m,\pm} \frac{\partial Z }{\partial S^y} + \frac{b^2_t}{a^2 -(G+a^2) c^2_s} \times \nonumber \\
  \times \bigg\{ (1-c^2_s) a (G+a^2) \frac{ B^x B^y}{\rho h} + \frac{ B^2 u^y - b^0 B^y}{ \rho h u^0} (\lambda_{m, \pm} u^x - u^0) G \bigg\} {}
\nonumber \\
{} + b^0_t \bigg\{ \frac{2(G+a^2) B^x  }{u^0 -\lambda_{m, \pm} u^x} \left( a u^y u^0 + \frac{ B^y \mathcal{B} }{\rho h } \right) - \frac{ B^2 u^y -b^0 B^y}{ \rho h u^0} \, \left(\frac{\cal B}{a}\right)_{m,\pm} \, G \bigg\} {}
\nonumber \\
{} + g_1 \bigg\{ u^0 H_{m, \pm} \left(1+(u^y)^2 + \frac{B^y b^y}{\rho h u^0} \right) - (u^0 - \lambda_{m, \pm} u^x) \bigg( u^y B^y \frac{b^2}{\rho h} + b^0 \Big(1 + (u^y)^2\Big) \bigg) \bigg\} {}
\nonumber \\
{} + g_2 \bigg\{ u^0 H_{m, \pm} \left(u^y u^z + \frac{B^y b^z}{\rho h u^0} \right) - (u^0 - \lambda_{m, \pm} u^x) \bigg( u^z \left( b^y u^0 + B^y \frac{b^2}{\rho h} \right) - B^z u^y \bigg) \bigg\},
\end{eqnarray}

\begin{eqnarray}
  L_{m ,\pm, S^z} = C_{m,\pm} \frac{\partial Z }{\partial S^z} + \frac{b^2_t}{a^2 -(G+a^2) c^2_s}\times \nonumber \\
  \times\bigg\{ (1-c^2_s) a (G+a^2) \frac{ B^x B^z}{\rho h} + \frac{ B^2 u^z -b^0 B^z}{ \rho h u^0} (\lambda_{m, \pm} u^x -u^0) G \bigg\} {} 
\nonumber \\
{} + b^0_t \bigg\{ \frac{2(G+a^2) B^x  }{u^0 -\lambda_{m, \pm} u^x} \left( a u^z u^0 + \frac{B^z \mathcal{B} }{\rho h } \right) - \frac{ B^2 u^z -b^0 B^z}{ \rho h u^0} \,\left(\frac{\cal B}{a}\right)_{m,\pm} \, G \bigg\} {}
\nonumber \\
{} + g_1 \bigg\{ u^0 H_{m, \pm} \left( u^z u^y + \frac{B^z b^y}{\rho h u^0} \right) -(u^0 -\lambda_{m, \pm} u^x) \bigg( u^y \left( b^z u^0+  B^z \frac{b^2}{\rho h} \right) - B^y u^z ) \bigg) \bigg\} {}
\nonumber \\
{} + g_2 \bigg\{ u^0 H_{m, \pm} \left(1 +(u^z)^2 + \frac{B^z b^z}{\rho h u^0} \right) -(u^0 -\lambda_{m, \pm} u^x) \bigg(  u^z B^z \frac{b^2}{\rho h} + b^0 \Big(1+(u^z)^2 \Big) \bigg)  \bigg\},
\end{eqnarray}

\begin{eqnarray}
  L_{m , \pm,\tau} = C_{m,\pm} \frac{\partial Z }{\partial \tau} -
  \left( (u^0)^2 +\frac{B^2}{\rho h} \right) G
  \bigg\{ \frac{b^2_t }{ a^2 -(G+a^2)c^2_s }
  (\lambda_{m, \pm} u^x -u^0) - \, b^0_t \left(\frac{\cal B}{a}\right)_{m,\pm} \bigg\},
\end{eqnarray}
			      
\begin{eqnarray}
  L_{m ,\pm, B^y} = C_{m,\pm} \frac{\partial Z }{\partial B^y} + \frac{b^2_t}{a^2 -(G+a^2) c^2_s} \times \nonumber \\
  \times \bigg\{ (1-c^2_s) a (G+a^2) \left( \frac{ S^y B^x}{\rho h} - 2 B^y u^x u^0 \right) {}
\nonumber \\
{} + \left(\frac{-b^0 S^y + B^2 b^y}{\rho h \, u^0} + \Big( 2 (u^0)^2 - 1 \Big) B^y \right) G (\lambda_{m, \pm} u^x -u^0) \bigg\} {}
\nonumber \\ 
{} + b^0_t \Bigg\{ \frac{2 B^x (G+a^2)}{u^0 - \lambda_{m, \pm} u^x} \bigg(b^0 a u^y + 
\frac{ S^y \mathcal{B}}{\rho h} -2 u^0 B^y (\lambda_{m, \pm} + u^0 a)\bigg) \nonumber \\
 +\bigg(\frac{b^0 S^y - B^2 b^y}{\rho h u^0} - \Big(2 (u^0)^2 - 1\Big) B^y \bigg) \left(\frac{\cal B}{a}\right)_{m,\pm} \!\!\!\! G \Bigg\} {}
\nonumber \\
{} + g_1 \Bigg\{H_{m, \pm} \bigg(b^0 \Big(1 +(u^y)^2\Big) + \frac{ S^y b^y}{\rho h} -2 (u^0)^2 B^y u^y \bigg) - (u^0 -\lambda_{m, \pm} u^x) \bigg( {\cal E}u^0 
\nonumber \\
+ S^y u^y \left( 1 + \frac{b^2}{\rho h} \right) {}
{} + \frac{u^y b^0 }{u^0}  (b^0 u^y -B^y) \bigg)  -2 B^y \bigg( u^0 u^y b^0 + \frac{1- (u^0)^2}{u^0}  B^y \bigg) \Bigg\}  {}
  \nonumber \\
{} + g_2 \Bigg\{ H_{m, \pm} \bigg( \Big(b^0 u^y -2 (u^0)^2 B^y \Big) u^z + \frac{ S^y b^z}{\rho h} \bigg)  -(u^0 -\lambda_{m, \pm} u^x) \bigg( S^y u^z \left( 1 + \frac{b^2}{\rho h} \right) {}
\nonumber \\
{} + \frac{u^y b^0 }{u^0}  (b^0 u^z -B^z) \bigg) -2 B^y \bigg( u^0 u^z b^0 + \frac{1- (u^0)^2}{u^0}  B^z \bigg) \Bigg\} ,
\end{eqnarray}
		 
\begin{eqnarray}
  L_{m ,\pm, B^z} = C_{m,\pm} \frac{\partial Z }{\partial B^z} + \frac{b^2_t}{a^2 -(G+a^2) c^2_s} \times \nonumber \\
  \times \bigg\{ (1-c^2_s) a (G+a^2) \left( \frac{ S^z B^x}{\rho h} -2 B^z u^x u^0 \right) {}
\nonumber\\
{} + \left(\frac{-b^0 S^z + B^2 b^z}{\rho h \, u^0} + \Big( 2 (u^0)^2 - 1 \Big) B^z  \right) G (\lambda_{m, \pm} u^x -u^0) \bigg \} {}
\nonumber \\
{} + b^0_t \Bigg\{  \frac{2 B^x (G+a^2)}{u^0 -\lambda_{m, \pm} u^x} \bigg(b^0 a u^z + \frac{ S^z \mathcal{B}}{\rho h} -2 u^0 B^z (\lambda_{m, \pm} + u^0 a)\bigg) \nonumber \\ 
+ \bigg(\frac{b^0 S^z - B^2 b^z}{\rho h u^0} - \Big( 2 (u^0)^2 - 1 \Big) B^z \bigg) \left(\frac{\cal B}{a}\right)_{m,\pm} \!\!\!\! G \Bigg\} {}
\nonumber \\
{} + g_1 \Bigg\{ H_{m, \pm} \bigg( \Big(b^0 u^z -2 (u^0)^2 B^z\Big) u^y + \frac{ S^z b^y}{\rho h} \bigg)  -(u^0 -\lambda_{m, \pm} u^x) \bigg( S^z u^y \left( 1 + \frac{b^2}{\rho h} \right) {}
\nonumber \\
{} + \frac{u^z b^0 }{u^0}  (b^0 u^y -B^y) \bigg) -2 B^z \bigg( u^0 u^y b^0 + \frac{1- (u^0)^2}{u^0}  B^y \bigg) \Bigg\} {}
\nonumber \\   
{} + g_2 \Bigg\{ H_{m, \pm} \bigg(b^0 \Big(1 +(u^z)^2\Big) + \frac{ S^z b^z}{\rho h} -2 (u^0)^2 B^z u^z\bigg) - (u^0 -\lambda_{m, \pm} u^x) \bigg( {\cal E}u^0 
\nonumber \\
+ S^z u^z  \left( 1 + \frac{b^2}{\rho h} \right) {}
{} + \frac{u^z b^0 }{u^0}  (b^0 u^z -B^z) \bigg) -2 B^z \bigg( u^0 u^z b^0 + \frac{1- (u^0)^2}{u^0}  B^z \bigg) \Bigg\}. 
\end{eqnarray}  

  Our proposal of renormalization for Type II degeneracy is the following. For those magnetosonic vectors, one of each class, whose eigenvalue is closer to the corresponding Alfv\'en eigenvalue, we divide by $|b_t|$. Hence the following replacements in the components of the eigenvector (defined previously) must be performed
		  
\begin{equation} \label{eff1}
  \frac{ b_t^2 }{a^2 - (G+a^2) c^2_s}  \longrightarrow 
  \frac{ |b_t| }{a^2 -(G+a^2) c^2_s}, \qquad \qquad
  b^0_t  \longrightarrow   \frac{  b^0_t  }{|b_t|},
\end{equation}
			
\begin{equation}
  g_1 \longrightarrow \frac{ g_1 }{ |b_t| }, 
  \qquad \qquad 
  g_2 \longrightarrow \frac{ g_2 }{ |b_t| },
\end{equation}

\noindent 
where the expressions that must be used in the case of Type II degeneracy are defined in (\ref{ubt}) and (\ref{g12}).

  For the remaining two eigenvectors, those of each class whose eigenvalues are farther from the corresponding Alfv\'en eigenvalue, we propose to divide the original components by the expression $b_t^2 / (a^2 - (G + a^2) c_s^2)$, and make the substitution 

\begin{equation}
  \frac{ b_t^2 }{ a^2 - (G + a^2) c_s^2} = 
  \frac{\rho h a^2 - b^2 G }{ G (G+a^2) c_s^2}.
\end{equation}
   
\noindent
Hence the following replacements in the vector components shall be performed

\begin{equation} 
  \frac{ b_t^2 }{a^2 - (G+a^2) c^2_s} \longrightarrow  1,
  \qquad \qquad b^0_t \longrightarrow 
  \frac{ b^0_t G(G+a^2) c^2_s }{ \rho h a^2 -b^2 G },
\end{equation}
			
\begin{equation}
  g_1 \longrightarrow 
  \frac{ g_1 G(G+a^2) c^2_s }{ \rho h a^2 -b^2 G }, 
  \qquad \qquad 
  g_2 \longrightarrow
  \frac{ g_2 G(G+a^2) c^2_s }{ \rho h a^2 -b^2 G },
\end{equation} 

\noindent  
where the expressions that must be used in the case of Type II degeneracy are defined in (\ref{ubt}) and (\ref{g12}).

\subsubsection{Normalization factors}
Finally, for the implementation in the numerical code the left eigenvectors just defined must be multiplied by normalization factors in order to fulfill the condition ${\bf L}_i \, {\bf R}_j=\delta_{ij}$. 

\section{HRSC scheme for relativistic MHD}
\label{numflux}

  As established in the Introduction, HRSC schemes have become a standard tool in numerical relativistic hydrodynamics and magnetohydrodynamics. In this Section we describe the ingredients of our numerical algorithm, the most important being the linearized Riemann solver based on the renormalized spectral decomposition of the RMHD Jacobians described in the preceding sections. Although the specific expressions for the spectral decomposition have been worked out explicitly for the $x$-direction, symmetry arguments allows one to obtain the equivalent expressions for the remaining directions easily. The starting point are the equations of RMHD written as a first-order, flux-conservative, hyperbolic system of partial differential equations (\ref{eq:fundsystem}). All the ingredients described in this section have been implemented in the code RGENESIS (Leismann et al. 2005), that we have used to perform all the numerical tests and applications presented in this paper.

\subsection{Integral form of the RMHD equations}

 To apply this technique to the system (\ref{eq:fundsystem}) we need first to obtain an integral form of the equations. Integrating the system on the three-dimensional volume bounded by the surfaces ${\Sigma_{x},\Sigma_{x+\Delta x}}$, ${\Sigma_{y},\Sigma_{y+\Delta y}}$,
${\Sigma_{z},\Sigma_{z+\Delta z}}$, that connect two temporal slices, we obtain

\begin{eqnarray}
  \frac{d \bar{\bf F}^{0}}{dt} =     -\left(\int_{\Sigma_{x+\Delta x}}\!\!\!\!\!\!\!\!\!\widehat{\bf F}^{1} dy dz -\int_{\Sigma_{x}} \widehat{\bf F}^{1} dy dz\right)  -\left(\int_{\Sigma_{y+\Delta y}}\!\!\!\!\!\!\!\!\!\widehat{\bf F}^{2} dx dz -\int_{\Sigma_{y}} \widehat{\bf F}^{2} dx dz\right)  \nonumber \\
  -\left(\int_{\Sigma_{z+\Delta z}}\!\!\!\!\!\!\!\!\!\widehat{\bf F}^{3} dx dy -\int_{\Sigma_{z}} \widehat{\bf F}^{3} dx dy\right),
\label{eq:system}
\end{eqnarray}

\noindent
where

\begin{equation}
\bar{\bf F}^{0}=
\frac{1}{\Delta V}\int_{x}^{x+\Delta x} \int_{y}^{y+\Delta y}
\int_{z}^{z+\Delta z} {\bf F}^{0} dxdydz
\end{equation}

\noindent
and

\begin{equation}
\Delta V= \Delta x\Delta y\Delta z.
\end{equation}

\noindent
The carets appearing on the fluxes denote that these fluxes, which are calculated at cell interfaces where the flow conditions can be discontinuous, are obtained by solving Riemann problems between the corresponding numerical cells. These numerical fluxes are discussed below.

\subsection{Induction equation and divergence-free condition}

 Independently of the particular Riemann solver we use for calculating the numerical fluxes at cell interfaces, the main advantage of the numerical procedure just described to advance in time the system of equations, is that those variables which obey a conservation law are, by construction, conserved during the evolution as long as the balance between the fluxes at the boundaries of the computational domain is zero. Although this is one of the most important properties a hydrodynamical code should fulfill, there is an additional condition we should demand to any magnetohydrodynamical code: at any time in the numerical evolution, the divergence of the magnetic field must be zero. This condition is not satisfied by construction if we use equation~(\ref{eq:system}) to evolve the components of the magnetic field. In order to preserve the divergence free condition, we use the constraint transport method designed by \citet{evans88}, and extended to HRSC schemes by \citet{ryu:95}. The essential of this method is to use Stokes theorem after integration of the induction equation on the surfaces that bound our numerical cells. This allows one to obtain evolution equations for the magnetic flux at each cell interface in terms of the electromotive force around the contour defined by the boundary of the cell. Since the sign of the electromotive force changes if the direction of the contour is changed, the magnetic flux through the faces of a numerical cell will be constant in time. If initially this flux is zero, this condition will be fulfilled during the evolution and the divergence free condition numerically satisfied.

  Taking, for example, the magnetic flux $\Phi_{\Sigma_{z}}$, through the surface $\Sigma_{z}$, defined by $z={\rm const.}$, and the remaining two coordinates spanning the intervals from $x$ to $x+\Delta x$, and from $y$ to $y+\Delta y$, we have

\begin{equation}
\Phi_{\Sigma_{z}}=\int_{\Sigma_{z}} \vec{B} \cdot d\vec{\Sigma}.
\end{equation}
  
The discretized form of the induction equation based on the method just described is the following

\begin{equation}
 \frac{d \Phi_{\Sigma_{z}}}{dt } =
 \int_{\partial(\Sigma_{z})}
 \widehat{\Omega}_i dx^i.
\label{eq:mflux}
\end{equation}

\noindent
$\Omega_i=\varepsilon_{ijk}v^jB^k$ is the electric field with opposite sign, and the caret denotes again that quantities $\widehat{\Omega}_i$ are calculated at the edges of the numerical cells, where they can be discontinuous. At each edge, as we will describe below, these quantities are calculated using the solution of four Riemann problems between the corresponding faces whose intersection defines the edge. However, irrespective of the expression we use for calculating $\widehat{\Omega}_i$, the method to advance the magnetic fluxes at the faces of the numerical cells satisfies, by construction, the divergence constraint.

\subsection{Spatial and temporal order of accuracy}

 In order to increase the spatial accuracy of the numerical solution, the primitive variables defined in equation~(\ref{eq:primitive}) can be reconstructed at the cell interfaces before the actual computation of the numerical fluxes. We use standard second order (MINMOD) and third order (PPM) reconstruction procedures to compute the values of $p$, $\rho$, $v_i$ and $B^i$ ($i= 1,2,3$) at both sides of each numerical interface. However, when computing the numerical fluxes along a certain direction, we do not allow for discontinuities in the magnetic field component along that direction. Furthermore, the equations in integral form are advanced in time using the method of lines in conjunction with a second/third order, TVD Runge-Kutta method \citep{shu:88}.

\subsection{Numerical fluxes}
\label{ss:numflux}

  As discussed in the Introduction, there are several strategies to calculate the numerical fluxes in HRSC schemes. Our procedure is based on the construction of a linearized Riemann solver after the original work of Roe (1981). The main idea of Roe's approach is to linearize the hyperbolic system of equations and to use the analytical solution of this linearized system at each interface to obtain the fluxes across the interfaces. These fluxes are then employed in the discretized equations to advance in time the variables. Linearizing the system amounts to choose a constant Jacobian matrix at each interface. In  the original Roe's solver \citep{roe81}, this constant Jacobian matrix is built to give the exact solution of the nonlinear Riemann problem if a discontinuous wave is located at the interface. However, it is possible to relax the conditions imposed to the Roe's matrix and to build simply the Jacobian matrix for some intermediate state between the two states separated by the interface, and still producing accurate solutions. Riemann solvers following this approach are commonly known as Roe-type Riemann solvers. In our numerical implementation, we use the arithmetic mean between the variables at each side of the interface to obtain the intermediate state. The variables we use to obtain the intermediate state are $\rho, p, v^x, v^y, v^z, B^y$ and $B^z$. Knowing the intermediate state we obtain the Jacobian matrix, and the eigenvalues and right and left eigenvectors in conserved variables as described in Sects.~\ref{s:csrmhde}, \ref{s:recv} and \ref{s:lecv}. Contrary to the HLL strategy and its sequels, Roe-type Riemann solvers calculate the numerical fluxes consistently with the breakout of the original discontinuity in the full set of characteristic waves (full wave decomposition Riemann solver; FWD, in the next). Explicitly, numerical fluxes are computed according to

\begin{equation} \label{froeI}
  {\widehat{\bf F}}^{\rm FWD} = \frac{1}{2} \left[{\bf F}({\bf U}_{L}) + {\bf F}({\bf U}_{R}) - \sum_{p=1}^{7} |\widetilde{\lambda}^{(p)}| \widetilde{\alpha}^{(p)} \widetilde{\bf R}^{(p)}\right],
\end{equation}

\noindent
where

\begin{equation}
 \widetilde{\alpha}^{(p)} = \widetilde{\bf L}^{(p)} ({\bf U}_{R} - {\bf U}_{L}),
\label{froe2}
\end{equation}

\noindent
$\widetilde{\lambda}^{(p)}$, $\widetilde{\bf R}^{(p)}$ and $\widetilde{\bf L}^{(p)}$ being respectively the eigenvalues and right and left eigenvectors of the Jacobian matrix for the intermediate state.

  One of the problems of Roe-type Riemann solvers is that the entropy condition is not satisfied through transonic rarefactions. To solve this problem, an extra viscosity term is required at these points \citep{hh83,yee87} and this can be implemented by substituting the eigenvalues $\widetilde{\lambda}^{(p)}$ in equation~(\ref{froeI}) by $\max\Big(0, |\widetilde{\lambda}^{(p)}|, (\widetilde{\lambda}^{(p)} - \lambda_L^{(p)}), \, (\lambda_R^{(p)} - \widetilde{\lambda}^{(p)}) \Big)$, where $\lambda_L^{(p)}$ and $\lambda_R^{(p)}$ are the eigenvalues associated to the left and right states, respectively.

  In Sect.~\ref{s:csrmhde} we obtained the characteristic equation whose zeroes are the eigenvalues. In the case of the entropic and Alfv\'en eigenvalues, the solution of the characteristic equation leads to analytical expressions which can be used directly to compute them. Magnetosonic eigenvalues, however appear as roots of a quartic equation which is solved numerically. To solve the quartic equation, we use a Newton-Raphson algorithm to obtain the eigenvalues corresponding to the fast magnetosonic waves using the light speed ($+1$ and $-1$, in the units of the numerical code) as seeds. Once obtained these two eigenvalues, the quartic equation is reduced to a quadratic one which is solved analytically. This procedure is very accurate in the determination of the eigenvalues. Only when either the kinetic or the magnetic energy of the physical state is much larger than thermal energy, we find eigenvalues with large uncertainties (0.1 \% or larger) satisfying the quartic within the machine roundoff error ($10^{-16}$, in double precision). The analytical solution of the quartic does not represent an improvement, since the large number of operations involved tends to increase the numerical error.

  The flux formula (\ref{froeI}) is used to advance the hydrodynamic variables according to equation~(\ref{eq:system}) and to calculate the quantities $\widehat{\Omega}_i$ needed to advance in time the magnetic fluxes following equation~(\ref{eq:mflux}). At each edge of the numerical cell, $\widehat{\Omega}_i$ is written as an average of the numerical fluxes calculated at the interfaces between the faces whose intersection define the edge. Let us consider, for illustrative purposes, $\widehat{\Omega}_x$. If the indices $(j,k,l)$ denote the center of a numerical cell, an $x-$edge is defined by the indices $(j,k+1/2,l+1/2)$. By definition, $\Omega_x = (v^yB^z - v^zB^y)$. Since

\begin{equation}
\label{f1}
 F^y(B^z) = v^yB^z - v^zB^y
\end{equation}

\noindent
and

\begin{equation}
\label{f2}
 F^z(B^y) = v^zB^y - v^yB^z,
\end{equation}

\noindent
we can express $\widehat{\Omega}_x$ in terms of these fluxes as follows

\begin{eqnarray}
\label{oom}
  \widehat{\Omega}_{x\,j,k+1/2,l+1/2} = \frac{1}{4} [\widehat{F}^y_{j,k+1/2,l} + \widehat{F}^y_{j,k+1/2,l+1} - \widehat{F}^z_{j,k,l+1/2} - \widehat{F}^z_{j,k+1,l+1/2}],
\end{eqnarray}
where $\widehat{F}^y$($\widehat{F}^z$) refers to the numerical flux in the $y$
($z$) direction corresponding to the equation for $B^z$ ($B^y$).

  In our numerical scheme, we need also to know the value of the magnetic field at the center of the cells in order to obtain the primitive variables after each time step (see below) and to compute again the numerical fluxes of the
other conserved variables for the next time step. If $\widehat{B}^x_{j \pm 1/2,k,l}$ is the $x$-component of the magnetic field at the interface $(j\pm 1/2,k,l)$, then the $x$-component of the magnetic field at the center of the $(j,k,l)$ cell, $B^x_{j,k,l}$, is obtained by taking the arithmetic average of the corresponding fluxes, i.e.

\begin{eqnarray}
  B^x_{j,k,l} = \frac{1}{2} (\widehat{B}^x_{j-1/2,k,l} \Delta S^x_{j-1/2,k,l} + \widehat{B}^x_{j+1/2,k,l}\Delta S^x_{j+1/2,k,l})/\Delta S^x_{j,k,l} ,
\end{eqnarray}
where $\Delta S^x_{j\pm1/2,k,l}$ is the area of the interface surface between two adjacent cells, located at $x_{j\pm1/2}$ and bounded between $[y_{k-1/2},y_{k+1/2}]$ and $[z_{l-1/2},z_{l+1/2}]$. Analogous expressions for
$\widehat{\Omega}_{y\,j+1/2,k,l+1/2}$ and $\widehat{\Omega}_{z\,j+1/2,k+1/2,l}$, and $B^y_{j,k,l}$ and $B^z_{j,k,l}$ can be easily derived.

\subsection{Recovery of the primitive variables}

  After every integration time step, primitive variables are recovered following the procedure described in Sect.~2, as in Leismann et al. (2005).

\section{Numerical tests}
\label{s:numtetsts}

  In this Section, we evaluate the performance of the Riemann solver presented in Sect.~\ref{ss:numflux} by means of a number of one dimensional shock tube problems, as well as, some two dimensional standard tests previously computed in the literature.

\subsection{One dimensional tests}
\label{sec:1D}

  For easier comparison with previous works in the field, we perform the same one dimensional tests as proposed by MUB09, with initial states to the left and to the right of the discontinuity given in Table~1. The initial jump is set at $x=0.5$ in a domain
spanning the range $[0,1]$, which is covered by $N_x$ uniform numerical zones (see Table~1). A constant $\gamma$-law with
$\gamma=5/3$ is assumed for the equation of state. We perform all the tests with a CFL number equal to 0.8. By default, RMHD equations are integrated using a second order Runge-Kutta integration with no spatial reconstruction in order to assess the performance of our FWD Riemann solver. We will refer to this basic setting as our first-order scheme (in analogy with MUB09).

  To show the numerical accuracy of our Roe-type Riemann solver, we also compute the numerical solution with both HLL \citep{Leismann05} and HLLC \citep{MB06} solvers. Tests with the newer HLLD solver are not performed but our results are directly compared with those presented in MUB09 for the same numerical tests. The discrete errors in L-1 norm are computed as

\begin{equation}\label{eq:L1-err}
 \varepsilon_{\rm L1} = \frac{1}{N_x} \sum_{i=1}^{N_x} 
                        \left|V^{\rm an}_i - V_i\right| \, ,
\end{equation}

\noindent
where $V_i$ is the first-order numerical solution (density or magnetic field) of our scheme and $V^{\rm an}_i$ is the analytic solution at $x_i$, computed with the exact Riemann solver of \cite{GR06}.

\subsubsection{Contact and Alfv\'en discontinuities}
\label{sec:contacts}

\begin{figure*}
\begin{center}
\plotone{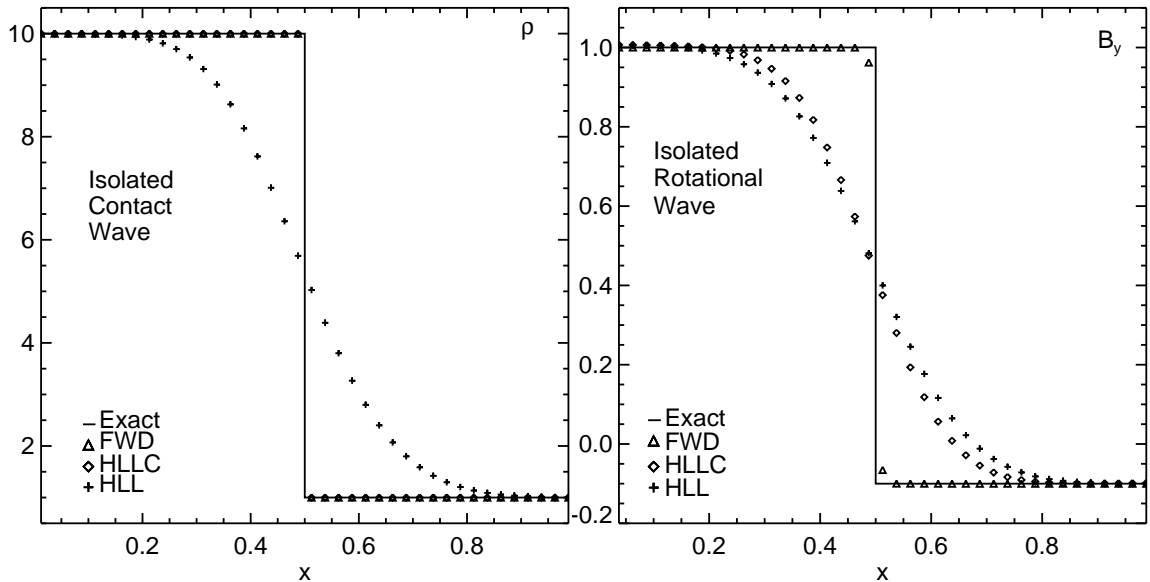}
\caption{Density and $y$ component of magnetic field from an isolated contact (left panel) and a rotational (right panel) wave at $t=1$ (as in MUB09). The different symbols show results computed with the FWD solver (triangles), the HLLC solver (diamonds) and the simpler HLL solver (plus signs). Note that both FWD and HLLC solvers are able to capture exactly an isolated contact wave, keeping it perfectly sharp without producing any grid diffusion effect. HLLC can capture the contact wave but not the rotational discontinuity, whereas HLL
spreads both of them on several grid zones. The Roe-type solver captures with very small diffusion the rotational wave.}
\label{fig:isolated}
\end{center}
\end{figure*}

  In an isolated contact wave, only the rest-mass density is discontinuous. In Fig.~\ref{fig:isolated} (left), we show the solution of the first Riemann problem of Table~1 at $t=1$ as computed with the HLL, HLLC and FWD solvers. Both HLLC and FWD solvers show no smearing of the initial profile, while HLL displays the largest numerical difussion. This behavior is expected for the HLLC solver, which is specifically built to include a contact wave in the numerical solution \citep{MB06}. Also in the FWD solver the use of the full spectral information of the RMHD system yields perfectly sharp contact waves.

  We set up an isolated rotational wave as in MUB09, and show the variation of $B_y$ across it. The right panel of Fig.~\ref{fig:isolated} shows that both HLL and HLLC solvers are unable to capture the initially specified jump (see Table~1), while the FWD solver captures it with very little diffusion. In this regard the Roe-type solver compares reasonably well with the HLLD solver of MUB09, which exhibits no
diffusion at all at this kind of discontinuities.

\begin{figure*}
\begin{center}
\plotone{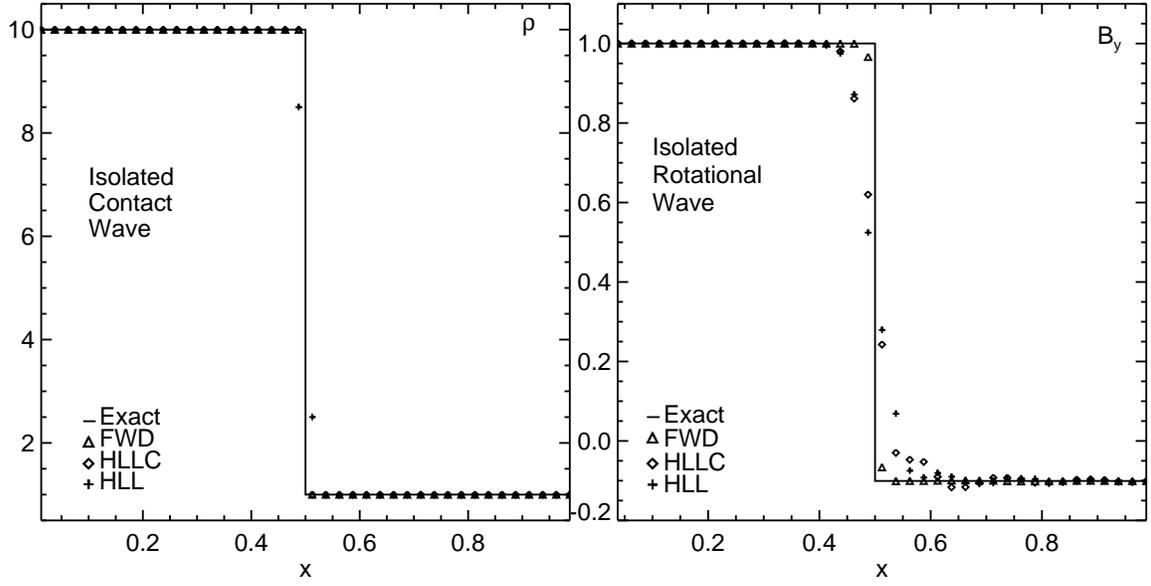}
\caption{Same as in Fig.~\ref{fig:isolated} but using a third order accurate Runge-Kutta time integrator and PPM spatial reconstruction. The diffusion of the HLL solver across the contact wave is drastically reduced. The same comment applies for both HLLC and HLL across the rotational wave. We note the small oscillations behind the contact (in the HLL and HLLC cases) resulting from the higher order of the method.}
\label{fig:isolated_PPM}
\end{center}
\end{figure*}

  We shall note that, in spite of the fact that HLL smears both contact and rotational discontinuities, in practical applications, the numerical diffusion is not as bad as one may guess by extrapolating the results of Fig.~\ref{fig:isolated}. If we use a third order accurate scheme both in space and time (using PPM for the spatial interpolation, and a third order Runge-Kutta time integrator \footnote{In the following we refer to this combination as the third order scheme}, the numerical diffusion is drastically reduced (Fig.~\ref{fig:isolated_PPM}). Even HLL is able to resolve the contact discontinuity in a couple of points (Fig.~\ref{fig:isolated_PPM} left), and both HLL and HLLC need {\em only} 4 to 5 numerical cells to resolve the rotational wave.

\subsubsection{Shock Tube 1: The Brio \& Wu test}
\label{ss:st1}

  \cite{Balsara01} proposed a relativistic extension of the classical magnetized shock tube of \cite{BW88}, which has also been considered in a number of recent papers (e.g., \citealp{delzanna03,Leismann05,MB06,MUB09}). In this case, the adiabatic index is $\gamma=2$, and the initial piecewise discontinuous state breaks into a left-going fast rarefaction, a left-going compound wave (see Sect.~\ref{s:convex}), a contact discontinuity, a right-going slow shock and a right-going fast rarefaction wave.

\begin{figure}
\begin{center}
\plotone{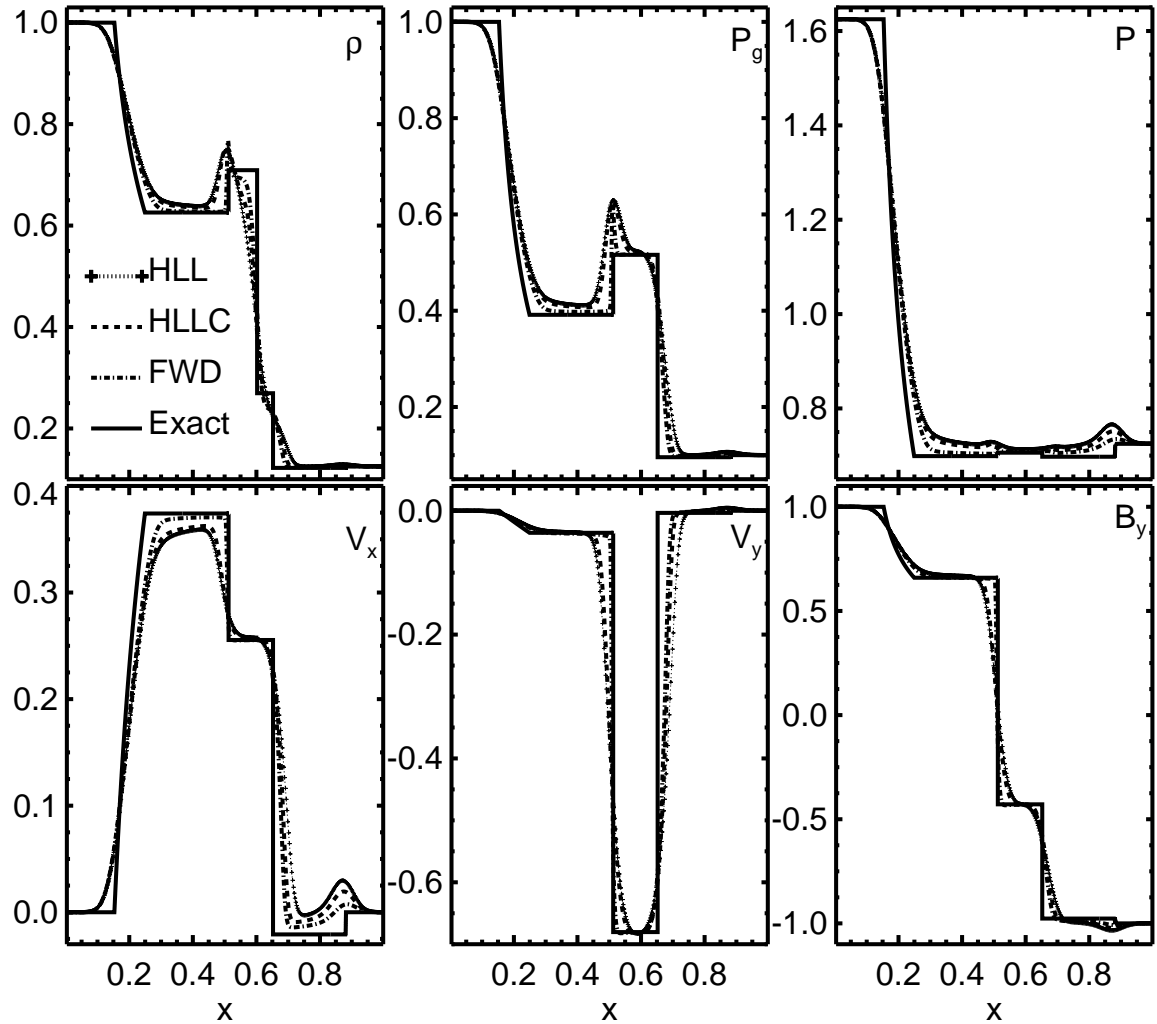}
 \caption{Relativistic Brio-Wu shock tube test at $t=0.4$. Computations are carried on $400$ zones using the FWD (dashed-dotted line), HLLC (dashed line), HLL (dotted-crossed line) and exact (solid line) Riemann solver, respectively.  The top panel shows, from left to right, the rest-mass density, gas pressure, total pressure. In the bottom panel the $x$ and $y$ components of velocity and the $y$ component of magnetic field are shown. The compound wave in the numerical solution between the fast left-propagating rarefaction and the contact discontinuity (at $x \approx 0.51$ in the plots) is not captured in the analytical solution provided by Giacomazzo \& Rezzolla (2006) that avoids this kind of solution by construction.}
\label{fig:shock_tube1}
\end{center}
\end{figure}

\begin{figure}
\begin{center}
\plotone{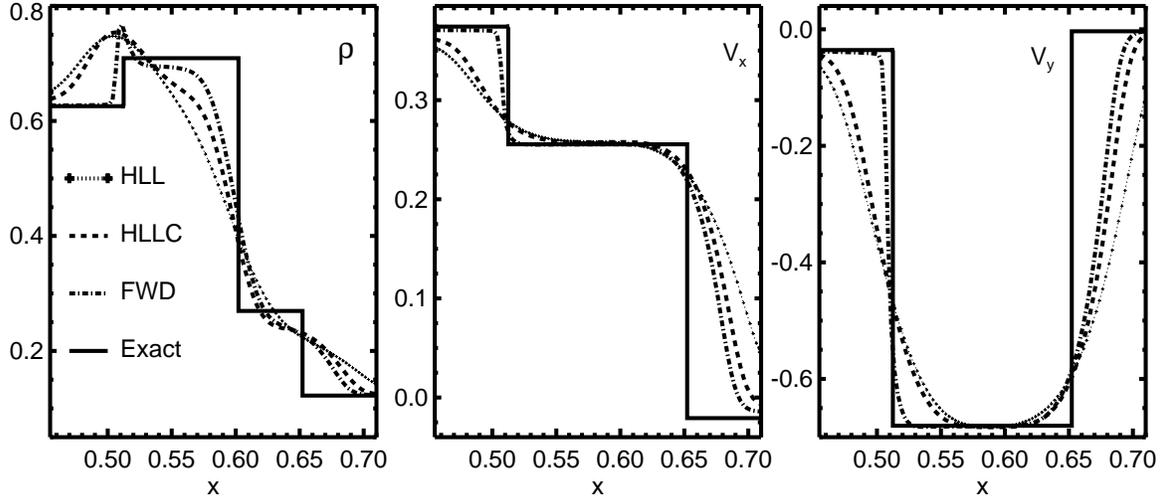}
\caption{Zoom of the central region of Fig.\ref{fig:shock_tube1}, focusing on the compound wave (at $x\simeq0.51$), the contact discontinuity (at $x\simeq0.60$) and the right-going slow shock (at $x\simeq0.65$).  From left to right, we show the rest-mass density and the two components of velocity.}
\label{fig:shock_tube1_zoom}
\end{center}
\end{figure}

  The final state of this test using the first order scheme is shown in Figs.~\ref{fig:shock_tube1}~and~\ref{fig:shock_tube1_zoom}. Our FWD solver performs clearly better than HLL or HLLC, and the quality of the results compares fairly well with the HLLD solver of MUB09. The superior quality of the results using the Roe-type solver is more obvious in the central part of the Riemann fan (Fig.~\ref{fig:shock_tube1_zoom}), which is much better resolved than with the HLL or HLLC solvers. A more precise quantification of how better is the overall solution can be seen from Fig.~\ref{fig:errors} (upper left), where the L1-norm errors associated to the FWD solver are, systematically $\sim 54$\% and $\sim 40$\% smaller than those associated to the HLL and HLLC solvers, respectively.

  The computational cost of the FWD solver is higher than that of the other two solvers, being the ratio of CPU times for this test and the first order scheme $t_{\rm HLL}:t_{\rm HLLC}:t_{\rm FWD} = 1:1.02:2.92$. Using a the third-order scheme, the CPU times scale as $t_{\rm HLL}:t_{\rm HLLC}:t_{\rm FWD} = 1:1.05:2.48$.

\subsubsection{Shock Tube 2: Non-planar Riemann problem}
\label{sec:st2}

  In this test, proposed by \cite{Balsara01} and also performed by, e.g., \cite{Leismann05,MB06}, and \cite{MUB09}, the transverse magnetic field rotates across the discontinuity by $\simeq 0.55\pi$, and the Riemann fan is composed of three left-going waves (fast shock, Alfv\'en wave, and slow rarefaction), a contact discontinuity (at $x\simeq0.475$ when $t=0.55$), and three right-going waves (slow shock, Alfv\'en wave, and fast shock). Because of the small relative velocity of the different waves emerging from the break-up of the initial discontinuity, which generates extremely narrow structures, this test is rather demanding for the different numerical algorithms which resolve the wave structure either totally (as it is the case of the FWD solver) or partially (as in the case of HLLC). 

\begin{figure}
\begin{center}
\plotone{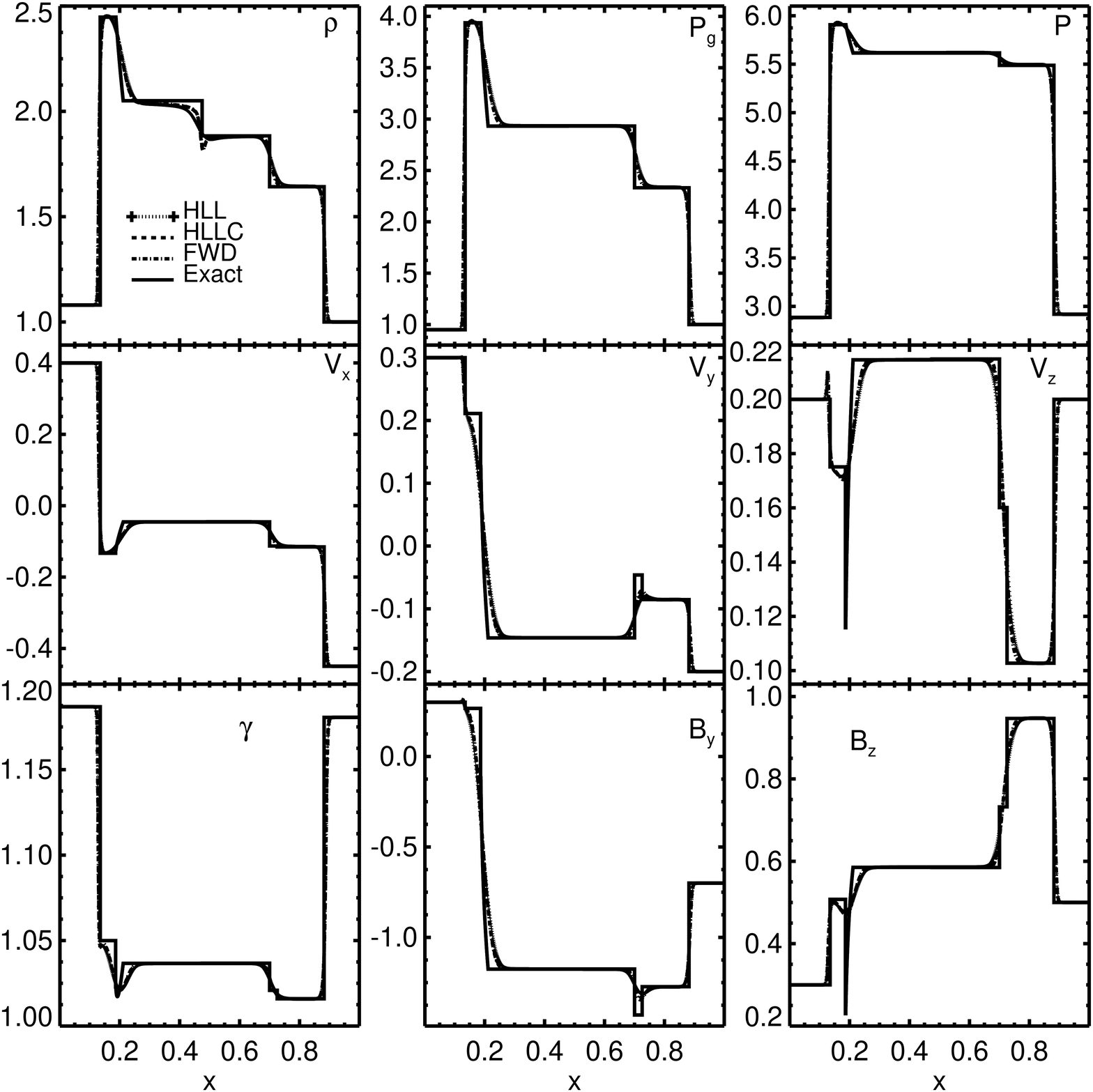}
\caption{Second shock tube problem.  From left to right, the top panel shows rest-mass density, gas and total pressure.  The middle panel displays the three components of velocity. The bottom panel shows the Lorentz factor and the transverse components of magnetic field.  Solid, dashed-dotted, dashed and dotted lines are used to identify results computed with the exact, FWD, HLLC and HLL solvers, respectively.}
\label{fig:shock_tube2}
\end{center}
\end{figure}

  As we show in Fig.~\ref{fig:shock_tube2}, all the approximate solvers yield quantitatively similar results across the fastest waves (independent on whether they are heading to the left or to the right), since the numerical flux has to be consistent with the upwind update of the state vector, which happens either if $\tilde{\lambda}^p>0$, $\forall p$, in which case $\widehat{\bf F}={\bf F}({\bf U}_L)$, or if $\tilde{\lambda}^p<0$, $\forall p$, in which case $\widehat{\bf F}={\bf F}({\bf U}_R)$ (see, e.g., \citealp{Aloy99a}).

\begin{figure}
\begin{center}
\plotone{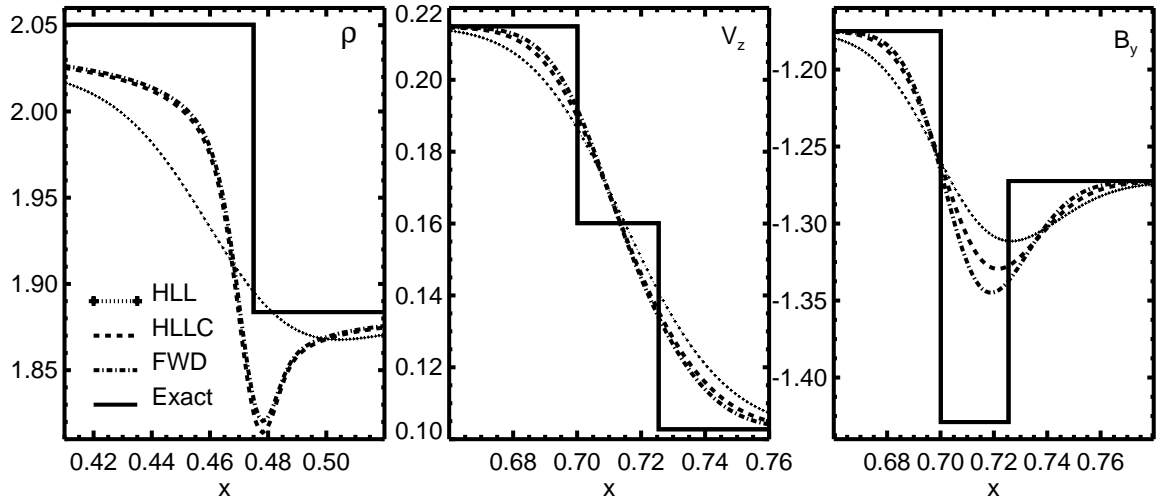}
\caption{Left panel: closer look of the rest-mass density of Fig.~\ref{fig:shock_tube2} around the contact wave. Middle and right panels: enlargements of the $z$ component of velocity and $y$ component of magnetic field around the right-going slow shock and Alfv\'en discontinuity, respectively.}
\label{fig:shock_tube2_zoom}
\end{center}
\end{figure}

  More obvious differences are noticeable across the contact discontinuity (Fig.~\ref{fig:shock_tube2_zoom} left), where both HLLC and FWD solvers are closer to the exact solution than the HLL solver. However, because of the extra dissipation produced by the rotation of the transverse components of the magnetic field, an unphysical undershoot is produced in the rest-mass density (Fig.~\ref{fig:shock_tube2_zoom} left). 

  The narrow structure generated between the slow shock and the rotational wave ($0.7 \lesssim x \lesssim 0.725$) is much sharply resolved with the HLLC and FWD solvers than with HLL, but the superiority of the Roe-type solution is obvious in the middle and right panels of Fig.~\ref{fig:shock_tube2_zoom}. For the working resolution of 800 uniform zones, none of the three schemes represents adequately the extremely narrow structure confined between the left-going Alfv\'en and slow rarefaction ($0.185\lesssim x \lesssim 0.19$). Indeed, this is to be expected, since resolving this region is a challenge, not only for RMHD codes, but also for the exact solver of
\cite{GR06} (actually, this is the reason for the impossibility of obtaining an exact solution with an accuracy better then $3.4\times10^{-4}$; c.f., \citealp{GR06}).

  The overall similar quality of the results obtained for this test using the HLLC and the FWD solvers is also reflected in the L1-norm errors, which are virtually the same in both cases, independent of the numerical resolution employed (Fig.~\ref{fig:errors} upper right). As expected, HLL deviates more from the exact solution and possesses L1-norm errors $\sim 24$\% and $\sim 19$\% larger than those of the FWD and HLLC solvers, respectively. 

  For future reference, we provide the ratio of CPU times for this test using the first order scheme, which are $t_{\rm HLL}:t_{\rm HLLC}:t_{\rm FWD} =1:1.03:3.15$.

\subsubsection{Shock Tube 3: Colliding streams}

\begin{figure}
\begin{center}
\plotone{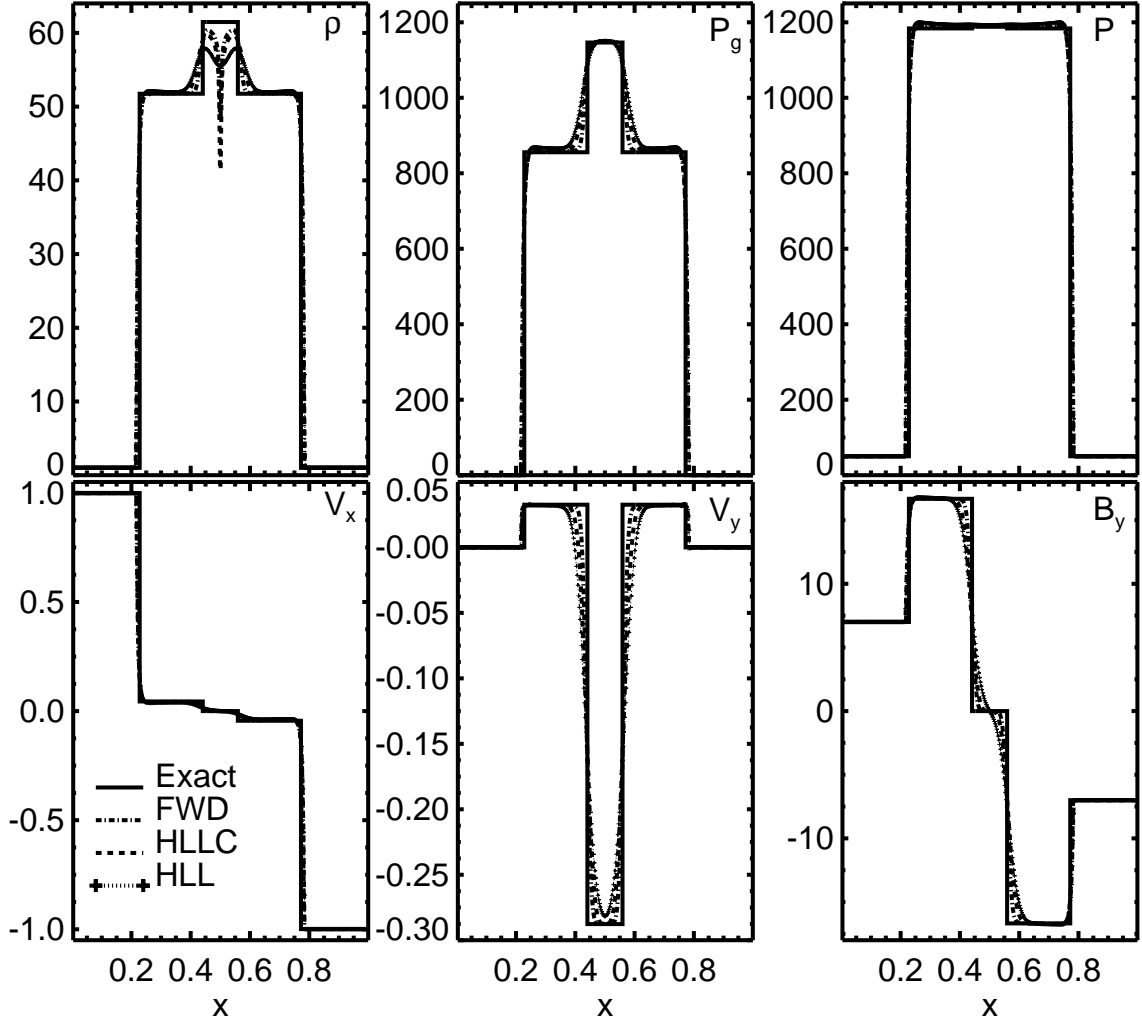}
\caption{Results of the shock tube 3 (Table~1) using different solvers. From top to bottom, left to right, the panels show rest-mass density, gas pressure, total pressure, $x$ and $y$ components of velocity and $y$ component of magnetic field. The exact solution of this Riemann problem is plotted with a solid line. Dash-dotted, dashed and dotted lines refer to the results obtained with FWD, HLLC and HLL solvers, respectively.}
\label{fig:shock_tube3}
\end{center}
\end{figure}

\begin{figure}
\begin{center}
\plotone{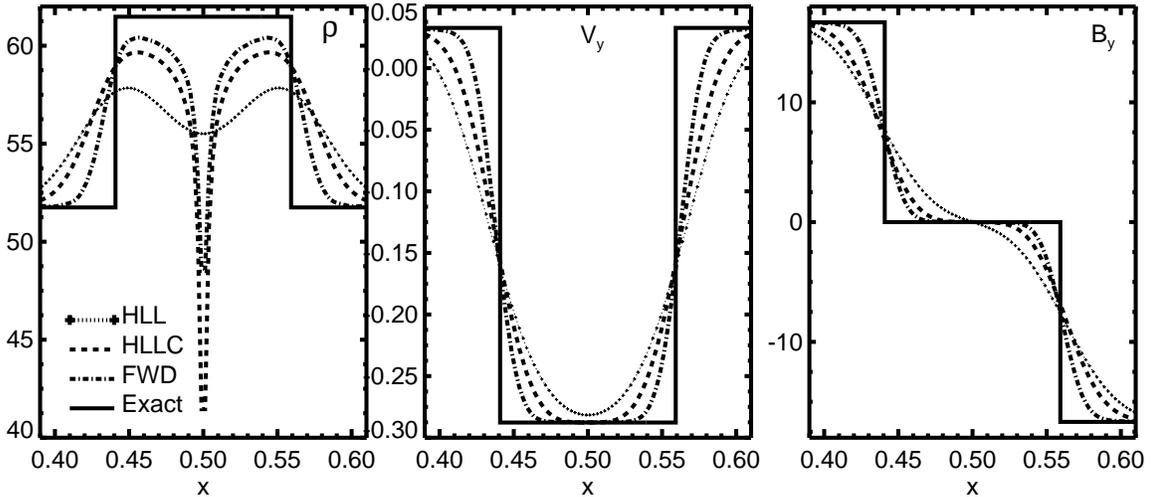}
\caption{Closer view of the central region of Fig.\ref{fig:shock_tube3}. Type lines have the same meaning as in Fig.\ref{fig:shock_tube2_zoom}. The
rest-mass density is represented in the left panel, where the wall
heating pathology is notorious. Central and right panels show the
profiles transverse components of the velocity and of the magnetic field, respectively.}
\label{fig:shock_tube3_zoom}
\end{center}
\end{figure}

  Figure~\ref{fig:shock_tube3} shows the results of a strong relativistic shock reflection test with two streams approaching each other at a velocity $|v^x|= 0.999$. The test was proposed by \cite{Balsara01}, and it has also been used for code validation in, e.g., \cite{delzanna03,Leismann05,MB06}, and MUB09. This setup results in two fast and two slow shocks. At the initial collision point ($x = 0.5$) a certain amount of ``wall heating'' occurs, which is a numerical pathology of approximate Riemann solvers (e.g. \citealp{Noh87,DM96}). The problem arises because an excess of entropy is generated at the collision point at $t = 0$, which can diffuse numerically only slowly, because the fluid is at rest at that point. Higher order reconstruction schemes help to confine the problem to a small number of points initially, but generate less diffusion. Hence the ``hole'' in the rest mass density is deeper with the third order scheme than with the first order one. In a close analogy to the HLLD solver of MUB09, the FWD solver displays a density undershooting which is smaller than that corresponding to the HLLC solver (the error with respect to the analytic solution at $x=0.5$ is 21\% and 32\% for the FWD and the HLLC solvers, respectively; Fig.~\ref{fig:shock_tube3_zoom}). The larger numerical diffusion of HLL yields a relative error at the former point of only 10\%, but this comes at the cost of requiring more zones to resolve both slow shocks. 

  The smallest L1-norm errors correspond systematically to the FWD solver, followed by the HLLC solver and, finally, the largest errors are obtained using the HLL solver (Fig.~\ref{fig:errors} bottom left). In this particular test, the CPU times scale as $t_{\rm HLL}:t_{\rm HLLC}:t_{\rm FWD} =1:1.06:2.91$.

\subsubsection{Shock Tube 4: Generic Alfv\'en test}
\label{ss:st4}

  \cite{GR06} proposed the so-called {\em Generic Alfv\'en} test, which MUB09 have also adopted as a benchmark for their HLLD solver. The initial discontinuity results into a contact discontinuity which separates a fast rarefaction wave (at $x\simeq0.05$), a rotational wave (at $x\simeq0.44$) and a slow shock (at $x\simeq0.46$), from a slow shock (at $x\simeq0.56$), an Alfv\'en wave (at  $x\simeq0.57$) and a fast shock (at $x\simeq0.97$).

\begin{figure}
\begin{center}
\plotone{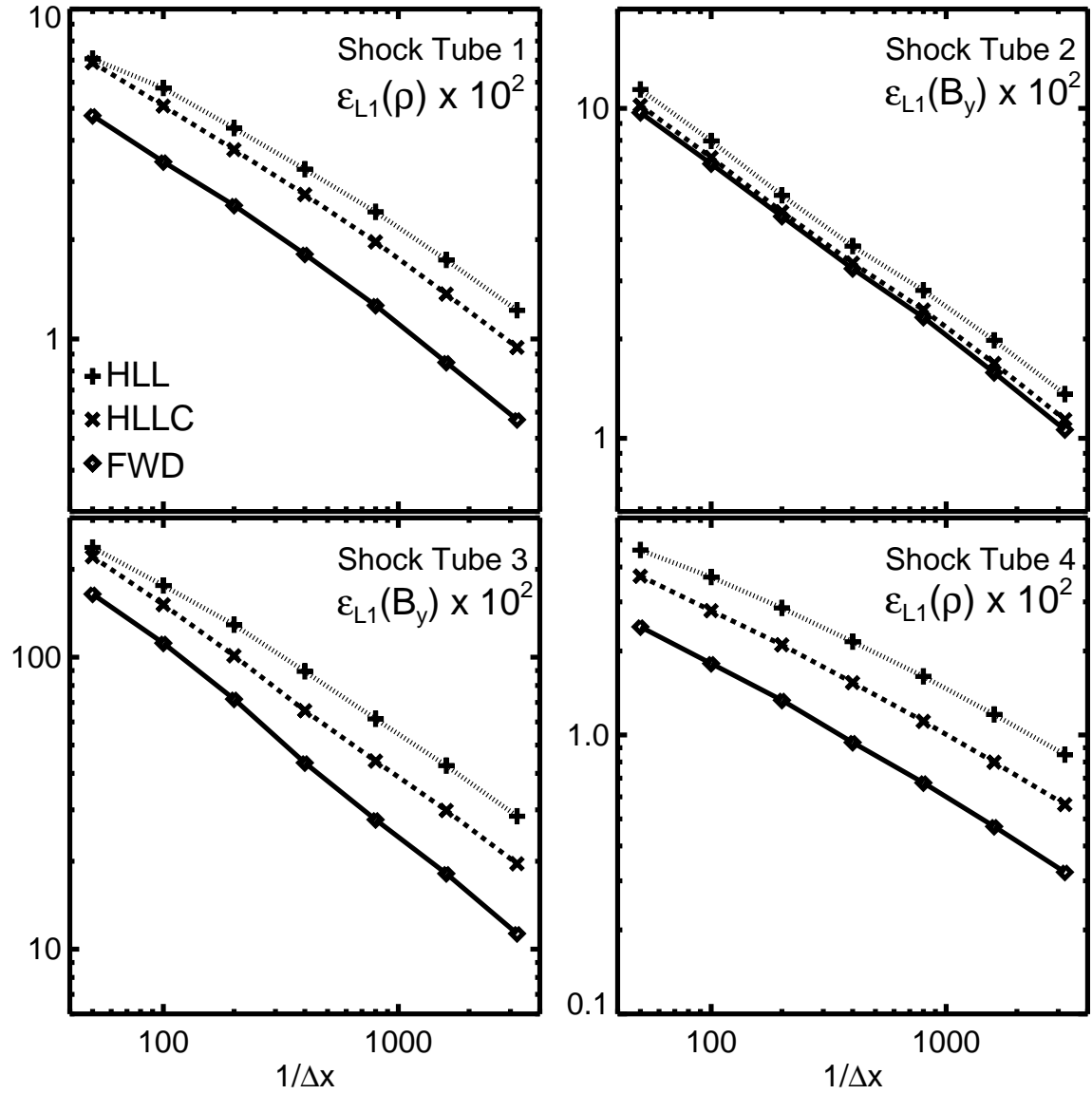}
\caption{L-1 norm errors (in $10^2$) for the four shock tube problems of Table~1 as function of the grid resolution $\Delta x=1/N_x$. The different combinations of lines and symbols refer to FWD (solid, diamonds), HLLC (dashed, crosses) and HLL (dotted, plus signs).} 
\label{fig:errors}
\end{center}
\end{figure}

\begin{figure}
\begin{center}
\plotone{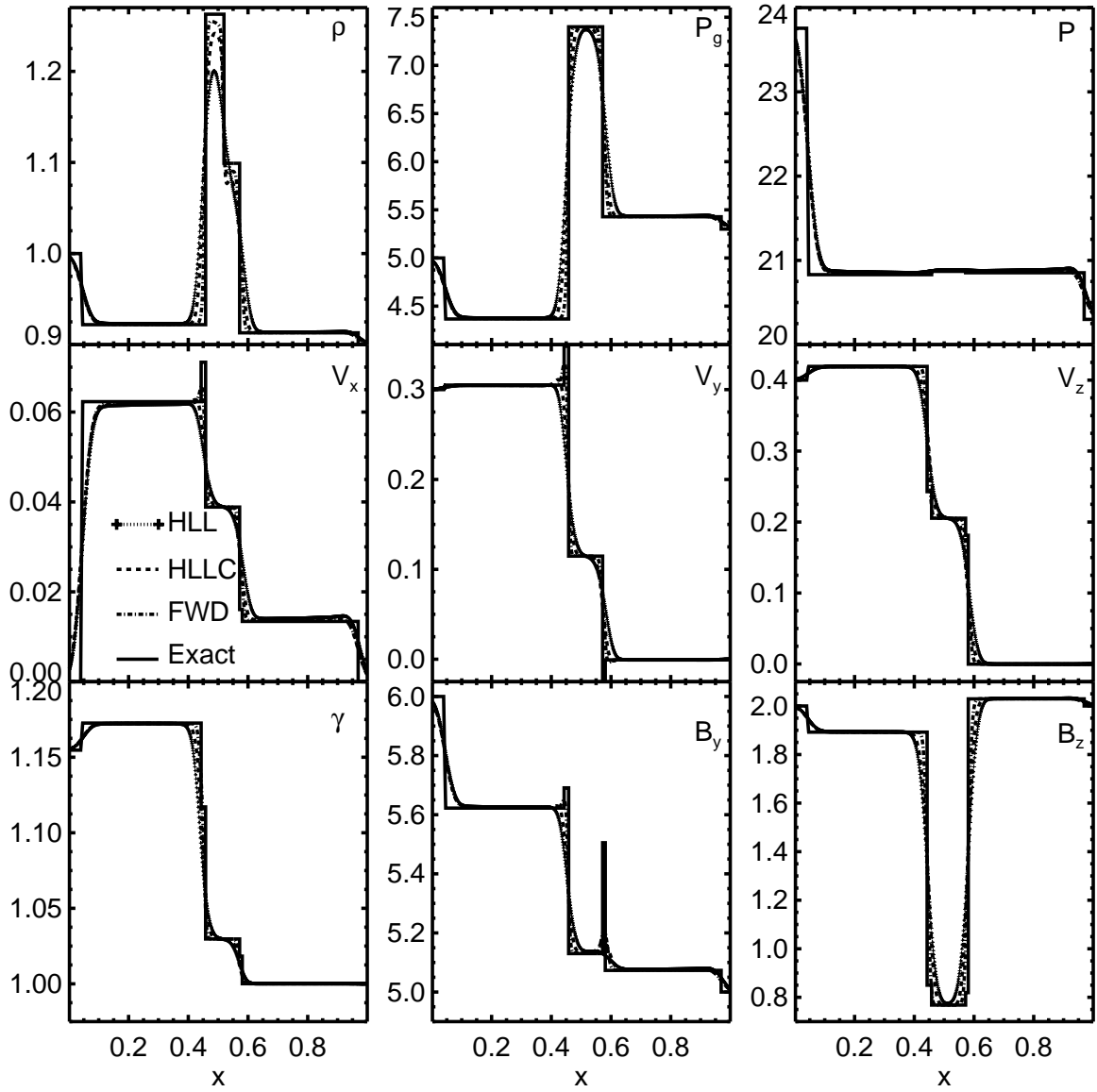}
\caption{Same as Fig.~\ref{fig:shock_tube2}, but for the general Alfv\'en test (Table~1).}
\label{fig:shock_tube4}
\end{center}
\end{figure}

\begin{figure}
\begin{center}
\plotone{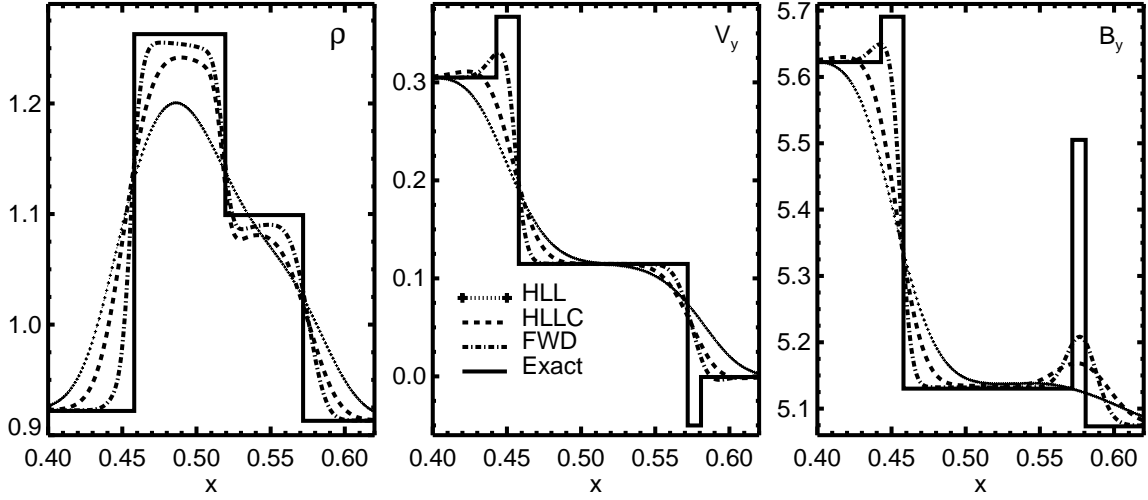}
\caption{Enlargement of the central region of Fig.\ref{fig:shock_tube4} where the profiles of the rest-mass density (left), $v^y$ (middle) and $B^y$ are displayed. The contact wave at $x\simeq 0.52$ separating the two slow shocks is better captured by the FWD solver. Both HLLC and FWD solvers undershoot the true solution to the right of the contact wave. The Alfv\'en modes at $x\approx 0.44$ and $x\approx 0.58$ are not fully smeared at this resolution using the FWD solver, although they are obviously unresolved using 800 uniform numerical zones.}
\label{fig:shock_tube4_zoom}
\end{center}
\end{figure}

  The exact solution of this test, together with the results for the FWD, HLLC and HLL solvers are shown in Fig.~\ref{fig:shock_tube4}. The finest structure in this test, associated with the rotational discontinuities which travel very close to the slow shocks, is barely resolved by the first order scheme using the Roe-type solver at the working resolution of 800 uniform numerical zones (Fig.~\ref{fig:shock_tube4_zoom}). This is the reason for the considerably smaller L1-norm errors produced by the FWD solver as compared with HLL and HLLC (Fig.~\ref{fig:errors} bottom right).

  For completeness, we also provide with the CPU times needed to run this test for the different solvers $t_{\rm HLL}:t_{\rm HLLC}:t_{\rm FWD} = 1:1.03:3.01$, and with the ratio of L1-norm errors obtained with the HLL, HLLC and FWD solvers, at the maximum resolution employed in Fig.~\ref{fig:errors}, which is $1:0.66:0.38$.

\subsubsection{Ultrarelativistic shell}

  With the previous one dimensional tests, we have demonstrated the effectiveness and strength of accurate Riemann solvers in the description of complex relativistic magnetized flows, although, as we pointed out in Sect.~\ref{sec:contacts}, the use of high-order schemes tends to reduce the differences between simpler, more diffusive Riemann solvers (like HLL) and more elaborate ones (like our FWD solver). However, it remains true that, in multidimensional applications, even in combination with high order schemes, it may pay off to use more computationally expensive solvers, which reduce substantially the numerical errors without the need of employing huge, perhaps in practice unreachable, numerical resolutions. 

  In this section we will show that the new Roe-type solver can dodge the difficulties inherent to some realistic astrophysical flows, where the other two solvers fail. For this purpose, we have set up a magnetized ultrarelativistic shell in spherical coordinates, which moves radially with a Lorentz factor $W_0 = 15$. The magnetization parameter is $\sigma_0 \equiv b^2/4\pi\rho = 0.1$, corresponding to an azimuthal (toroidal) magnetic field $B^\phi = 14.56$ perpendicular to the purely radial velocity (note that in Table~1, where the shell state is denominated {\em M}, the vector components $(x,y,z)$ correspond, in this test, to the spherical components $(r,\phi,\theta)$). With this set up we model the ejecta in gamma-ray burst afterglows as in \cite{MGA09}, where a pure one-dimensional treatment is adequate because of the ultrarelativistic speed of propagation of the ejecta, in spite of the fact that the ejecta is axially symmetric about the direction of propagation, but it may be heterogenous in the $\theta$-direction. Therefore, we consider here the evolution of a representative hollow wedge of the ejecta located at a small polar angle $\theta_0$ from the spherical axis. The shell has an initial radial thickness $\Delta_0 = 10^{15}\,$cm ($\Delta_0 = 0.1$ in the code units), and moves in a uniform medium whose density is much smaller than that of the shell (state {\rm R} in Table~1). Behind the shell (state {\rm L} in Table~1), the medium is almost as light as that ahead, but moves at the same speed than the overdense shell.

\begin{figure}
\begin{center}
\plotone{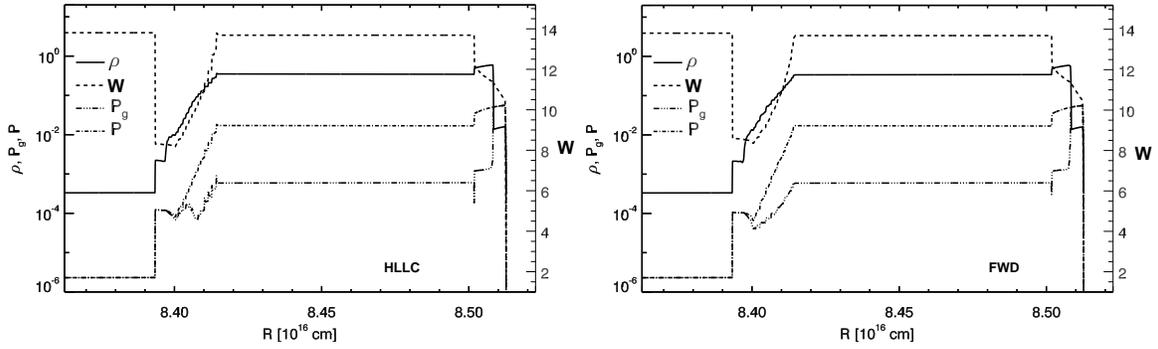}
\caption{Ultrarelativistic shell at $t=3.5$ (simulation units, which correspond to $1.1\times106\,$s) run with 32000 numerical zones
 (Table~1). The left and right panels show the results using HLLC and FWD solvers, respectively. Different line styles are used to display a number of physical variables.}
 \label{fig:shell}
\end{center}
\end{figure}

  Because of the numerical difficulty of the proposed set up, we have to run this test with 32000 uniform numerical zones in the radial direction, with the third-order scheme and a CFL number $0.2$. The results at $t = 3.5$ (about $10^6$ timesteps) can be seen in Fig.~\ref{fig:shell}, just before the HLLC scheme fails to continue the evolution. At this stage the breakup of the initial sharp edges of the shell has produced a (reverse-)shock-contact-(forward-)shock structure ($8.501\times10^{16}\,$cm\,$<R<8.512\times 10^{16}\,$cm) at the front edge and a shock-contact-rarefaction at the rear one ($8.393\times10^{16}\,$cm\,$<R<8.413\times 10^{16}\,$cm)\footnote{Note that, since the magnetic field is perpendicular to the velocity, there are only three waves emerging from the discontinuities -and not seven as in non-degenerate RMHD-.}. We point out that, because of the spherical expansion, the intermediate states between the two shocks in the front of the shell are not uniform.

  The reason for the failure of the HLLC solver is the high frequency noise produced in the rear contact discontinuity, which extends with time to the whole rarefied shell (see the gas pressure in the region $8.401\times10^{16}\,$cm\,$<R<8.413\times 10^{16}\,$cm; Fig.~\ref{fig:shell}). This noise is practically nonexistent when we use the FWD solver. Another difference is noticeable at the reverse shock ($R\simeq 8.501\times10^{16}\,$cm), where the unphysical undershooting is deeper when using the HLLC solver than when the FWD solver is employed.

  The test we have presented here argues in favor of the use of more elaborate (although more expensive computationally) Riemann solvers for ultrarelativistic applications. 

\subsection{Two-dimensional tests}
\label{sec:2D}

  We employ the framework provided by the RGENESIS code \citep{Leismann05} to set up a pair of two dimensional test problems, with the aim of illustrating the multidimensional performance of our new Roe-type solver and the proposed renormalised full wave spectral decomposition. For the multidimensional tests we use the third-order scheme and the constraint transport method to update the magnetic field flux. Although there is no analytic solution to verify against, our results can be compared with those produced by other authors. As these tests encompass all the degeneracies of the RMHD eigensystem, they are a challenge for our linearized, Roe-type Riemann solver.

\subsubsection{Cylindrical explosion}

  The cylindrical explosion test consists of a strong shock propagating into a magnetically dominated medium. Results for different cylindrical explosion problems in RMHD have been published by, e.g., \cite{Dubal91,vanputten95,komissarov99,delzanna03,Leismann05,MB06}. We have chosen a setup very similar to that in \cite{komissarov99}: a cylinder of high pressure and density is located in the center of a square Cartesian grid, which initially contains a uniform, strong magnetic field. Simulations have been carried out on grids of $400 \times 400$ and $800 \times 800$ zones, spanning a region of $12\times12$ units of area. In the center of the grid there is a circle of radius 0.8 where $\rho = 10^{-2}$ and $p = 1$. Between a radius of 0.8 and 1.0 the values exponentially decrease to those of the homogeneous ambient medium ($\rho= 10^{-4}$ and $p = 3 \times 10^{-5}$ ). Initially, the magnetic field is $B_x = 1$, and the velocity is zero everywhere. We have chosen this particular explosion set up among several other possible because it is the strongest cylindrical explosion in \cite{komissarov99}. Several other possible initial states have computed with the new Roe-type solver in \cite{Anton08}.

  The test runs with different solvers have been performed using our third-order scheme and a Courant number $0.4$, for the highest resolution runs, and $0.6$ for the lower resolution ones. To compute the most extreme of the \cite{komissarov99} tests, the energy fix proposed by \cite{MB06} (cf., their equation~76) has been used.

\begin{figure*}
\begin{center}
\plotone{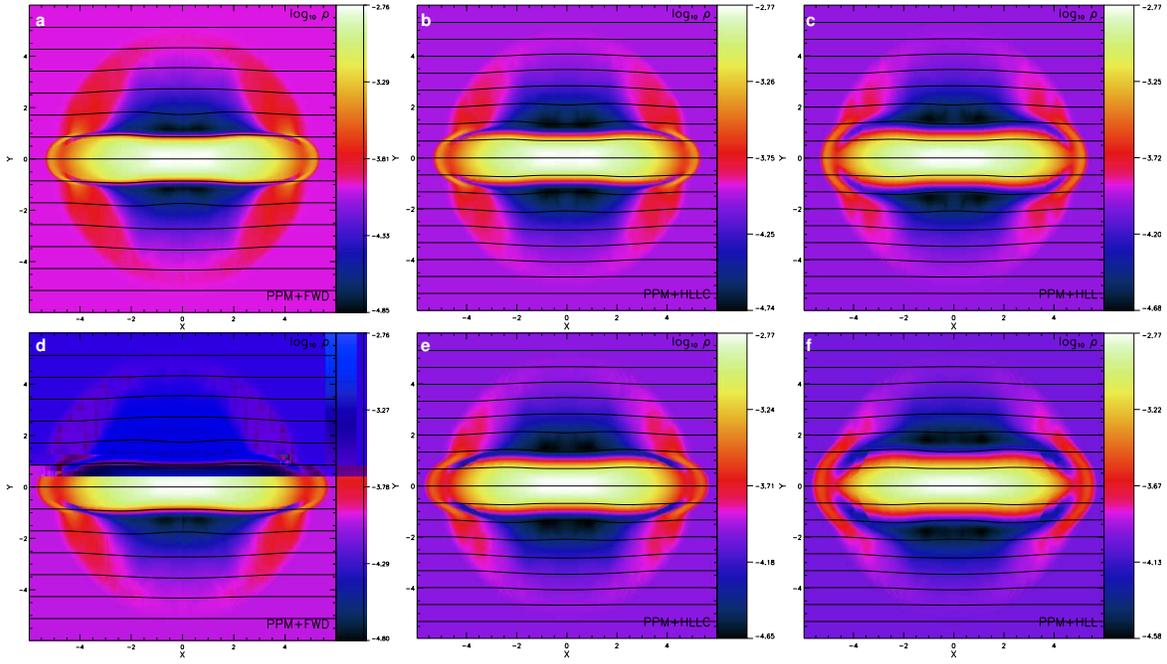}
\caption{The rest-mass density distribution of the two-dimensional cylindrical explosion problem at $t=4.0$. The upper (bottom) panels have been run with a grid of $800 \times 800$ ($400 \times 400$) numerical zones, a CFL number $0.4$ ($0.6$), PPM spatial reconstruction and a third-order accurate Runge-Kutta integration scheme. Left, middle and right panels show the results obtained using FWD, HLLC and HLL solvers, respectively.}
\label{fig:KT23}
\end{center}
\end{figure*}

  The explosion of the central overpressured region is notably confined by the presence of the strong magnetic field, which hinders the expansion of the fluid in the $y-$direction, and yields a pair of twin jets propagating along the $x-$direction. The outer fast shock has an almost spherical shape, with a hump along the $x$-direction, where the maximum Lorentz factor of the expanding fluid is reached\footnote{One could associate these humps with the bow shocks of both relativistic, oppositely-moving, twin jets.}. This maximum Lorentz factor attained by the models at $t=4$ depends on the numerical resolution and on the solver, and ranges from $W_{\rm max}=3.5$ to $4.03$, with HLL yielding the  smaller values of $W_{\rm max}$ and the FWD solver the largest ones. On the other hand, at higher resolution $W_{\rm max}$ is about a $10\%$ larger than at low resolutions. Figure~\ref{fig:KT23} shows that the FWD solver produces the sharpest and best resolved features in the distribution of the rest-mass density at $t=4$. However, the results we obtain using HLLC and FWD solvers are very similar, particularly at high resolution, the only noticeable differences being in the narrow horizontal region of high density (along which the explosion happens) and the shape of the bow shocks. Compared with HLL, the FWD solver delivers sharper profiles in the discontinuities even with half the resolution (compare in Fig.~\ref{fig:KT23} panels c and d).

  In this test, whose dynamics is dominated by the fast shocks emerging from the break-up of the initial discontinuity the ratios of CPU times are $t_{\rm HLL}:t_{\rm HLLC}:t_{\rm FWD} = 1:1.07:1.52$ and $t_{\rm  HLL}:t_{\rm HLLC}:t_{\rm FWD} = 1:1.07:1.86$ at resolutions of $400\times 400$ and $800\times 800$, respectively. This means that the FWD solver is less demanding computationally than in all the 1D cases analyzed in Sect.~\ref{sec:1D}. The advantage of using the FWD solver over the HLLC one is questionable, provided the similitude of their respective results. However, it is possible to find a trade-off between the numerical efficiency that the HLL solver yields and the accuracy that the Roe-type solver lends.

\subsubsection{The rotor problem}

  \cite{delzanna03} adapted the rotor problem of classical MHD shown by \cite{BS99} and \cite{Toth00} to the relativistic case. In a domain $[-0.5,0.5]\times[-0.5,0.5]$ in the $xy$-plane, a circular region of radius $r_c = 0.1$ rotates rigidly with an angular velocity $\omega_c = 9.95$, i.e., with linear velocities $(v^x,v^y)=\omega_c(-y,x)$. The central circle is homogeneous, with a density $\rho_c = 10$. The medium surrounding the central circle is also homogeneous and has a density $\rho = 1$. The thermal pressure, the magnetic field and the adiabatic index are constant everywhere, $p_c=1$, $(B^x,B^y)=(1,0)$, $\gamma=5/3$.

  We carry our computations until $t=0.4$ at resolutions of $128^2$, $256^2$, $512^2$ and $1024^2$ employing HLL, HLLC and FWD solvers. We use the third-order algorithm with a Courant number $0.6$. We point out that this test has a resolution-dependent complexity since the maximum Lorentz factor in the set up of the initial model depends on the resolution. Theoretically, the maximum Lorentz factor in the initial set up is reached at the circumference that bounds the central circular region (i.e., at $r=r_c$), where $W_{\rm max}^{\rm th}=(1-r_c^2\omega^2)^{-1/2}$. The finer the resolution, the closer is the maximum Lorentz factor in the grid to $W_{\rm max}^{\rm th}$, since the center of the cells crossed by the circle $r=r_c$ is closer to $r_c$.

\begin{figure*}
\begin{center}
\plotone{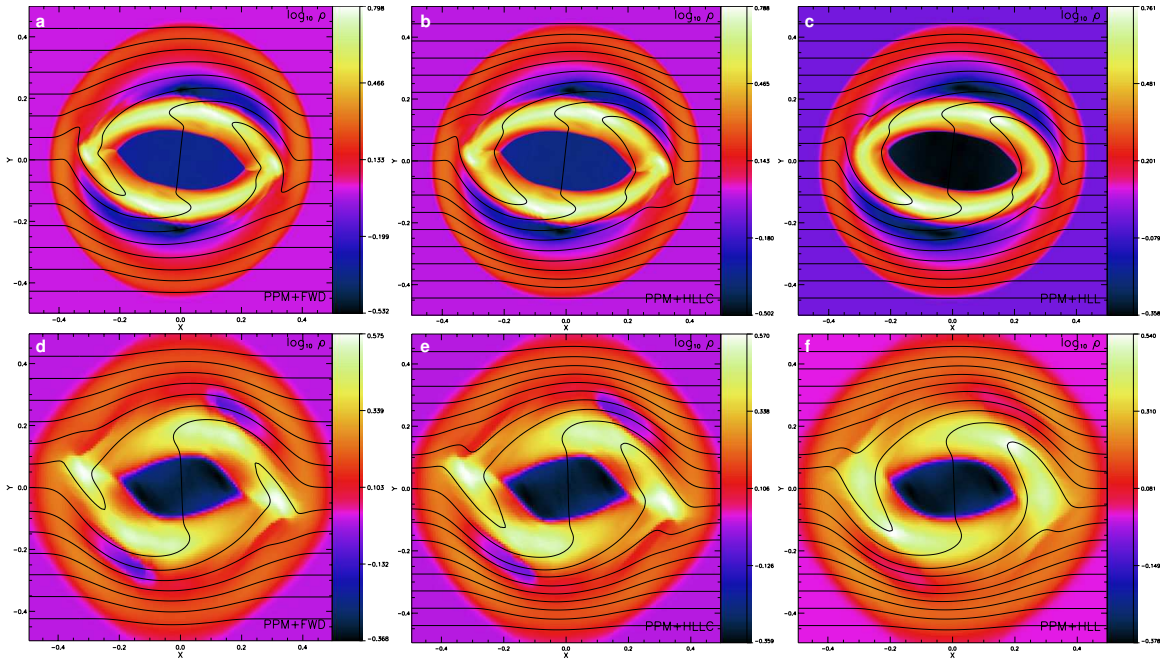}
 \caption{The rest-mass density distribution of the two-dimensional
   rotor problem at $t=4.0$. The upper (bottom) panels have been run
   with a resolution of $1024^2$ ($128^2$). Left, middle and right
   panels show the results obtained using FWD, HLLC and HLL solvers, respectively.}
 \label{fig:rotor}
\end{center}
\end{figure*}

  Even at the coarsest resolution employed ($128^2$), both solvers, HLLC (Fig.~\ref{fig:rotor}e) and FWD (Fig.~\ref{fig:rotor}d), capture adequately the two rotational discontinuities that are generated from the points $(x,y)=(0,\pm 0.1)$, where the magnetic field is initially parallel to the velocity field. The HLL solver smooths out such discontinuities, even at the highest resolution we have considered ($1024^2$; Fig.~\ref{fig:rotor}a).

  Both FWD and HLLC solvers yield very similar results independent of the numerical resolution, but HLLC is between a $40\%$ and a $70\%$ faster at high resolution. We note that increasing the resolution brings a reduced ratio $t_{\rm FWD}/t_{\rm HLL}$ (Table~2), which implies that it is more favorable to use a more accurate solver (FWD) at moderate or high resolution than a more diffusive one (HLL).

\section{Summary and conclusions}

  In the first part of the paper, we present renormalized sets of right and left eigenvectors of the flux vector Jacobians of the RMHD equations, which are regular and span a complete basis in any physical state including degenerate ones. Our starting point are the expressions of right and left eigenvectors in covariant variables (Sects.~\ref{ss:renom_eigen} and \ref{ss:renorm_leigen}). The renormalization procedure relies in the characterization of the degeneracy types in terms of the normal and tangential components of the magnetic field in the fluid rest frame (Sect.~\ref{ss:degs}). Type I (II) degeneracy occurs whenever the magnetic field is purely tangential (normal) to the Alfv\'en wave front. Then, proper expressions of the renormalized right and left eigenvectors in conserved variables are obtained through the corresponding matrix transformations (Sects.~\ref{s:recv} and \ref{s:lecv}). Our work completes previous analysis (Balsara 2001, Komissarov 1999, Koldoba et al. 2002) that present different sets of right eigenvectors for non-degenerate and degenerate states, and can be regarded as a relativistic generalization of the work performed by Brio \& Wu (1988) in classical MHD. Our theoretical analysis is completed with a brief note about the non-convex character of the relativistic MHD (Sect.~\ref{s:convex}).

  The second part of the paper deals with the development of a linearized Riemann solver based on the renormalized spectral decomposition of the equations just obtained. Here our work departs from previous works along the same line (Balsara 2001, Komissarov 1999, Koldoba et al. 2002) in that we build the numerical fluxes from the (renormalized) analytic right and left eigenvectors in conserved variables. Technical details of the numerical code are given in Sect.~\ref{numflux}. Intensive testing against one- and two-dimensional numerical problems (see Sect.~\ref{s:numtetsts}) allows us to conclude that our solver is very robust. When compared with a family of simpler solvers  (HLL, HLLC, HLLD) that avoid the knowledge of the full characteristic structure of the equations in the computation of the numerical fluxes, our solver turned out to be less diffusive than HLL and HLLC. This is clearly seen in the tests presented in Sect.~\ref{sec:contacts}, with both first and third order algorithms, that involve the description of isolated contact and Alfv\'en discontinuities. This is an expected result since Roe-type Riemann solvers are based in the full wave decomposition of Riemann problems. This conclusion is confirmed by the numerical results shown through Sects.~\ref{ss:st1}-\ref{ss:st4}. Comparison with HLLD solver is not as favorable.  Here, the expected superiority of Roe-type Riemann solvers with respect to central TVD schemes (see, e.g., the discussion in Toro 1997) is put into question since the increased accuracy of the five-wave HLLD solver on one hand, and the large amount of operations involved in the computation of numerical fluxes in the FWD solver on the other hand, lead to comparable numerical accuracies. This large amount of operations needed by the FWD solver makes it, in principle, less efficient computationally than those of the HLL family. However the relative efficiency of the FWD solver increases in multidimensional simulations since the number of zones needed to achieve a given accuracy is much smaller in the case of the FWD solver than in that of more diffusive solvers.

  It is well know, that choosing the adequate restrictions on characteristic variables serves optimally the purpose of setting boundary conditions. The computation of characteristic variables needs of the knowledge of the complete set of left and right eigenvectors. Thus, in spite of the efficiency and complexity arguments brandished in the previous paragraph, we point out that the set of computed left and right eigenvectors has the potentiallity of being used to properly set boundary conditions in a number of numerical applications. Work along this line will be presented elsewhere.

\section*{Acknowledgements}

  This research has been supported by the Spanish Ministerio de Educaci\'on y Ciencia and the European Fund for Regional Development through grants AYA2007-67626-C03-01 and AYA2007-67626-C03-02 and by the Generalitat Valenciana (grant Prometeo-2009-103). P.~M. acknowledges the support obtained through the grant CSD-2007-00050. M.~A.~A. is a Ram\'on y Cajal Fellow of the Spanish Ministry of Education and Science. The authors thankfully acknowledge the computer resources technical expertise and assistance provided by the Barcelona Supercomputing Center. We also thank the computer time obtained through the Spanish Supercomputing Network.

\vspace{6cm}

\appendix
\section{Characteristic wavespeeds diagrams and degeneracies}
\label{app_A}

  Contrarily to what happens in the case of (classical and relativistic) hydrodynamics, in a homogeneous magnetofluid at rest the presence of a magnetic field leads to a dependence on direction in the propagation speed of waves. This dependence can be visualized by means of the characteristic wavespeeds diagram, equivalent to the phase speed diagrams introduced by Friedrichs (see \citep{JT64}). These diagrams show the normal speed of planar wavefronts propagating in different directions, the speed is given by the distance between the origin and the normal speed surface along the corresponding direction.

  In a recent paper, Keppens and Meliani (2008) have revisited the theory for linear RMHD wave propagation showing the equivalence with the characteristic speed approach, and the connection between the phase and group speed diagrams by means of a Huygens construction, in arbitrary reference frames. In this Appendix, we discuss our results obtained in the representation of the characteristic speeds for fluids under different thermodynamical conditions, focussing in the layout of degeneracies as a function of the fluid's motion. Our results can be directly compared with those shown in Keppens and Meliani (2008).

\begin{figure*}
\begin{center}
\plotone{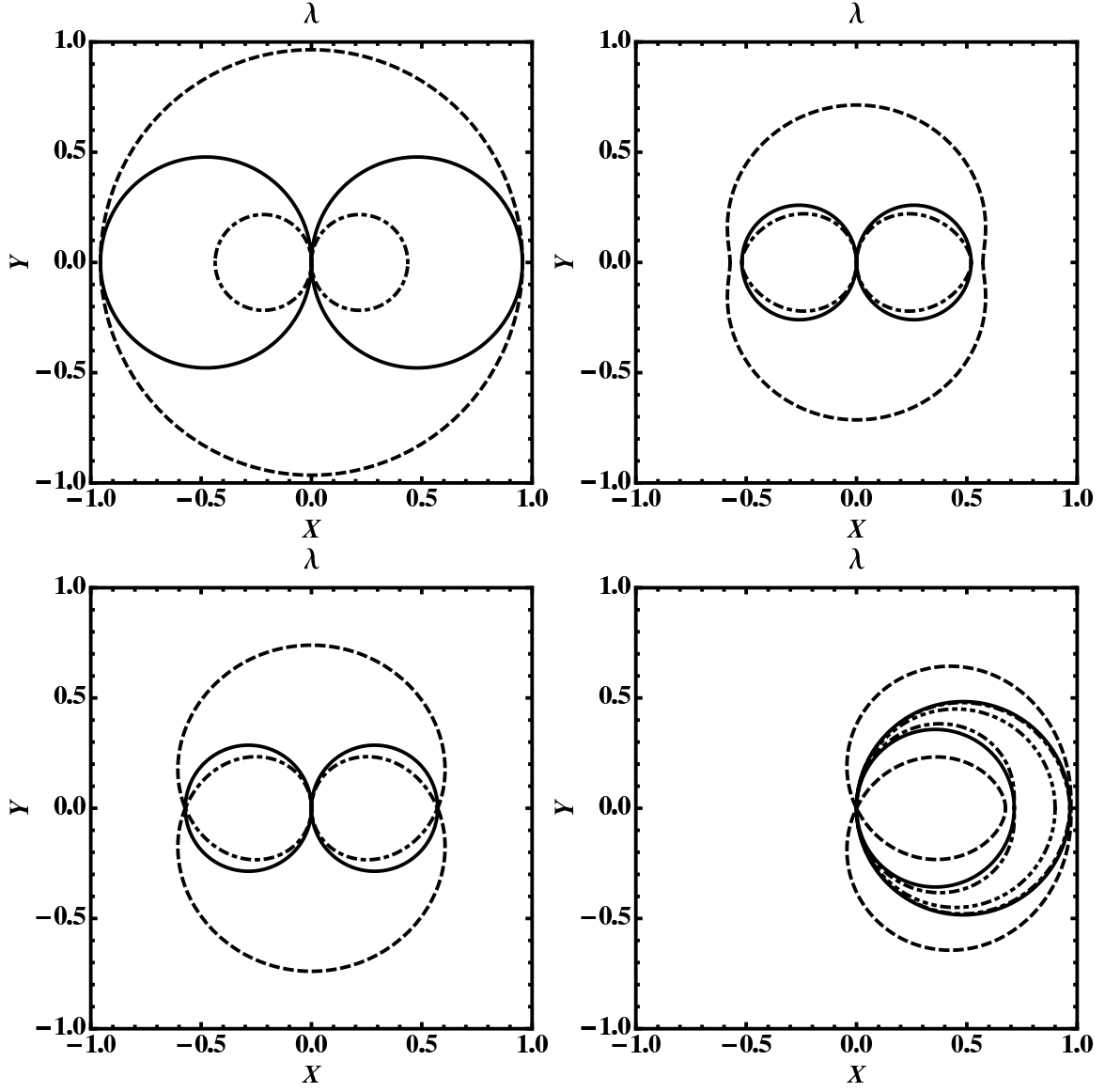}
\caption{Characteristic wavespeeds. Ideal gas with $\gamma = 4/3$.
Dashed lines correpond to fast magnetosonic waves, continuous lines to Alfv\'en wavespeeds, and dash-dotted lines to slow magnetosonic waves. Top-left panel: Fluid at rest with $\rho = 1.0$, $\varepsilon = 1.0$, $B^x = 5.0$, $B^y = 0.0$, $B^z = 0.0$. Top-right panel: Fluid at rest with  $\rho = 1.0$, $\varepsilon = 50.0$, $B^x = 5.0$, $B^y = 0.0$, $B^z = 0.0$. Bottom-left panel: Fluid at rest with $\rho = 1.0$, $\varepsilon = 37.864$, $B^x = 5.0$, $B^y = 0.0$, $B^z = 0.0$. In these three panels, the curve associated to the entropy wave degenerates in a point located at the origin, and Type I degeneracy is along $y$ axis, whereas the three subcases of Type II degeneracy are along the $x$ axis. Bottom-right panel: Fluid state as in the top-right panel but moving along the $x$-axis at a speed $v^x = 0.9$. Entropic wavespeed corresponds to dash-triple-dotted line. Type I and II degeneracies are again along $y$ and $x$ axis, respectively.}
\label{f:app:1-4}
\end{center}
\end{figure*}

  Top-left panel of Fig.~\ref{f:app:1-4} displays a cut in the $xy$-plane of the characteristic wavespeed surfaces corresponding to a homogeneous magnetized fluid at rest. In this case, the surfaces exhibit two symmetries. The first one is a rotation symmetry around the direction of the magnetic field, which justifies our choice of displaying a plane containing the magnetic field (along the $x$-axis in the figure). The second one is a mirror symmetry across the orthogonal plane to the magnetic field. The two types of degeneracy present in classical and relativistic MHD are found on this orthogonal plane (Type I) and along the magnetic field direction (Type II).  Top-left panel of Fig.~\ref{f:app:1-4} shows a particular subcase of Type II degeneracy, namely the one in which the speeds of the Alfv\'en waves coincide with the corresponding fast magnetosonic waves. Top-right and bottom-left panels of the same figure display the remaining two subcases (see the corresponding figure captions for the fluid conditions). In the top-right panel the speeds of Alfv\'en waves along the magnetic field direction coincide with those of slow magnetosonic waves. The triple degeneration case in which Alfv\'en, slow and fast magnetosonic waves propagate at the same speed along the magnetic field direction is shown in the bottom-left panel.

\begin{figure}
\begin{center}
\plotone{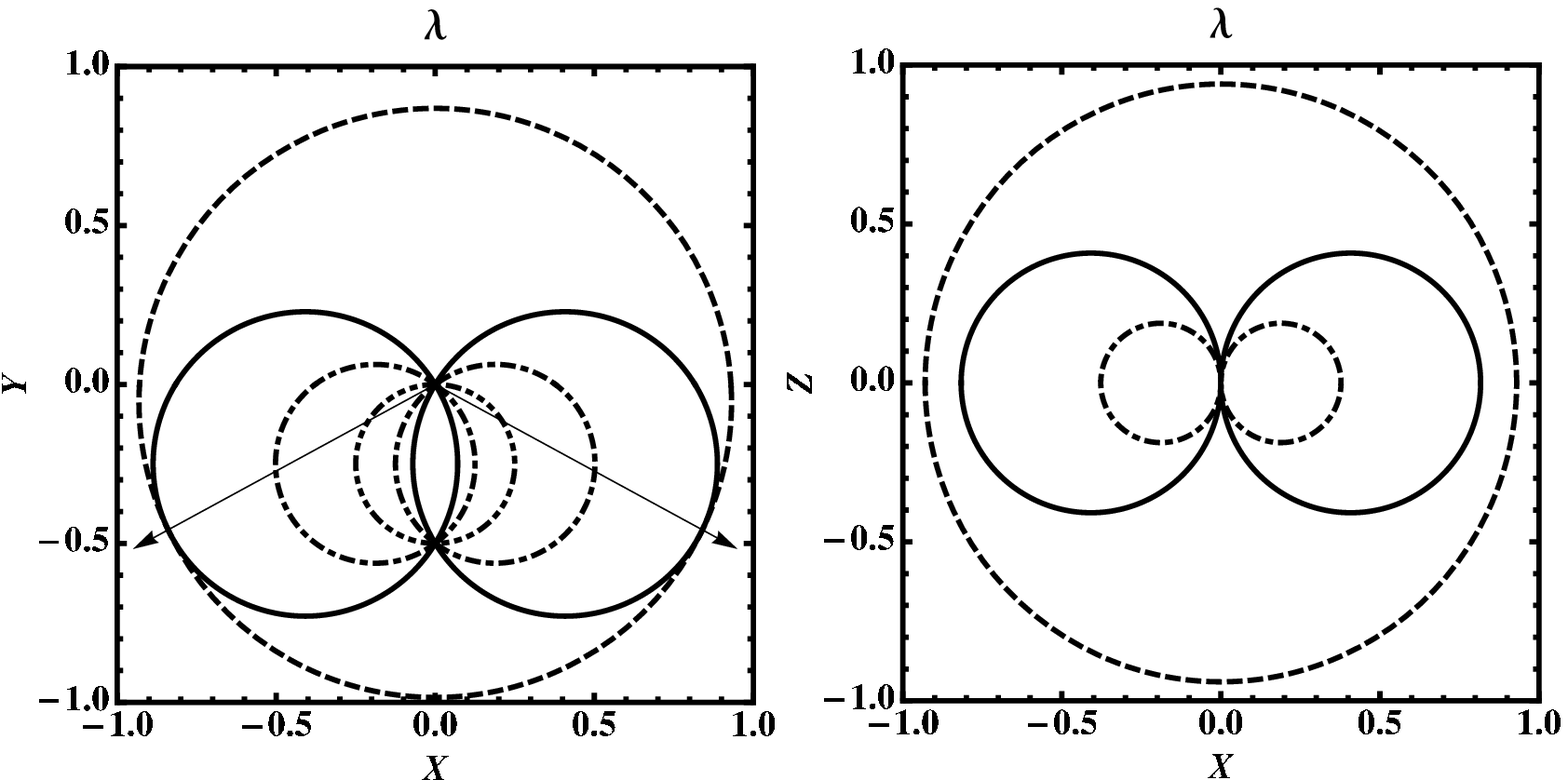}
\caption{Characteristic wavespeeds. Ideal gas with $\gamma = 4/3$, $\rho = 1.0$, $\varepsilon = 1.0$, $v^x = 0.0$, $v^y = -0.50$, $v^z = 0.0$, $B^x = 5.0$, $B^y = 0.0$, $B^z = 0.0$. Line types as in Fig.~\ref{f:app:1-4}. Left panel: $xy$-plane. Type I degeneracy is along $y$ axis. Type II degeneracy is along the directions pointed by the arrows. Right panel: $xz$-plane. Type I degeneracy is along $z$ axis. There is no Type II degeneracy on this plane.}
\label{f:app:5-6}  
\end{center}
\end{figure}

  For fluids in motion, the diagrams loose the symmetry properties described in the previous paragraph. In the case of motion along the direction of the magnetic field, as in the bottom-right panel of Fig.~\ref{f:app:1-4}, the diagram still displays the rotation symmetry around the magnetic field direction and the degeneracies of types I and II still occur in the orthogonal plane to the magnetic field and along the magnetic field, respectively. When the fluid velocity is not aligned with the magnetic field, the rotation symmetry around the magnetic field direction is lost. Figure~\ref{f:app:5-6} displays cuts of the characteristic wavespeeds surfaces on the $xy$ and $xz$ planes  for a magnetized fluid under suitable thermodynamical conditions moving along the $y$-axis (the magnetic field is still along the x axis). In this particular case in which the motion is orthogonal to the magnetic field direction, the mirror symmetry across the $yz$-plane is retained. Type I degeneracy is also found on this plane. Concerning Type II degeneracy, it is remarkable to note that the degeneracy is not present in the $xz$-plane, whereas in the $xy$-plane appears along two different directions. Moreover, contrarily to the classical MHD case, along each of these directions only one of the Alfv\'en waves is degenerate with the corresponding (fast) magnetosonic wave.

\begin{figure}
\begin{center}
\plotone{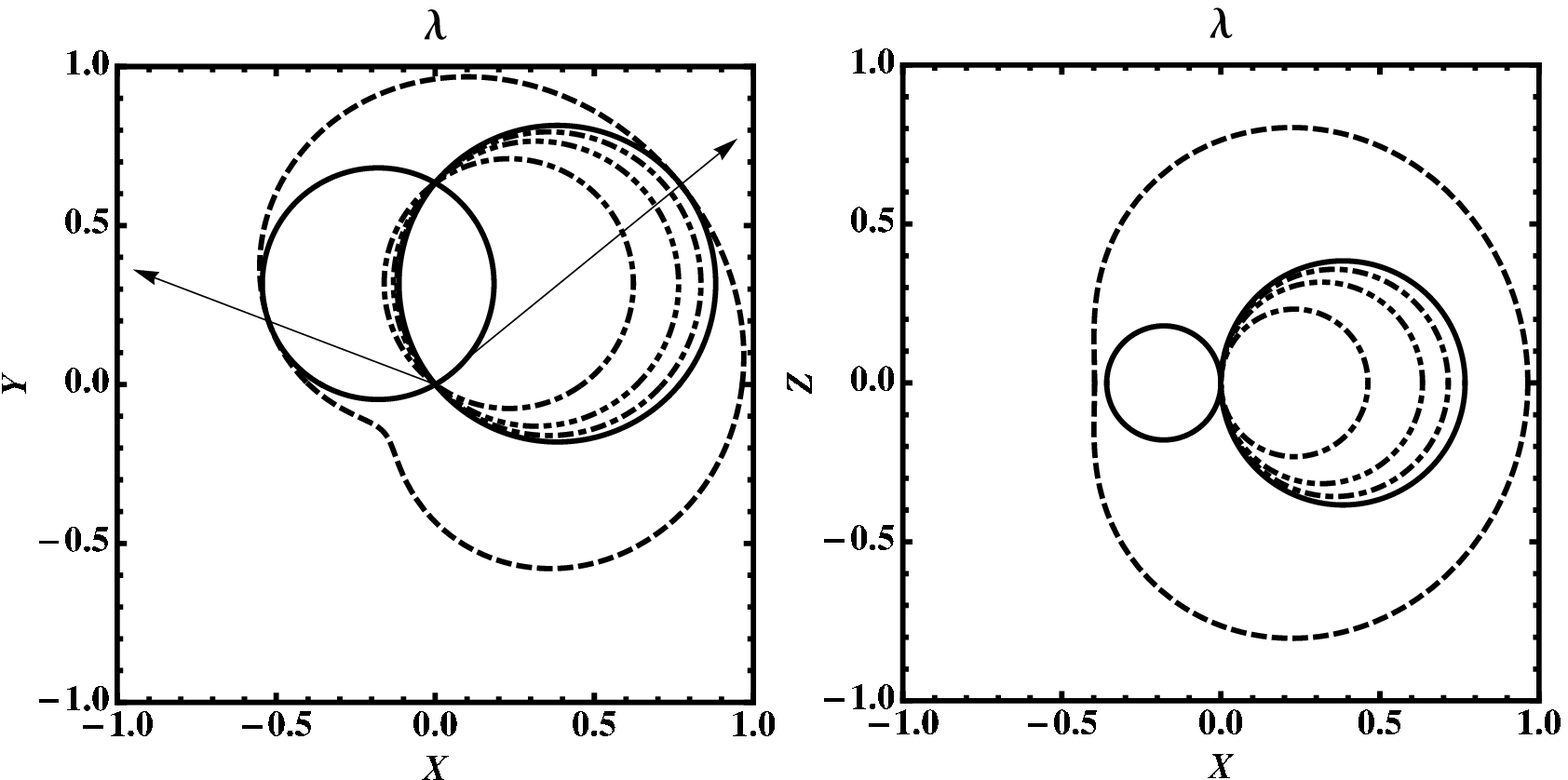}
\caption{Characteristic wavespeeds. Ideal gas with $\gamma = 4/3$, $\rho = 1.0$, $\varepsilon = 1.0$, $v^x = 0.634$, $v^y = 0.634$, $v^z = 0.0$, $B^x = 5.0$, $B^y = 0.0$, $B^z = 0.0$. Line types as in Fig.~\ref{f:app:1-4}. Left panel: $xy$-plane. Type I degeneracy is along $y$ axis. Type II degeneracy is along the directions pointed by the arrows. Right panel: $xz$-plane. Type I degeneracy is along $z$ axis. There is no Type II degeneracy on this plane.}
\label{f:app:7-8}
\end{center}
\end{figure}

  In the most general case (Fig.~\ref{f:app:7-8}) a mirror symmetry across the plane defined by the magnetic field and fluid velocity vectors (the $xy$-plane in these figures) is still retained. Again, Type II degeneracy appears along two different directions and only one of the Alfv\'en waves is affected by the degeneracy.

\section{Proof of the proportionality between the magnetosonic right eigenvector and the modulus of the tangential magnetic field}
\label{app_B}

  In this appendix we will proof that the components of the magnetosonic eigenvector, equation~(\ref{msrenorm}), are proportional to $|b_t|$. For a magnetosonic wave, and similarly to the Alfv\'en wave case, we can decompose the magnetic field in perpendicular and parallel part to the magnetosonic front. If we do so, $b^\alpha$ is written as

\begin{equation}
  b^\alpha=\frac{{\cal B}}{G+a^2}(\phi^\alpha+a u^\alpha)+b_t^\alpha,
\end{equation}

\noindent
and, from here,
 
\begin{equation} \label{apB:b2}
  b^2=\frac{{\cal B}^2}{G+a^2}+b_t^2.
\end{equation}    

\noindent
Substituting this decomposition in the quartic equation, ${\cal N}_4=0$, it is straightforward to obtain the following relation,

\begin{equation} \label{tranv1}
  \frac{a^2}{G} - \frac{{\mathcal B }^2}{\rho h (G +a^2)} = \frac{a^2  b_t^2}{ \rho h  \Big( a^2 -(G+a^2) c_s^2 \Big) }.
\end{equation}

    Let us now transform the components of the magnetosonic eigenvectors (equation~\ref{msrenorm}) using the above expressions. For the first four components we have

\begin{eqnarray} \label{ec011}
  \frac{a}{G}  ( \phi^{\nu} + a u^{\nu}) - \frac{ b^{\nu}}{\rho h} \left(\frac{\cal B}{a}\right) = \frac{a \,  b_t^2 }{  \rho h \Big(a^2  -c_s^2  (G +a^2) \Big) } ( \phi^{\nu} + a u^{\nu}) -\frac{ ({\cal B}/a) }{\rho h}   b^{\nu}_t .
\end{eqnarray}  

Similarly, for the following four components

\begin{eqnarray} 
  \left(\frac{\cal B}{a}\right)  \left( \phi^{\nu} \frac{a}{G} + \left(\frac{2 a^2}{G} - \frac{b^2}{\rho h} \right) u^{\nu} \right) - b^{ \nu} \left( 1 + \frac{ a^2}{G} \right) = \nonumber \\
  \left(\frac{\cal B}{a}\right) \frac{ b_t^2 \, (G+a^2 ) c_s^2 u^{\nu} }{\rho h \Big(a^2 -(G+a^2) c_s^2\Big) } - \left(1 + \frac{a^2}{G} \right) b_t^{\nu}.  
\end{eqnarray}

And for the ninth component of the eigenvector,         

\begin{equation}   \label{laC}
  b^2 - \rho h  \frac{a^2}{G} =  - b_t^2 \, \frac{ (G+a^2)c_s^2 }{ a^2 -(G+a^2) c_s^2} \quad.
\end{equation}

Since the tangential magnetic field $b_t^\nu$ can be written as the product of its modulus and a unitary vector, the proportionality is proven.

\bibliographystyle{apj}
\bibliography{ms}

%
%

%
%

%
%
%
%
%
%
\clearpage
%
%
%
%
%
%
%
%
%
\clearpage
%

%
%

%
%
\clearpage
%
\setlength\tabcolsep{1.0mm}
\begin{deluxetable}{lccccccccccc}
\tabletypesize{\scriptsize}
\tablecaption{Initial conditions for the test problems of Sect.~\ref{sec:1D}\label{tab:IVP}}
\tablewidth{0pt}
\tablehead{
  \colhead{Test}  & \colhead{State} & \colhead{$\rho$} &
  \colhead{$p_g$} & \colhead{$v^x$}  & \colhead{$v^y$}  &
  \colhead{$v^z$} & \colhead{$B^x$}  & \colhead{$B^y$}  &
  \colhead{$B^z$} & \colhead{Time}   & \colhead{Zones}
}
\startdata 
\multirow{2}{*}{Contact Wave}  
  & L  & $10$ & $1$ & $0$  & $0.7$ & $0.2$ & $5$ & $1$ & $0.5$ & \multirow{2}{*}{1} & \multirow{2}{*}{40}\\ 
  & R  &  $1$ & $1$ & $0$  & $0.7$ & $0.2$ & $5$ & $1$ & $0.5$ &  &\\ \hline
\multirow{2}{*}{Rotational Wave}  & 
    L  & $1$ & $1$ & $0.4$      & $-0.3$      & $0.5$      & $2.4$ & $1$    & $-1.6$ & \multirow{2}{*}{1} & \multirow{2}{*}{40}\\ 
  & R  & $1$ & $1$ & $0.377347$ & $-0.482389$ & $0.424190$ & $2.4$ & $-0.1$ & $-2.178213$ &   &\\ \hline
\multirow{2}{*}{Shock Tube 1}  & 
    L  & $1$     & $1$   & $0$  & $0$ & $0$ & $0.5$ & $ 1$ & $0$ & \multirow{2}{*}{0.4} & \multirow{2}{*}{400}\\ 
  & R  & $0.125$ & $0.1$ & $0$  & $0$ & $0$ & $0.5$ & $-1$ & $0$ &   &\\ \hline
\multirow{2}{*}{Shock Tube 2}  & 
    L  & $1.08$ & $0.95$ & $0.4$   & $0.3$  & $0.2$ & $2$ & $0.3$  & $0.3$ & \multirow{2}{*}{0.55} & \multirow{2}{*}{800}\\ 
  & R  & $1$    & $1$    & $-0.45$ & $-0.2$ & $0.2$ & $2$ & $-0.7$ & $0.5$ &  &\\ \hline
\multirow{2}{*}{Shock Tube 3}  & 
    L  & $1$ & $0.1$ & $ 0.999$ & $0$ & $0$ & $10$ & $ 7$  & $7$  & \multirow{2}{*}{0.4} & \multirow{2}{*}{400}\\ 
  & R  & $1$ & $0.1$ & $-0.999$ & $0$ & $0$ & $10$ & $-7$  & $-7$ &   &\\ \hline
\multirow{2}{*}{Shock Tube 4}  & 
    L  & $1$   & $5$   & $0$ & $0.3$ & $0.4$ & $1$ & $6$ & $2$ & \multirow{2}{*}{0.5} & \multirow{2}{*}{800}\\ 
  & R  & $0.9$ & $5.3$ & $0$ & $0$   & $0$   & $1$ & $5$ & $2$ &   &\\ \hline
  \multirow{2}{*}{UR shell}  & 
  L  & $10^{-3}$   & $10^{-5}$   & $0$ & $0$ & $0$ & $0$ & $0$ & $0$ &
  \multirow{2}{*}{3.5} & \multirow{2}{*}{32000}\\ 
  & M  & $1$ & $2.5\times10^{-3}$ & $0.997775$ & $0$   & $0$   & $0$ & $4.60706$ & $0$ &   &\\
  & R  & $3.297\times10^{-4}$ & $8.2424\times10^{-7}$ & $0.997775$ & $0$   & $0$   & $0$ & $0$ & $0$ &   &\\ 
\enddata
\tablecomments{The last two columns give, respectively, the time at which the solution is shown in the figures and the number of numerical zones used in the computation. The last test is set up in spherical coordinates, and the vector components $(x,y,z)$ correspond to the spherical ones $(r,\phi,\theta)$}
\end{deluxetable}
%
\clearpage
%
\setlength\tabcolsep{1.5mm}
\begin{deluxetable}{ccc}
\tablecaption{CPU times for the two-dimensional rotor problem.\label{tab:CPU-times-rotor}}
\tablewidth{0pt}
\tablehead{
\colhead{Resolution} & 
\colhead{$t_{\rm HLLC} / t_{\rm HLL}$} & 
\colhead{$t_{\rm FWD}/ t_{\rm HLL}$}
}
\startdata
$128^2$    & 1.75  & 2.53 \\ 
$256^2$    & 1.45  & 2.84 \\ 
$512^2$    & 1.08  & 1.50 \\ 
$1024^2$   & 1.13  & 1.93 \\
\enddata
\tablecomments{Execution times for test cases run with the HLLC and FWD solvers are normalized to those corresponding to the HLL solver. In all cases the third-order scheme with a Courant number 0.3 has been used}
\end{deluxetable}

\end{document}